\documentclass[aps,twocolumn,superscriptaddress,floatfix,showpacs,amsmath,longbibliography,nofootinbib,amssymb]{revtex4-2}
\usepackage[colorlinks=true,citecolor=blue,linkcolor=blue]{hyperref}
\usepackage[usenames]{color}
\usepackage{subfigure}
\usepackage{dcolumn}
\usepackage{mathrsfs}
\usepackage{bm}
\usepackage{amsmath,amssymb,epsfig,float,graphics}
\newcommand{\bk}{\mathbf{k}}
\newcommand{\br}{\mathbf{r}}
\newcommand{\bR}{\mathbf{R}}
\newcommand{\cA}{\mathcal{A}}
\newcommand{\cH}{\mathcal{H}}
\DeclareMathOperator{\erf}{erf}

\begin{document}
\baselineskip=0.45 cm

\title{Harvesting entanglement from the Lorentz-violating quantum field vacuum in a dipolar Bose-Einstein condensate}

\author{Zehua Tian}
\email{tzh@hznu.edu.cn}
\affiliation{School of Physics, Hangzhou Normal University, Hangzhou, Zhejiang 311121, China}

\author{Weiping Yao}
\email{yao11a@126.com}
\affiliation{Department of physics and electrical engineering, Liupanshui Normal University, Liupanshui 553004, Guizhou, China}

\author{Xiaobao Liu}
\affiliation{Department of physics and electrical engineering, Liupanshui Normal University, Liupanshui 553004, Guizhou, China}

\author{Mengjie Wang}
\affiliation{Department of
Physics, Key Laboratory of Low Dimensional Quantum Structures and
Quantum Control of Ministry of Education, and Synergetic Innovation
Center for Quantum Effects and Applications, Hunan Normal
University, Changsha, Hunan 410081, China}

\author{Jieci Wang}
\email{jcwang@hunnu.edu.cn}
\affiliation{Department of
Physics, Key Laboratory of Low Dimensional Quantum Structures and
Quantum Control of Ministry of Education, and Synergetic Innovation
Center for Quantum Effects and Applications, Hunan Normal
University, Changsha, Hunan 410081, China}

\author{Jiliang Jing}
\affiliation{Department of
Physics, Key Laboratory of Low Dimensional Quantum Structures and
Quantum Control of Ministry of Education, and Synergetic Innovation
Center for Quantum Effects and Applications, Hunan Normal
University, Changsha, Hunan 410081, China}

\begin{abstract}
We theoretically propose an experimentally viable scheme to explore the transfer of nonclassical correlations from a dipolar Bose-Einstein condensate (BEC) to
a pair of impurities immersed in it. Operating at ultra-low temperature, density fluctuations of the dipolar BEC emulate a vacuum field with 
Lorentz-violating dispersion, while the two impurities function as Unruh-DeWitt detectors for the BEC quasiparticles. We study the harvesting of 
entanglement from the quantum vacuum of this analogue Lorentz-violating quantum field by spatially separated Unruh-DeWitt detectors. 
Our analysis reveals key parameter dependencies that optimize the harvesting of entanglement. In particular, 
unlike the Lorentz-invariant case, smoother detector switchings does not enhance the entanglement harvesting 
efficiency from the Lorentz-violating quantum field vacuum. Moreover, the strength of the Lorentz-invariant violation 
can shift the optimal energy structure of the detectors for harvesting entanglement from the Lorentz-violating quantum field vacuum---a clear deviation from 
the Lorentz-invariant scenario. As a fundamental quantum mechanical setup,
 our quantum fluid platform provides an experimentally realizable testbed for examining the entanglement harvesting protocol from an effective Lorentz-violating quantum field vacuum using a pair of impurity probers, which may also has potential implications for exploring the Lorentz-invariant violation in quantum field theory.
\end{abstract}

\baselineskip=0.45 cm
\maketitle
\newpage

\section{Introduction} 
Vacuum of a free quantum field does not mean ``nothing", rather, it exhibits rich physical phenomena, e.g., including the presence of entanglement \cite{Entanglement1, VALENTINI1991321, Entanglement2, PhysRevA.71.042104, PhysRevA.75.052307, PhysRevD.79.044027, Salton_2015}. 
Within the relativistic framework, this vacuum entanglement serves as a fundamental ingredient in several intriguing phenomena, such as Unruh effect \cite{PhysRevD.14.870} and Masahiro Hotta's
quantum energy teleportation \cite{doi:10.1143/JPSJ.78.034001, hotta2011, PhysRevA.89.012311, PhysRevA.82.042329, PhysRevLett.130.110801, PhysRevApplied.20.024051, PhysRevD.107.L071502, PhysRevA.110.052424, Wang2024quantumenergy}.  
It also lies at the core of a number of enduring open problems in theoretical physics, including the black hole information loss paradox  \cite{preskill1992blackholesdestroyinformation, giddings1995black, Mathur_2009, RAJU20221} 
and its proposed solutions—most notably, the ``black hole firewalls" and the principle of black hole complementarity \cite{PhysRevD.48.3743, firewalls, PhysRevLett.110.101301}. In particular, theoretical studies have predicted that such entanglement can be transferred to two or more spatially separated local Unruh-DeWitt detectors that interact with the vacuum state \cite{VALENTINI1991321, PhysRevA.71.042104, PhysRevA.75.052307}. Moreover,
this extraction of entanglement still remains possible even when the detectors are spacelike separated \cite{ PhysRevA.71.042104, PhysRevA.75.052307, PhysRevD.79.044027, Salton_2015, PhysRevD.92.064042}.
This process, known as \emph{entanglement harvesting} in the Relativistic Quantum Information (RQI) protocol, has recently attracted extensive 
attention for exploring various physics \cite{Wu-etanglement, PhysRevD.110.105016, Wu2, PhysRevD.106.025010, PhysRevD.106.045005, PhysRevD.106.076002, PhysRevD.105.085012, PhysRevD.104.025001, PhysRevD.103.065013, PhysRevD.97.125011, zhao}---for instance, to explore the structure of the background spacetime \cite{PhysRevD.79.044027, PhysRevD.93.044001, Martin2012, Martin2014, PhysRevD.97.105030, Tian-Casimir, Tian-Casimir-Polder, PhysRevD.111.104052, Yu-Zhang, PhysRevD.108.085025}. Although the process of entanglement harvesting is theoretically well-understood, so far its experimental implementation remains unrealized.

On the other hand, Lorentz invariance (LI), one of the fundamental symmetries of relativity, is expected to break down in many theories of quantum gravity 
at sufficiently high energies \cite{AmelinoCamelia:2008qg}. Such LI violation can significantly alter the properties of the quantum vacuum, leading to intriguing 
phenomena absent in the LI case---
such as challenging the equivalence principle \cite{PhysRevD.104.124001}, exciting uniformly moving detectors in the Minkowski vacuum \cite{Kajuri_2016, PhysRevLett.116.061301, KAJURI2018412, PhysRevD.97.025008, PhysRevD.103.085014, wu2024geometricphaseassisteddetection}, and resulting in a non-thermal Unruh effect \cite{Agullo_2010, PhysRevD.92.024018, PhysRevLett.123.041601, Hossain_2016, Unruh10, GUP, PhysRevD.106.L061701, xu2025momentumresolvedprobinglorentzviolatingdispersion}. Any discovery of LI violation would thus signal physics beyond the standard model, motivating extensive 
searches across diverse physics systems  \cite{wu2024geometricphaseassisteddetection, PhysRevD.55.6760, Zhang:2023wwk, Quan:2025tgz, PhysRevD.58.116002, Mattingly:2005re, RevModPhys.83.11}.
Recently, Unruh-DeWitt detector \cite{PhysRevD.14.870, birrell_davies_1982, 10.1143/PTP.88.1, RevModPhys.80.787, Hu_2012} has been used to explore the various properties of vacuum in Lorentz-violating quantum field theory \cite{Kajuri_2016, KAJURI2018412, PhysRevD.97.025008, PhysRevLett.116.061301, PhysRevLett.123.041601, PhysRevD.106.L061701}, while the role of LI violation in entanglement harvesting remains largely unexplored \cite{liu2025harvestingcorrelationsbtzblack}. Moreover, experimental progress in this direction has so far remained elusive.

\begin{figure}
\centering
\includegraphics[width=0.28\textwidth]{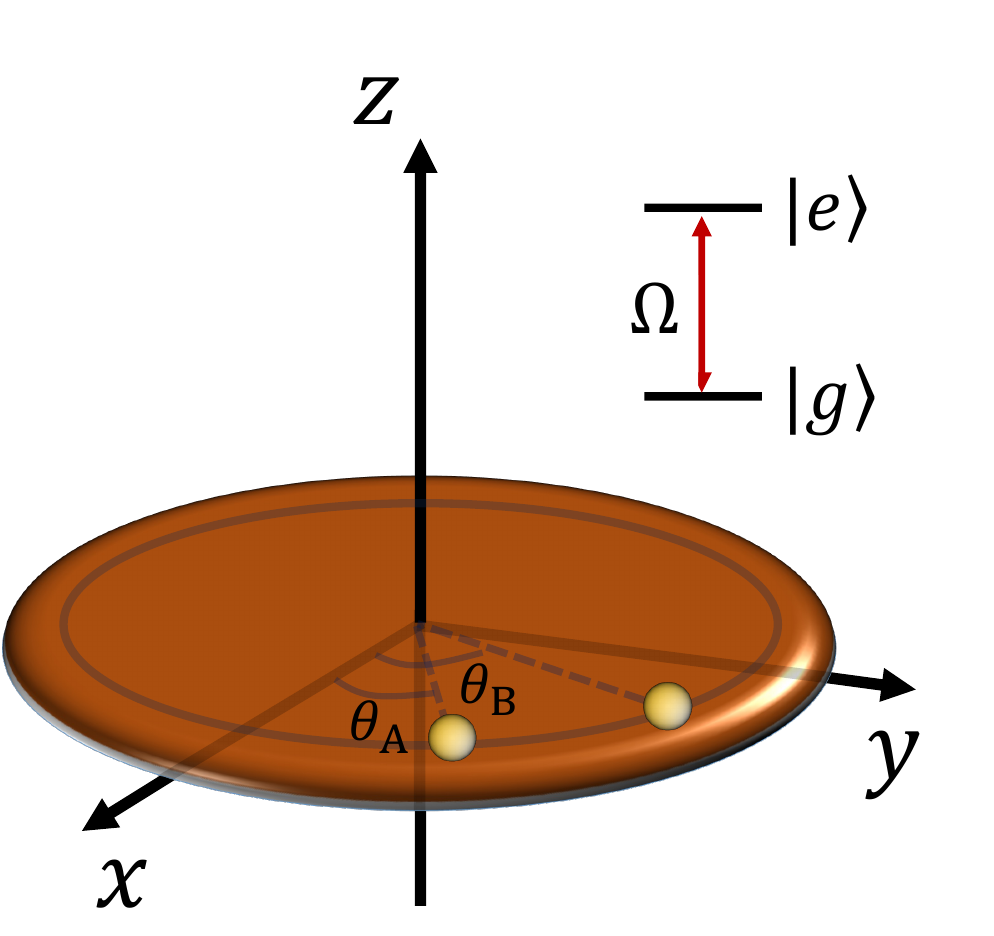}
\caption{Schematic of two impurities with effective internal frequency $\Omega$ immersed in a quasi-two-dimensional dipolar condensate. $|e\rangle$ and $|g\rangle$ denote the effective excitation and ground state of the impurity, respectively. The two impurities act as two 
Unruh-DeWitt detectors, denoted as A and B, at separate locations, $(r, \theta_\text{A})$ and $(r, \theta_\text{B})$, respectively.}\label{fig1}
\end{figure}

In this paper, we aim at closing this gap by probing the relevant fundamental properties of
Lorentz-violating quantum field vacuum states, using an experimentally feasible platform consisting of a dipolar BEC \cite{BARANOV200871} 
and two immersed impurities \cite{PhysRevLett.94.040404, PhysRevLett.91.240407}. 
Concretely, we show that the density fluctuations in the dipolar BEC---featuring a
roton spectrum \cite{PhysRevLett.98.030406, PhysRevA.97.063611, PhysRevLett.118.130404, PhysRevA.73.031602, PhysRevLett.90.250403}
due to the dipole-dipole interaction (DDI) between atoms (see below for a detailed discussion)---constitute the analog Lorentz-violating quantum field. Furthermore, the two impurities, analogously dipole coupled to the density fluctuations in the condensate, 
act as two Unruh-DeWitt detectors at separate locations coupling to the Lorentz-violating quantum field, as depicted in Fig. \ref{fig1}.
We will demonstrate the harvesting of entanglement from the quantum vacuum of the analogue Lorentz-violating quantum field to the spatially separated 
Unruh-DeWitt detectors.
We explore how the Lorentz-violating strength and other dependent parameters of the detectors, such as their spatial distance, affect the entanglement harvesting, with a particular focus on comparing the relevant results to the LI vacuum scenario to advance the understanding of LI violation physics.

Our proposed quantum fluid platform offers a tabletop, experimentally realizable test bed to examine the entanglement harvesting protocol of Lorentz-violating physics, thereby providing crucial guidance for exploring the LI violation in quantum field theory.
Our study may strengthen the connection between experimentally implemented particle detectors in dipolar BEC and the idealized Unruh-DeWitt detector models commonly employed in RQI, as well as between the Lorentz-violating quasiparticle experimentally studied in dipolar BEC and various  Lorentz-violating quantum field predicted  by possible quantum gravity theories. Furthermore, continuous, nondestructive measurement of a BEC via local probes is exploited here, thus our approach offers a complementary alternative to existing destructive measurements that aim at mapping out the global entanglement structure of BEC. It is also worth noting that the Unruh-DeWitt detector model has already been constructed in dipolar BEC to explore various physical effects arising from  Lorentz-violating quantum fields \cite{PhysRevD.103.085014, PhysRevD.106.L061701}.

Our paper is constructed as follows. In Sec. \ref{section2} we introduce our model---the dipolar BEC, and
demonstrate how a Lorentz-violating quantum field can be analogously constructed from the density fluctuations of the condensate. In Sec. \ref{section3}
we present the basic formulas governing the quantum state parameters of 
two Unruh-DeWitt detectors, which are simulated by two impurities immersed in the dipolar BEC. 
In Sec. \ref{section4} we investigate entanglement harvesting for Unruh-DeWitt detectors in the vacuum of the analogue Lorentz-violating quantum field. 
The experimental feasibility of the relevant entanglement harvesting physics within the current technologies of 
dipolar BEC is discussed in Sec. \ref{section5}. Finally, a summary of the main results of our work is provided in Sec. \ref{section6}.

\section{Analogue Lorentz-violating quantum field in dipolar Bose-Einstein condensate} \label{section2}
We consider an interacting Bose gas consisting of atoms or molecules with mass $m$, whose Lagrangian density reads $(\hbar=1)$
\begin{eqnarray}\label{Lagrangian}
\nonumber
\mathcal{L}&=&\frac{i}{2}(\Psi^\ast\partial_t\Psi-\partial_t\Psi^\ast\Psi)-\frac{1}{2m}|\nabla\Psi|^2-V_\text{ext}|\Psi|^2
\\
&&-\frac{1}{2}|\Psi|^2\int\,d^3\bR^\prime\,V_\text{int}(\bR-\bR^\prime)|\Psi(\bR^\prime)|^2.
\end{eqnarray}
Here the system is assumed to be trapped by an external potential of the form $V_\text{ext}(\bR)=m\omega^2\br^2/2+m\omega^2_zz^2/2$, with 
$\bR=(\br, z)$ being spatial three-dimensional coordinates. Moreover, we also assume that over the whole time evolution
the gas is strongly confined in the $z$ direction, with aspect ratio $\kappa=\omega_z/\omega\gg1$. The last term in Eq. \eqref{Lagrangian} denotes 
the two-body interaction, containing two terms---the contact interaction and the dipolar interaction. Specifically, this total interaction term is given by 
\begin{eqnarray}\label{two-body-interaction}
V_\text{int}(\bR-\bR^\prime)=g_c\delta^3(\bR-\bR^\prime)+V_\text{dd}(\bR-\bR^\prime),
\end{eqnarray}
where $g_c$ is the contact interaction coupling, and 
\begin{eqnarray}
V_\text{dd}(\bR-\bR^\prime)=\frac{3g_d}{4\pi}\frac{[1-3(z-z^\prime)^2/|\bR-\bR^\prime|^2]}{|\bR-\bR^\prime|^3}
\end{eqnarray}
describes the dipolar interaction with coupling constant $g_d$. It is worth noting that these two parameters $g_c$ and $g_d$ can be controlled as required in the experiment. 
In addition, the dipoles here have been assumed to be polarized along the $z$ direction (or perpendicular to the $x$-$y$ plane) by an external field. 
As shown in the following, it is the interaction between atoms or molecules (including both the contact and dipolar interaction) that leads to 
the Lorentz-violating quasiparticle spectrum, similar to the case in Ref. \cite{EDWARDS2018319}. Therefore, in principle, through controlling the parameters $g_c$ and $g_d$,
various types of Lorentz-violating quasiparticle spectrum could be implemented experimentally. In order to ensure the stability in the DDI-dominated regime \cite{PhysRevA.73.031602}, we require the system to remain sufficiently close to the quasi-two-dimensional (quasi-2D) regime throughout the process considered.
In such case, we assume that in the $z$ direction, the condensate density has a Gaussian form,
$\rho_z(z)=(\pi\,d^2_z)^{-1/2}\exp[-z^2/d^2_z]$, with $d_z=\sqrt{1/m\omega_z}$. Therefore, integrating out the $z$ dependence, we can obtain 
the effective quasi-2D interaction, 
$V^\text{2D}_\text{int}(\br-\br^\prime)=\int\,dzdz^\prime\,V_\text{int}(\bR-\bR^\prime)\rho_z(z)\rho_z(z^\prime)$ \cite{PhysRevLett.118.130404}.

In the 2D case, the reduced atomic field operator $\hat{\psi}$ can be decomposed as 
\begin{eqnarray}
\hat{\psi}=\psi_0(1+\hat{\phi}),
\end{eqnarray}
where $\psi_0=\sqrt{\rho_0}e^{i\theta_0}$, and correspondingly the 2D condensate density reads $|\psi_0(\br)|^2=\rho_0\simeq\mathrm{const}$. Besides,
the operator $\hat{\phi}$ denotes the perturbations (excitations) on the top of the condensate. 
The dynamics of such fluctuation field is governed by the Bogoliubov-de Gennes equation  \cite{2001camw.book1C, PhysRevA.97.063611},
\begin{eqnarray}\label{Bogoliubov-eq}
\nonumber
i\partial_t\hat{\phi}&=&-\frac{1}{2m}\nabla^2_\br\hat{\phi}
+\rho_0\int\,d^2\br^\prime\,V^\text{2D}_\text{int, 0}(\br-\br^\prime)
\\
&&\times\big[\hat{\phi}(\br^\prime)+\hat{\phi}^\dagger(\br^\prime)\big].
\end{eqnarray}
Here the condensate has been assumed to be 
static, i.e., $\mathbf{v}=\frac{1}{m}\nabla_\br\theta_0=0$.
Solving Eq. \eqref{Bogoliubov-eq}, one can find that the corresponding density fluctuations in Heisenberg representation can be written as 
\begin{eqnarray}\label{DF}
\nonumber
\delta\hat{\rho}(t,\br)&\simeq&\rho_0(\hat{\phi}+\hat{\phi}^\dagger)
\\  \nonumber
&=&\sqrt{\rho_0}\int[d\bk/(2\pi)^2](u_\bk+v_\bk)\times[\hat{b}_\bk(t)e^{i\bk\cdot\br}
\\
&&+\hat{b}^\dagger_\bk(t)e^{-i\bk\cdot\br}].
\end{eqnarray}
It closely resembles the quantized Lorentz-violating field in terms of bosonic operators 
$\hat{b}_\bk(t)=\hat{b}_\bk\,e^{-i\omega_\bk\,t}$ satisfying the usual Bose commutation rules 
$[\hat{b}_\bk, \hat{b}_{\bk^\prime}^\dagger]=(2\pi)^2\delta^2(\bk-\bk^\prime)$, that destroy a collective excitations of the condensate with wave vector $\bk$. 
Besides, the parameters $u_\bk$ and $v_\bk$ in Eq. \eqref{DF} denote Bogoliubov parameters, which are given by
\begin{eqnarray}
\nonumber
u_\bk&=&(\sqrt{\cH_\bk}+\sqrt{\cH_\bk+2\cA_\bk})/2(\cH^2_\bk+2\cH_\bk\cA_\bk)^{1/4},
\\
v_\bk&=&(\sqrt{\cH_\bk}-\sqrt{\cH_\bk+2\cA_\bk})/2(\cH^2_\bk+2\cH_\bk\cA_\bk)^{1/4},
\end{eqnarray}
satisfying $u_\bk^2-v_\bk^2=1$.
Here $\cH_\bk=k^2/2m$ and $\cA_\bk=\rho_0V^\text{2D}_\text{int, 0}(k)$ \cite{PhysRevA.97.063611}, with $k=|\bk|$. The quasiparticle frequency 
$\omega_\bk=\sqrt{\cH^2_\bk+2\cH_\bk\cA_\bk}$, and the Fourier transformation of the effective quasi-2D interaction is given by \cite{PhysRevA.73.031602}
\begin{eqnarray}
V^\text{2D}_\text{int, 0}(k)=g^\text{eff}_0(1-\frac{3R}{2}kd_zw[\frac{kd_z}{\sqrt{2}}]), 
\end{eqnarray}
with $w[x]=\exp[x^2](1-\erf[x])$,
an effective contact coupling $g^\text{eff}_0=\frac{1}{\sqrt{2\pi}d_z}(g_c+2g_d)$, and the dimensionless ratio being defined as
\begin{eqnarray}
R=\sqrt{\pi/2}/(1+g_c/2g_d).
\end{eqnarray}
In experiment, the contact interaction and the dipolar interaction can be tunable via Feshbach resonance \cite{PhysRevLett.81.69, Inouye1998Observation} and rotating polarizing field \cite{PhysRevLett.89.130401}, thus the parameter $R$ could be tunable experimentally. 
It ranges from $R=0$ (when $g_d/g_c\rightarrow0$, i.e., contact dominance), to $R=\sqrt{\pi/2}$ (when $g_d/g_c\rightarrow\infty$, i.e., DDI dominance).

\begin{figure}
\centering
\includegraphics[width=0.3\textwidth]{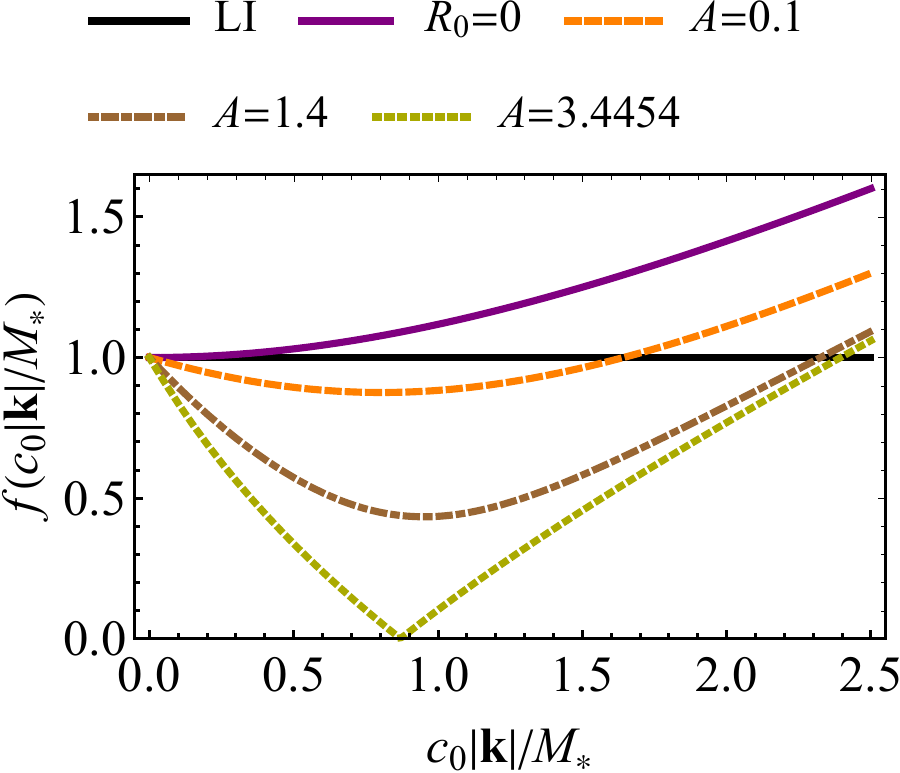}
\caption{The dimensionless function $f$ shown in \eqref{Dispersion} as a function of $c_0|\bk|/M_\ast$.
LI denotes Lorentz invariant case with $f=1$, such that $\omega_\bk=c_0|\bk|$. $R=0$ denotes the contact interaction case, where $f$ is independent of $A$. 
For DDI dominance case, we take $R=\sqrt{\pi/2}$, in such case $f$ could dip below 1 for an interval of $c_0|\bk|/M_\ast$. Note that 
$f$ becomes negative when $A>A_c=3.4454$, which means the spectrum of quasiparticle 
becomes unstable.}\label{fig2}
\end{figure}

The explicit dispersion relation of the density fluctuations shown in Eq. \eqref{DF} that closely resemble a Lorentz-violating scalar field 
reads 
\begin{eqnarray}\label{Dispersion}
\nonumber
\omega_\bk&=&c_0|\bk|\sqrt{1-\frac{3R}{2}\sqrt{A}\frac{c_0|\bk|}{M_\ast}w\bigg[\sqrt{\frac{A}{2}}\frac{c_0|\bk|}{{M_\ast}}\bigg]
+\frac{1}{4}\frac{c^2_0|\bk|^2}{M^2_\ast}}
\\
&=&c_0|\bk|f(c_0|\bk|/M_\ast).
\end{eqnarray}
Here $c_0=\sqrt{g^\text{eff}_0\rho_0/m}$ denotes the speed of sound, the dimensionless parameter
\begin{eqnarray}
A=g^\text{eff}_0\rho_0/\omega_z
\end{eqnarray}
represents the effective chemical potential as measured relative to the transverse trapping,  and 
\begin{eqnarray}
M_\ast=mc^2_0
\end{eqnarray}
is the analog energy scale of Lorentz violation, only well below which, i.e., only when $c_0|\bk|/M_\ast\ll1$, the dispersion relation in Eq. \eqref{Dispersion} can be approximately reduced to the LI case $(f(c_0|\bk|/M_\ast)\simeq1)$. In particular, by appropriately setting the relevant parameters $A$ and $R$, 
the dispersion could be analogously superluminal $(f(c_0|\bk|/M_\ast)>1)$ and subluminal $(f(c_0|\bk|/M_\ast)<1)$.  
Here we need to note that if our condensate system is considered to be in the DDI dominance regime, and in such case 
$A$ should be assumed to be not larger than the critical value $A_c=3.4454$, or the corresponding spectrum of quasiparticle becomes unstable \cite{PhysRevA.97.063611, PhysRevA.73.031602, PhysRevLett.118.130404}. Therefore, in order to keep the stability of the spectrum of quasiparticle in the 
DDI dominance regime, $A\leq\,A_c$ will be taken throughout the whole paper and this condition 
requires that the frequency of the confinement along the $z$-direction satisfies $\omega_z\geq\frac{2mg^2_d\rho^2_0}{\pi\,A^2_c}$.

In Fig. \ref{fig2}, we plot the function $f(c_0|\bk|/M_\ast)$ shown in Eq. \eqref{Dispersion} as a function of $c_0|\bk|/M_\ast$.
We can find that for the DDI dominance case $(R=\sqrt{\pi/2})$, the analogous subluminal spectrum develops a roton minimum 
for sufficiently large $A$, and the LI is strongly broken near $c_0|\bk|/M_\ast\simeq0.9$ \cite{PhysRevA.97.063611}. Alternatively, 
$f(c_0|\bk|/M_\ast)$ could dip below $1$ for an interval of $c_0|\bk|/M_\ast$. Moreover, the larger $A$ is, the further the Lorentz-violating spectrum deviates from the LI one. 
For the contact interaction case, i.e., $R=0$ case,  
$f(c_0|\bk|/M_\ast)$ is bigger than 1, which denotes the analogous superluminal spectrum case. 
We note that based on the similarity between the density fluctuations in Eq. \eqref{DF} and a Lorentz-violating scalar field, there has been colorful 
application to the exploration of Lorentz-violating physics in various areas \cite{PhysRevD.103.085014, PhysRevD.106.L061701, PhysRevLett.118.130404, PhysRevA.97.063611, PhysRevD.107.L121502, Rana_2023, 10.21468/SciPostPhysCore.6.1.003}. In particular, by appropriately tuning both the parameters $R$ and $A$, a variety of nonlinear trans-Planckian dispersions have been simulated to probe Planck-scale effects in cosmology recently \cite{chandran2025expansioncontractiondualitybreakingplanckscale}. Go beyond the LI vacuum case \cite{PhysRevD.92.064042}, in what follows, we will analyze 
the harvesting of entanglement from a Lorentz-violating quantum field vacuum to Unruh-DeWitt particle detectors.

\section{Basic formulas for entanglement harvesting of two Unruh-DeWitt detectors} \label{section3}
In order to probe quantum field, Unruh-DeWitt detector  \cite{PhysRevD.14.870, Dewitt1979General, birrell_davies_1982, 10.1143/PTP.88.1, RevModPhys.80.787, Hu_2012} has been introduced. It, for convenience, usually is modeled as a two-level (1 and 2) quantum system, such as an atom, and interacts with 
the field to be probed. 
This detector model captures the essential features of the light-matter interaction 
when angular momentum interchange is negligible \cite{PhysRevD.87.064038, PhysRevA.89.033835}. To realize such analogue physics in the ultracold atomic systems, we, inspired by the seminal atomic quantum dot idea introduced in Refs. \cite{PhysRevLett.94.040404, PhysRevLett.91.240407}, model impurity consisting of a two-level ($1$ and $2$) atom as the Unruh-DeWitt detector, and assume it is immersed in 
the quasi-2D dipolar BEC discussed above (see Fig. \ref{fig1}). We also assume that the impurity is externally imposed by a  tightly confining trap potential whose locality could be relatively moved, so that we can set the location of impurity and focus only on their internal degrees of freedom. The concrete formula 
of the analog between these two models has been introduced in Refs. \cite{PhysRevD.103.085014, PhysRevD.106.L061701, PhysRevLett.118.045301, PhysRevResearch.2.042009}.

Here we will consider two identical such kind of detectors immersed in the quasi-2D dipolar BEC. The spacetime trajectory of the detector, $(t_\text{D}(\tau), \br_\text{D}(\tau))$ with $\text{D}\in\{\text{A}, \text{B}\}$ denoting two detectors, is parameterized by its proper time $\tau$. Then,
the interaction Hamiltonian between the detectors and the analogue field in the interaction picture reads 
\begin{eqnarray} \label{detector-interaction3}
\nonumber
\hat{H}_I=\sum_{\text{D}=\text{A}, \text{B}}\big[g(\tau)(\sigma^+_\text{D}e^{i\Omega\tau}+\sigma^-_\text{D}e^{-i\Omega\tau})\delta\hat{\rho}(t_\text{D}(\tau), \br_\text{D}(\tau))\big],
\\
\end{eqnarray}
where $g(\tau)=\lambda\chi(\tau)=\lambda\exp[-\tau^2/(2\sigma^2)]$ with $\lambda$ being the coupling strength, and $\chi(\tau)=\lambda\exp[-\tau^2/(2\sigma^2)]$
is the Gaussian switching function with parameter $\sigma$ controlling the duration of the interaction. Besides, $\Omega$ represents the detector's energy-level gap 
between its ground state $|g_\text{D}\rangle$ and its excitation state $|e_\text{D}\rangle$ $(\text{D}\in\{\text{A}, \text{B}\})$. $\sigma^+_\text{D}$ and 
$\sigma^-_\text{D}$ denote the ladder operators of the detector, and $\delta\hat{\rho}(t_\text{D}(\tau), \br_\text{D}(\tau))$ is the density fluctuations of dipolar BEC
shown in Eq. \eqref{DF} that resembles the quantized Lorentz-violating field.

We take the initial state of joint system (the impurities and the density fluctuations, or the detectors and the quantum field) to be $|g_A, g_B\rangle\otimes|0\rangle$, i.e.,
the two detectors are prepared in their ground state and the field is in the vacuum state $|0\rangle$. With the interaction Hamiltonian in Eq. \eqref{detector-interaction3},
then, based on the perturbation theory, the density matrix for the final state 
of the detectors  in the basis $\{|g_\text{A}\rangle|g_\text{B}\rangle, |g_\text{A}\rangle|e_\text{B}\rangle, |e_\text{A}\rangle|g_\text{B}\rangle, |e_\text{A}\rangle|e_\text{B}\rangle\}$ can be obtained by tracing over the field degree of freedom to leading order in the coupling strength \cite{PhysRevD.92.064042, PhysRevD.93.044001, Supplemental-Material}
\begin{eqnarray}\label{final-state}
\rho_\text{AB}=\begin{pmatrix}
1-P_\text{A}-P_\text{B}  & 0 & 0 & X    \\
0& P_\text{B} & C   &   0      \\
0& C^\ast & P_\text{A}   &   0    \\
X^\ast     &0    &0 &     0
\end{pmatrix}+\mathcal{O}(\lambda^4),
\end{eqnarray}
where
the transition probability $P_\text{D}$ reads 
\begin{eqnarray}\label{probability-P}
\nonumber
P_\text{D}=\lambda^2\iint\,d\tau\,d\tau^\prime\chi(\tau)\chi(\tau^\prime)e^{-i\Omega(\tau-\tau^\prime)}\mathcal{W}(t_\text{D},\br_\text{D}, t^\prime_\text{D}, \br^\prime_\text{D}),
\\
\end{eqnarray}
with $\text{D}\in\{\text{A}, \text{B}\}$, and the quantities $C$ and $X$, which characterize correlations, are given by 
\begin{eqnarray}\label{Ccorrelation}
\nonumber
C=\lambda^2\iint d\tau\chi(\tau^\prime)\chi(\tau^\prime)e^{-i\Omega(\tau-\tau^\prime)}\mathcal{W}(t_\text{A},\br_\text{A}, t^\prime_\text{B}, \br^\prime_\text{B}),
\\
\end{eqnarray}
and
\begin{eqnarray}\label{Xcorrelation}
\nonumber
X=-\lambda^2\iint d\tau d\tau^\prime\chi(\tau)\chi(\tau^\prime)e^{-i\Omega(\tau+\tau^\prime)}\big[\theta(t-t^\prime)
\\ 
\times\mathcal{W}(t_\text{A}, \br_\text{A}, t_\text{B}^\prime, \br^\prime_\text{B})+\theta(t^\prime-t)\mathcal{W}(t_\text{B}^\prime, \br^\prime_\text{B}, t_\text{A}, \br_\text{A})\big].
\end{eqnarray}
Here $\theta$ represents the Heaviside theta function. $\mathcal{W}(t, \br, t^\prime, \br^\prime)=\langle0|\delta\hat{\rho}(t,\br)\delta\hat{\rho}(t^\prime,\br^\prime)|0\rangle$ is the Wightman function associated with field.

In order to quantify the entanglement among the detectors resulting from their interaction with the Lorentz-violating quantum field, we employ the concurrence as a measurement of entanglement, which has been introduced in Ref. \cite{PhysRevLett.80.2245}. For the above final state \eqref{final-state} of the detectors, the corresponding concurrence yields \cite{PhysRevD.93.044001}
\begin{eqnarray}\label{concurrence}
\mathcal{C}[\rho_\text{AB}]=2\max\bigg[0, |X|-\sqrt{P_\text{A}P_\text{B}}\bigg]+\mathcal{O}(\lambda^4).
\end{eqnarray}
We can see from this concurrence \eqref{concurrence} that the entanglement harvested by the two detectors from the field is determinated 
by the competition between two quantities: the off-diagonal matrix element $|X|$, which characterizes the nonlocal correlations between the detectors,
and geometric mean of the detector's transition probabilities $P_\text{A}$ and $P_\text{B}$ (the ``noise"). Note that the matrix element $C$ does not enter 
Eq. \eqref{concurrence}, and thus has no contribution to the entanglement harvesting.  Furthermore, since two identical detectors are considered, $P_\text{A}=P_\text{B}$,
and the concurrence is simply determined by $|X|-P_\text{D}$.

\section{Entanglement harvesting for detectors in Lorentz-violating field vacuum}\label{section4}
In this section, we are going to analyze how the LI violation affects the entanglement harvesting for two Unruh-DeWitt detectors in a Lorentz-violating quantum field vacuum.
For this purpose, we first need to calculate $P_\text{D}$ and $X$ in the expression of concurrence shown in Eq. \eqref{concurrence}.

We assume that that two identical detectors are static as shown Fig. \ref{fig1}, then their corresponding trajectories in the polar coordinates can be written as
\begin{eqnarray}\label{trajectories}
\nonumber
r_\text{A}&=&r, \theta_\text{A}=\theta_A, t_\text{A}=\tau,
\\
r_\text{B}&=&r, \theta_\text{B}=\theta_\text{B}, t_\text{B}=\tau.
\end{eqnarray}
We assume $\theta_\text{B}-\theta_\text{A}=\delta\theta$, and $L=|\br_\text{B}-\br_\text{A}|$ represents the constant interdetector spatial distance as measured
in the laboratory reference frame. It represents one of the key controllable parameters that affects the entanglement harvesting.

On the other hand, the Wightman function for vacuum analogue Lorentz-violating field reads \cite{Supplemental-Material}
\begin{eqnarray}\label{WF}
\nonumber
\mathcal{W}(t, \br, t^\prime, \br^\prime)=\frac{\rho_0}{(2\pi)^2}\int\,d^2\bk(u_{k}+v_{k})^2e^{-i\omega_\bk(t-t^\prime)+i\bk\cdot(\br-\br^\prime)}.
\\
\end{eqnarray}
Substituting Eqs. \eqref{trajectories} and \eqref{WF} into Eqs. \eqref{probability-P} and \eqref{Xcorrelation}, one can obtain 
\begin{eqnarray}\label{PExpression}
P_\text{D}=\lambda^2\rho_0\sigma^2\int^\infty_0k(u_k+\nu_k)^2e^{-(\omega_\bk+\Omega)^2\sigma^2}dk,
\end{eqnarray}
and 
\begin{eqnarray}\label{XExpression}
\nonumber
X&=&-\lambda^2\rho_0\sigma^2\int^\infty_0k(u_k+\nu_k)^2e^{-(\omega^2_\bk+\Omega^2)\sigma^2}
\\
&&\times(1+i \mathrm{Erfi}[\sigma\omega_\bk])J_0(kL)dk,
\end{eqnarray}
where $L=|\br_\text{B}-\br_\text{A}|$ has been defined above. Seen from Eqs. \eqref{PExpression} and \eqref{XExpression}, numerical integration is needed for
the evaluation of entanglement shown in \eqref{concurrence}.

\begin{figure}
\centering
\includegraphics[width=0.48\textwidth]{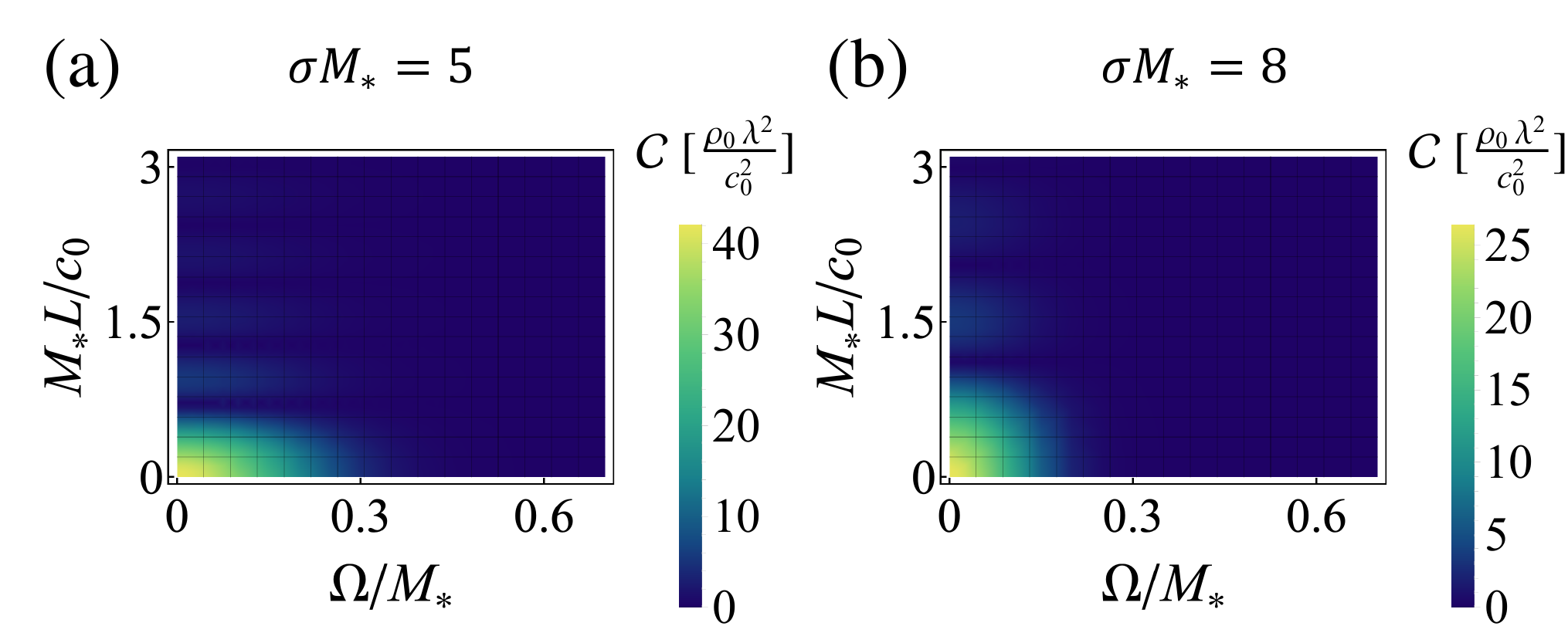}
\caption{Concurrence shown in Eq. \eqref{concurrence} as a function of the energy-level spacing $\Omega/M_\ast$ and the interdetector distance $M_\ast L/c_0$ for the Lorentz-invariant vacuum case.
The units of the concurrence is $\rho_0\lambda^2/c^2_0$, and the switching-function width $\sigma M_\ast=5$ and $8$ respectively corresponds to (a) and (b).}\label{fig3}
\end{figure}

\begin{figure*}
\centering
\includegraphics[width=0.86\textwidth]{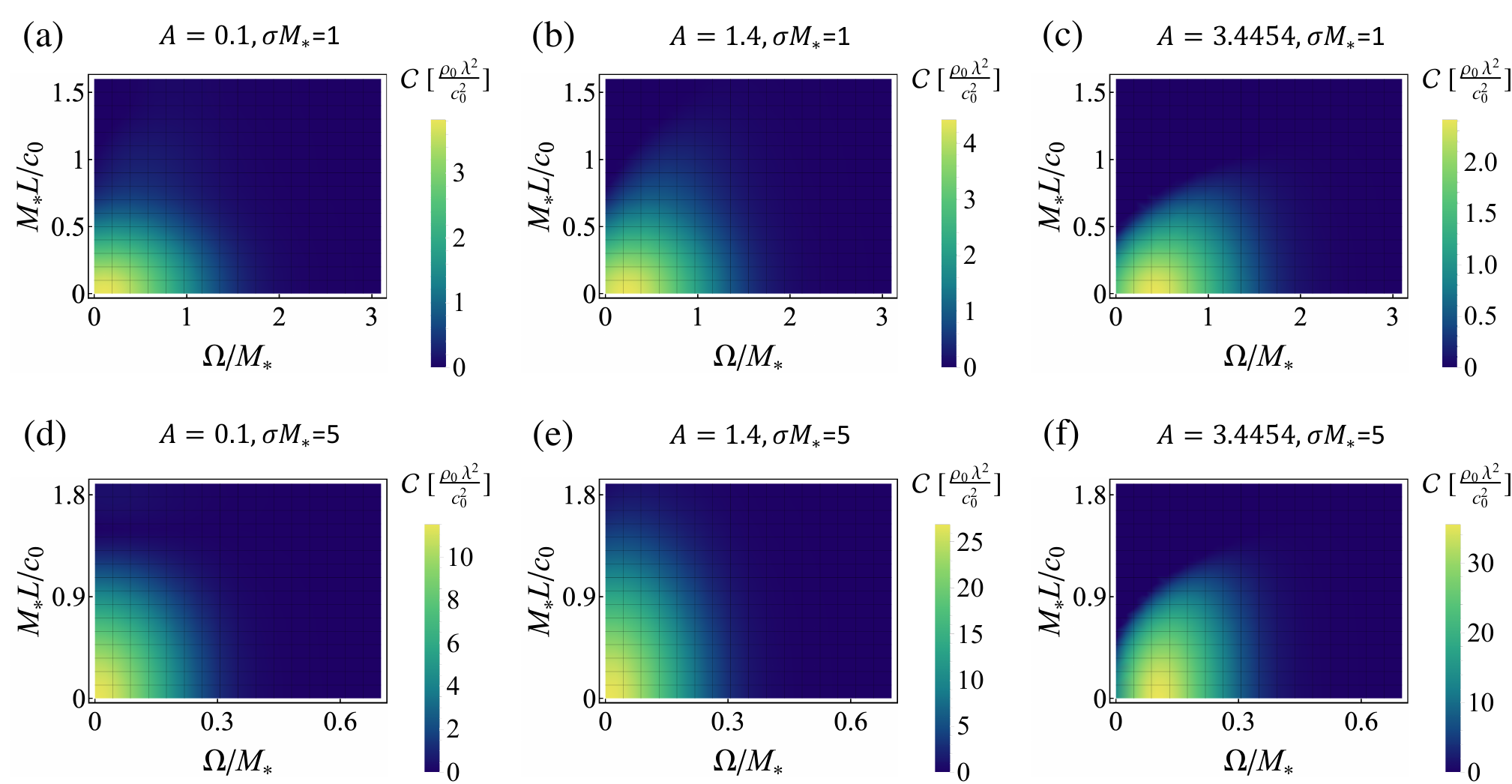}
\caption{Concurrence shown in Eq. \eqref{concurrence} as a function of the energy-level spacing $\Omega/M_\ast$ and the interdetector distance $M_\ast L/c_0$ for the Lorentz-violating case.
The units of the concurrence is $\rho_0\lambda^2/c^2_0$. The switching function width $\sigma M_\ast=1$ corresponds to (a), (b) and (c) cases, and $\sigma M_\ast=5$ corresponds to (d), (e) and (f) cases. We take $R=\sqrt{\pi/2}$ in the dispersion \eqref{Dispersion} such that the strength of LI violation increases with the increase of $A$. }\label{fig4}
\end{figure*}

\subsection{Lorentz invariant case}
If $R=0$, i.e., only contact interaction between the atoms or molecules in the Bose gas exists, the dispersion relation given in Eq. \eqref{Dispersion}
can be rewritten as 
\begin{eqnarray}\label{Dispersion2}
\omega^2_\bk=c^2_0|\bk|^2+\epsilon^2|\bk|^4,
\end{eqnarray}
where $\epsilon=c^2_0/2M_\ast$. This dispersion becomes relevant when the quartic term in \eqref{Dispersion2} becomes comparable to 
the quadratic term. In particular, at their equivalence point there is a crossover scale $k_\text{c}=|\bk|_\text{c}=c_0/\epsilon$, which means that a transition from the linear (phononic) band to the dispersive regime (quasiparticles) exists. Note that if $k_\text{c}$ here is sufficiently larger than the inverse switching 
function width $k_0=1/\sigma$, the contributions from high-$|\bk|$ to the integrands in \eqref{PExpression} and \eqref{XExpression} will be suppressed. Therefore,
if the switching function width for a detector is sufficiently larger than $1/k_\text{c}$, the effects of dispersion for harvesting are negligible. Usually, one can consider 
the long wavelength (phononic) regime, neglecting the quantum pressure term $\sim\triangledown^2\delta\rho$, and thus $\omega_\bk\approx c_0|\bk|$, as done in Refs. \cite{PhysRevLett.101.110402, PhysRevLett.125.213603, PhysRevLett.118.130404, tianzehua, chandran2025expansioncontractiondualitybreakingplanckscale}. Which in this issue allows us to study the entanglement harvesting from the LI quantum field vacuum.

We now analyze the effects of the detectors' energy-level spacing, $\Omega/M_\ast$, and the interdetector distance, $M_\ast L/c_0$, on the entanglement harvested between the two detectors for the Lorentz-invariant field case. Note that here both of the energy-level spacing and the interdetector distance are dimensionless. Seen from Fig. \ref{fig3}, when fixing the switching-function width $\sigma M_\ast$ there are several separated regions of the detectors' energy-level spacing and of the interdetector distance only in which the detectors can harvest entanglement from the quantum field vacuum. This results from the damped oscillation of Bessel function via the interdetector distance $L$ in Eq. \eqref{XExpression}.  Besides, there is an optimal parameter region where the most entanglement can be harvested from the quantum field vacuum. Specifically, this region corresponds to the area of both the relatively small detectors' energy-level spacing and small interdetector distance.
Furthermore, when both of the $\Omega/M_\ast$ and  $M_\ast L/c_0$ approache to zero, the maximum entanglement can be harvested.
We can also find that the width, $\sigma M_\ast$, in the Gaussian distribution for the switching function could affect the bound of both the energy-level spacing and 
the interdetector distance, beyond which entanglement can not be harvested from the quantum field vacuum. 
When the switching-function width $\sigma M_\ast$ increases, the bound for $\Omega/M_\ast$ decreases. However, the bound for the $M_\ast L/c_0$ increases with the increase of $\sigma M_\ast$. Besides, the amount of the harvested entanglement between the two Unruh-DeWitt detectors becomes smaller when the switching-function width $\sigma M_\ast$ increases. Therefore, for the Lorentz-invariant quantum field case, the optimal region to harvest entanglement from the 
field vacuum using the Unruh-DeWitt detectors is the region around the original of both the detectors' energy-level spacing and the interdetector distance.

\begin{figure*}
\centering
\includegraphics[width=0.8\textwidth]{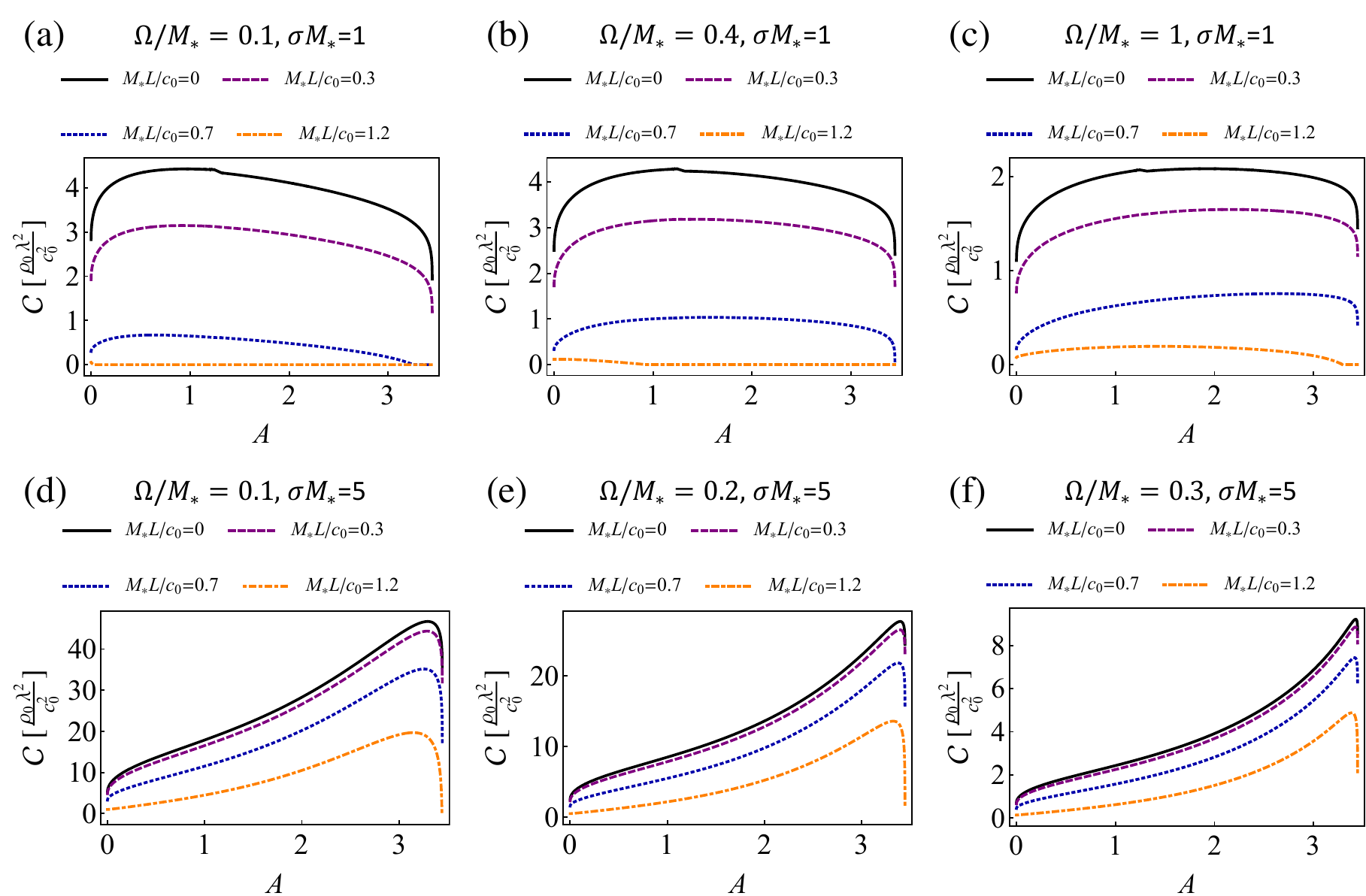}
\caption{Concurrence shown in Eq. \eqref{concurrence} as a function of the strength of the LI violation, $A$.
The units of the concurrence is $\rho_0\lambda^2/c^2_0$. Here the detector's energy-level spacing $\Omega/M_\ast$ is fixed, while the interdetector distance $M_\ast L/c_0$ varies. We take $R=\sqrt{\pi/2}$ in the dispersion \eqref{Dispersion} such that the strength of LI violation increases with the increase of $A$. The switching function width $\sigma M_\ast=1$ corresponds to (a), (b) and (c) cases, and $\sigma M_\ast=5$ corresponds to (d), (e) and (f) cases.}\label{fig5}
\end{figure*}

\subsection{Lorentz-violating case}
As discussed above, the density fluctuations shown in Eq. \eqref{DF} closely resemble a Lorentz-violating scalar field with an
explicit dispersion relation given in Eq. \eqref{Dispersion}. This analogue has been fruitful applications in the exploration of Lorentz-violating quantum field theory 
in curved spacetime, e.g., probing the scale invariance of the inflationary power spectrum \cite{PhysRevLett.118.130404}, studying impact of trans-Planckian excitations on black-hole radiation \cite{PhysRevD.107.L121502}, demonstrating the Lorentz-violation induced nonthermal Unruh effect \cite{PhysRevD.106.L061701}, and so on. In what follows, we will study how the LI violation affects the entanglement harvesting from the quantum field vacuum using this platform. Note that in Ref. \cite{chandran2025expansioncontractiondualitybreakingplanckscale} a variety of specific nonlinear trans-Planckian dispersions have been simulated through appropriately tuning both of the parameters $R$ and $A$ introduced above. We here fixed $R=\sqrt{\pi/2}$ and take various $A\in[0, 3.4454]$. As such, the obtained dispersion in Eq. \eqref{Dispersion} is subluminal, and correspondingly its departure from the Lorentz-invariant one increases with the increase of $A$. Which has been commonly used to explore the trans-Planckian physics of quantum field theory in different scenarios \cite{PhysRevA.97.063611, PhysRevLett.118.130404, PhysRevD.107.L121502, PhysRevD.106.L061701, PhysRevD.103.085014}.

\begin{figure*}
\centering
\includegraphics[width=0.8\textwidth]{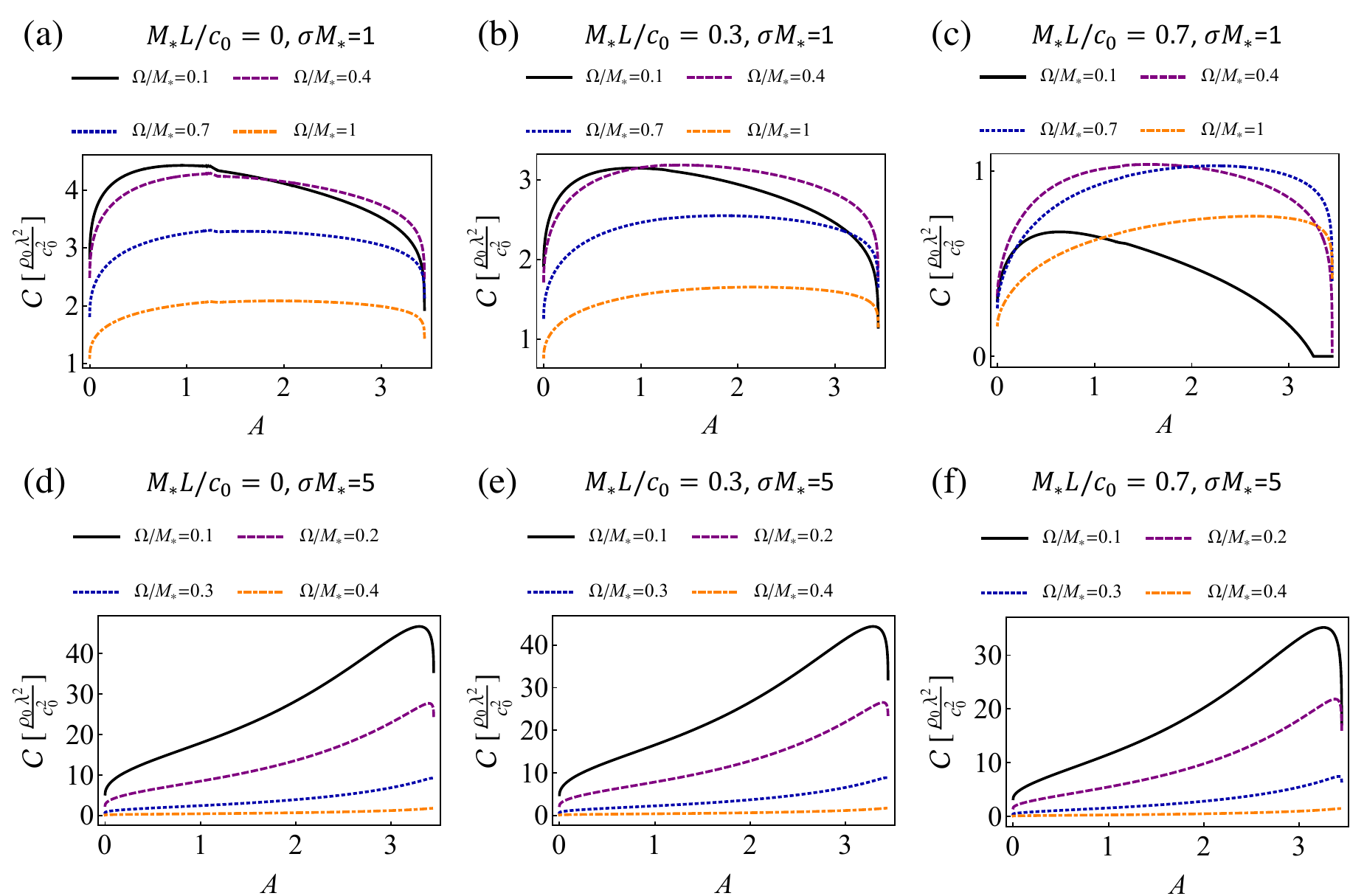}
\caption{Concurrence shown in Eq. \eqref{concurrence} as a function of the strength of the LI violation $A$.
The units of the concurrence is $\rho_0\lambda^2/c^2_0$. Here the interdetector distance $M_\ast L/c_0$  is fixed, while the detector's energy-level spacing $\Omega/M_\ast$ varies. We take $R=\sqrt{\pi/2}$ in the dispersion \eqref{Dispersion} such that the strength of LI violation increases with the increase of $A$. The switching function width $\sigma M_\ast=1$ corresponds to (a), (b) and (c) cases, and $\sigma M_\ast=5$ corresponds to (d), (e) and (f) cases.}\label{fig6}
\end{figure*}

In Fig. \ref{fig4}, we fix the parameter $R=\sqrt{\pi/2}$, vary $A$ and plot the concurrence shown in \eqref{concurrence} as a function of the energy-level spacing 
$\Omega/M_\ast$ and the interdetector distance $M_\ast L/c_0$. It is found that when the LI violation is weak, the optimal entanglement-harvesting 
region still concentrate in the area of relatively smaller $\Omega/M_\ast$ and $M_\ast L/c_0$. This consists with the Lorentz-invariant field case discussed above.
However, when $A$ increases, i.e., the LI violation becomes stronger, we can find that the optimal region of entanglement harvesting 
will deviate the origin of the energy-level spacing $\Omega/M_\ast$ and the interdetector distance $M_\ast L/c_0$. Specifically, with the increase of the strength
of the LI violation, the optimal parameter $\Omega/M_\ast$ for entanglement harvesting deviates the origin and increases monotonously.
However, the optimal parameter $M_\ast L/c_0$ for entanglement harvesting keeps the same as that of the Lorentz-invariant field case. Furthermore, 
the degree of LI violation also governs the parameter region in $(\Omega/M_\ast, M_\ast L/c_0)$ that is accessible for entanglement harvesting from the quantum field vacuum.
See from Fig. \ref{fig4} (a), (b) and (c), when the width of the switching function $\sigma M_\ast$ is small, the maximum entanglement that could be harvested first increases with the increase of the strength of LI violation and then decreases. However, when the width of the switching function $\sigma M_\ast$ is relatively larger, seen from Fig. \ref{fig4} (d), (e) and (f) the maximum entanglement that could be harvested always increases with the increase of the strength of LI violation. Besides, comparing  Fig. \ref{fig4} (a), (b) and (c) with Fig. \ref{fig4} (d), (e) and (f), we can find that when the switching function width $\sigma M_\ast$ increases, the maximum entanglement that could be harvested from the Lorentz-violating quantum field vacuum increases, after optimizing other parameters, such as $\Omega/M_\ast$ and $M_\ast L/c_0$. This result is quite different from that of Lorentz-invariant case where the bigger switching-function width will degrade the maximum entanglement that could be harvested by the 
Unruh-Dewitt detectors.

In order to see how the LI violation affects the entanglement harvesting more clearly,  we fix the parameters, such as the energy-level spacing 
$\Omega/M_\ast$, the interdetector distance $M_\ast L/c_0$, and the switching function width $\sigma M_\ast$, and plot 
the entanglement harvested as a function of the strength of the LI violation, $A$, in Figs. \ref{fig5} and \ref{fig6}. 
Seen from Fig. \ref{fig5} (a), (b) and (c) for relatively smaller switching-function width cases, we can find that for relatively smaller interdetector distance $M_\ast L/c_0$ case, the entanglement harvested first increases with the increase of the strength of the LI violation $A$, and the then decreases monotonously to a constant that depends on both the detector's energy-level spacing $\Omega/M_\ast$ and the interdetector distance $M_\ast L/c_0$. However, when $M_\ast L/c_0$ is relatively larger, the entanglement harvested decreases monotonously as the increases of $A$ for relatively smaller energy-level spacing cases, while still increases and then decreases monotonously for relatively larger energy-level spacing cases. Therefore, for the case of relatively smaller switching-function width $\sigma M_\ast$, the qualitative behavior of the entanglement harvest vs the strength of the LI violation depends on both the detector's energy-level spacing $\Omega/M_\ast$ and the interdetector distance $M_\ast L/c_0$. When the width of the switching function is relatively wider, we can see from Fig. \ref{fig5} (d), (e) and (f) that 
the entanglement harvested first increases monotonously and then decreases rapidly to a constant with the increase of the strength of the LI violation $A$. 
This behavior is different from the case with smaller width of the switching function shown in Fig. \ref{fig5} (a), (b) and (c), where the behavior of the entanglement harvested depends on both the detector's energy-level spacing $\Omega/M_\ast$ and the interdetector distance $M_\ast L/c_0$. 
Furthermore, we can also find that no matter for the small or large width of the switching function, the interdetector distance $M_\ast L/c_0$ may reduce the amount of the entanglement harvested. Therefore, the optimal interdetector distance for harvesting entanglement approaches to $M_\ast L/c_0=0$.

In Fig. \ref{fig6}, we fix the interdetector distance $M_\ast L/c_0$ and the width of switching function $\sigma M_\ast$, and plot the entanglement 
that could be harvested as a function of the strength of the LI violation $A$ via the various detector's energy-level spacing $\Omega/M_\ast$.
When the width of switching function $\sigma M_\ast$ is relatively smaller, we can see from Fig. \ref{fig6} (a), (b) and (c) that with the increase of the 
strength of the LI violation $A$, the entanglement that could be harvested from the Lorentz-violating quantum field vacuum increases monotonously first
and then decreases to a constant, which could be vanishing with properly choosing $M_\ast L/c_0$ and $\Omega/M_\ast$. 
In particular, there is a cross between the curves corresponding to various $\Omega/M_\ast$ cases with fixed $M_\ast L/c_0$. This indicates that the entanglement harvested is not a monotonous function of the detector's energy-level spacing $\Omega/M_\ast$ for fixed $M_\ast L/c_0$ and $A$. We can also find that when fixing the
$\Omega/M_\ast$ and $A$, the entanglement harvested decreases with the increase of the interdetector distance $M_\ast L/c_0$ by comparing Fig. \ref{fig6} (a), (b) and (c). When the width of switching function $\sigma M_\ast$ is relatively wider, we can see from Fig. \ref{fig6} (d), (e) and (f) that the 
entanglement harvested shows two kinds of behaviors as the increase of the strength of the LI violation: 
1) when the detector's energy-level spacing $\Omega/M_\ast$ is relatively smaller, it increases to a maximum and then decreases monotonously; 2)
when the detector's energy-level spacing $\Omega/M_\ast$ is relatively larger, it always increases monotonously. This is quite different from the case of 
relatively smaller width of the switching function. Furthermore, the wider width $\sigma M_\ast$ of the switching function may 
lead to more entanglement that could be harvested by the Unruh-DeWitt detectors from the Lorentz-violating quantum field vacuum.

We give a summary here. When one uses two Unruh-DeWitt detectors to harvest entanglement from quantum field vacuum,
the Lorentz-violating quantum field vacuum will lead to quite distinct behaviors compared with that induced by the Lorentz-invariant quantum field vacuum. 
Specifically, for the entanglement harvesting from the Lorentz-invariant quantum field vacuum, the optimal detector's energy-level spacing $\Omega/M_\ast$  
approaches to the vanishing region. However, for the Lorentz-violating quantum field vacuum case, the optimal detector's energy-level spacing $\Omega/M_\ast$ for the 
entanglement harvesting may have departure from the original point, and the stronger the LI violation is, the further the departure is.
Furthermore, the wider width $\sigma M_\ast$ of the switching function may 
lead to more entanglement that could be harvested by the Unruh-DeWitt detectors from the Lorentz-violating quantum field vacuum, which is completely opposite 
to that of the Lorentz-invariant quantum field vacuum case. These distinctions, in principle, could be used as the criteria to explore whether the quantum field is Lorentz invariant or not in practice.

\section{Experimental implementation}\label{section5}
Recent advances in ultra-cold atomic technology have enabled groundbreaking observations of magnetic DDI-dominated condensate \cite{BARANOV200871} 
using various atomic species \cite{Chromium, PhysRevLett.108.210401, PhysRevLett.107.190401, NP1, N1}; see also Ref. \cite{chomaz2022dipolar} for an extensive review and a comprehensive list of references. Furthermore, the realization of BEC composed of molecules with permanent electric dipoles \cite{doi:10.1021/cr300092g} represents a current frontier of research \cite{PhysRevLett.114.205302, PhysRevLett.116.205303, PhysRevLett.119.143001, De-Marco853, Son, N2}.
The presence of the DDI between atoms significantly influences the quasiparticle dispersion on the condensate. As a result, the excitation spectrum of a dipolar BEC deviates from the Lorentz-invariant form---a deviation that can be tuned by varying the ratio of contact to dipolar interaction strengths.
Notably, in the DDI-dominated regime, the spectrum even exhibits a roton minimum \cite{PhysRevA.97.063611, PhysRevLett.118.130404, PhysRevA.73.031602, PhysRevLett.90.250403, PhysRevLett.98.030406}, signaling an ultra-strong effective violation of LI for the analog quantum field. Such roton modes have been experimentally observed recently  \cite{chomaz2022dipolar, Chomaz, DDIexperiment, PhysRevLett.123.050402, PhysRevLett.122.183401, NP2}, underscoring the relevance of dipolar BECs as analog quantum simulators of Lorentz-violating field theories. In this context, the dipolar BEC could furnish a perfect platform for using its corresponding density fluctuations to emulate Lorentz-violating quantum field within experimentally accessible regimes. Furthermore, rapid progress in high-precision measurements of correlation functions 
\cite{PMID:23907531, PMID:21350171, thermal-Hawking-radiation1, thermal-Hawking-radiation2, PhysRevLett.127.060404, PhysRevLett.133.260201, N3, tamura2025observationmanybodycoherencequasionedimensional, Unruh-Simulation1}  
opens promising avenues for exploring the possible extraordinary propagation of quantum field arising from the broken LI.

In our proposal, an effective two-level quantum system serves as a probe to detect the density fluctuations in a dipolar BEC, which emulates 
a Lorentz-violating quantum field. The similar hybrid systems \cite{RevModPhys.91.035001, Grusdt_2025}, combining trapped ions and ultracold atomic gases, have recently emerged as a promising platform for fundamental research in quantum physics. Experimentally, single effective detector system, such as 
a trapped single ion  \cite{2010Natur.464..388Z, N4} and single electron \cite{2013Natur.502..664B}, that is coupled to a BEC has been reported recently.
Moreover, usually the dynamics of such impurities can be used to 
probe the density profile of a BEC \cite{PhysRevLett.105.133202, PhysRevLett.111.070401, PhysRevLett.109.235301, PhysRevLett.121.130403, PhysRevLett.120.193401, PhysRevLett.126.033401}, see Ref. \cite{Grusdt_2025} for an extensive review and a comprehensive list of references.
By designing an external trapping potential, one can tightly confine impurities and precisely control their positions, enabling
the study of entanglement harvesting from the quantum field vacuum at variable detector separations.
Thus, it is possible in principle to design a hybrid system combining two impurities embedded in a dipolar BEC,
allowing investigation of impurity dynamics mediated by roton modes arising from atomic DDI. 
This proposed ultra-cold atomic setup thereby functions as a programmable quantum simulator of the broken LI physics, as motivated in previous sections.

\section{Conclusions and discussions}\label{section6}
In this paper, we establish the connection between a pair of impurities in a dipolar BEC and a pair of idealized Unruh-DeWitt detectors interacting with
the Lorentz-violating quantum field. This mapping provides a synthetic quantum system enabling an experimental proof-of-principle demonstration of  the entanglement harvesting protocol from the Lorentz-violating quantum field vacuum. 
We systematically investigate how key parameters, such as the energy-level spacing and spatial distance, influence the amount of entanglement harvested from the Lorentz-violating quantum field vacuum. 
In particular, we analyze how the strength of the LI violation affects the harvested entanglement, revealing significant distinctions from the Lorentz-invariant case.
These include shifts in the optimal energy structure of detector and the feasible separation range for entanglement harvesting.
Our proposed quantum fluid platform provides an experimentally realizable testbed to examine the entanglement harvesting protocol in Lorentz-violating quantum field theories. The results also provide valuable guidance for explore LI violation in quantum field theory.

Our proposed quantum fluid platform can function as a quantum simulator \cite{RevModPhys.86.153} for studying quantum field theories.
It enables the exploration of entanglement harvesting from Lorentz-violating quantum field vacuum by various Unruh-DeWitt detectors, including 
inertial detectors with constant velocity \cite{PhysRevD.103.085014}, uniformly accelerated detectors \cite{PhysRevD.106.L061701}, those in black hole background \cite{PhysRevD.107.L121502}, and those in an expanding universe \cite{PhysRevLett.118.130404, Rana_2023}. In turn, our proposed scheme 
also offers a local and nondestructive measurement method for probing continuous variable entanglement in non-equilibrium ultra-cold atom systems.

~~~~~
\begin{acknowledgments}
ZT was supported by the National Natural Science Foundation of China under Grant No. 12575050, and the scientific research start-up funds of Hangzhou Normal University: 4245C50224204016. This research was supported by Hangzhou Leading Youth Innovation and Entrepreneurship Team project  under Grant No. TD2024005,
and the HZNU scientific research and innovation team project (TD2025003).
XL acknowledges the support by the National Natural Science Foundation of China (NSFC) under Grant No. 12065016 and the Discipline-Team of Liupanshui Normal University of China under Grant No. LPSSY2023XKTD11. 
M. Wang is supported by the National Natural Science Foundation of China under Grant Nos. 12475050, 11705054, by the Hunan Provincial Natural Science Foundation of China under Grant No. 2022JJ30367 and by the Scientific Research Fund of Hunan Provincial Education Department Grant No. 22A0039. 
J. Wang was supported by the National Natural Science Foundation of China under Grants No. 12475051, No. 12375051, and No. 12421005; the science and technology innovation Program of Hunan Province under grant No. 2024RC1050; the Natural Science Foundation of Hunan Province under grant No. 2023JJ30384; and the innovative research group of Hunan Province under Grant No. 2024JJ1006.
\end{acknowledgments}


\bibliography{Entanglement-LIV}

\begin{thebibliography}{138}%
\makeatletter
\providecommand \@ifxundefined [1]{%
 \@ifx{#1\undefined}
}%
\providecommand \@ifnum [1]{%
 \ifnum #1\expandafter \@firstoftwo
 \else \expandafter \@secondoftwo
 \fi
}%
\providecommand \@ifx [1]{%
 \ifx #1\expandafter \@firstoftwo
 \else \expandafter \@secondoftwo
 \fi
}%
\providecommand \natexlab [1]{#1}%
\providecommand \enquote  [1]{``#1''}%
\providecommand \bibnamefont  [1]{#1}%
\providecommand \bibfnamefont [1]{#1}%
\providecommand \citenamefont [1]{#1}%
\providecommand \href@noop [0]{\@secondoftwo}%
\providecommand \href [0]{\begingroup \@sanitize@url \@href}%
\providecommand \@href[1]{\@@startlink{#1}\@@href}%
\providecommand \@@href[1]{\endgroup#1\@@endlink}%
\providecommand \@sanitize@url [0]{\catcode `\\12\catcode `\$12\catcode
  `\&12\catcode `\#12\catcode `\^12\catcode `\_12\catcode `\%12\relax}%
\providecommand \@@startlink[1]{}%
\providecommand \@@endlink[0]{}%
\providecommand \url  [0]{\begingroup\@sanitize@url \@url }%
\providecommand \@url [1]{\endgroup\@href {#1}{\urlprefix }}%
\providecommand \urlprefix  [0]{URL }%
\providecommand \Eprint [0]{\href }%
\providecommand \doibase [0]{https://doi.org/}%
\providecommand \selectlanguage [0]{\@gobble}%
\providecommand \bibinfo  [0]{\@secondoftwo}%
\providecommand \bibfield  [0]{\@secondoftwo}%
\providecommand \translation [1]{[#1]}%
\providecommand \BibitemOpen [0]{}%
\providecommand \bibitemStop [0]{}%
\providecommand \bibitemNoStop [0]{.\EOS\space}%
\providecommand \EOS [0]{\spacefactor3000\relax}%
\providecommand \BibitemShut  [1]{\csname bibitem#1\endcsname}%
\let\auto@bib@innerbib\@empty
\bibitem [{\citenamefont {Summers}\ and\ \citenamefont
  {Werner}(1987)}]{Entanglement1}%
  \BibitemOpen
  \bibfield  {author} {\bibinfo {author} {\bibfnamefont {S.~J.}\ \bibnamefont
  {Summers}}\ and\ \bibinfo {author} {\bibfnamefont {R.}~\bibnamefont
  {Werner}},\ }\bibfield  {title} {\bibinfo {title} {Maximal violation of
  bell's inequalities is generic in quantum field theory},\ }\href
  {https://doi.org/10.1007/BF01207366} {\bibfield  {journal} {\bibinfo
  {journal} {Communications in Mathematical Physics}\ }\textbf {\bibinfo
  {volume} {110}},\ \bibinfo {pages} {247} (\bibinfo {year}
  {1987})}\BibitemShut {NoStop}%
\bibitem [{\citenamefont {Valentini}(1991)}]{VALENTINI1991321}%
  \BibitemOpen
  \bibfield  {author} {\bibinfo {author} {\bibfnamefont {A.}~\bibnamefont
  {Valentini}},\ }\bibfield  {title} {\bibinfo {title} {Non-local correlations
  in quantum electrodynamics},\ }\href
  {https://doi.org/https://doi.org/10.1016/0375-9601(91)90952-5} {\bibfield
  {journal} {\bibinfo  {journal} {Physics Letters A}\ }\textbf {\bibinfo
  {volume} {153}},\ \bibinfo {pages} {321} (\bibinfo {year}
  {1991})}\BibitemShut {NoStop}%
\bibitem [{\citenamefont {Reznik}(2003)}]{Entanglement2}%
  \BibitemOpen
  \bibfield  {author} {\bibinfo {author} {\bibfnamefont {B.}~\bibnamefont
  {Reznik}},\ }\bibfield  {title} {\bibinfo {title} {Entanglement from the
  vacuum},\ }\href {https://doi.org/10.1023/A:1022875910744} {\bibfield
  {journal} {\bibinfo  {journal} {Foundations of Physics}\ }\textbf {\bibinfo
  {volume} {33}},\ \bibinfo {pages} {167} (\bibinfo {year} {2003})}\BibitemShut
  {NoStop}%
\bibitem [{\citenamefont {Reznik}\ \emph {et~al.}(2005)\citenamefont {Reznik},
  \citenamefont {Retzker},\ and\ \citenamefont {Silman}}]{PhysRevA.71.042104}%
  \BibitemOpen
  \bibfield  {author} {\bibinfo {author} {\bibfnamefont {B.}~\bibnamefont
  {Reznik}}, \bibinfo {author} {\bibfnamefont {A.}~\bibnamefont {Retzker}},\
  and\ \bibinfo {author} {\bibfnamefont {J.}~\bibnamefont {Silman}},\
  }\bibfield  {title} {\bibinfo {title} {Violating Bell's inequalities in
  vacuum},\ }\href {https://doi.org/10.1103/PhysRevA.71.042104} {\bibfield
  {journal} {\bibinfo  {journal} {Phys. Rev. A}\ }\textbf {\bibinfo {volume}
  {71}},\ \bibinfo {pages} {042104} (\bibinfo {year} {2005})}\BibitemShut
  {NoStop}%
\bibitem [{\citenamefont {Silman}\ and\ \citenamefont
  {Reznik}(2007)}]{PhysRevA.75.052307}%
  \BibitemOpen
  \bibfield  {author} {\bibinfo {author} {\bibfnamefont {J.}~\bibnamefont
  {Silman}}\ and\ \bibinfo {author} {\bibfnamefont {B.}~\bibnamefont
  {Reznik}},\ }\bibfield  {title} {\bibinfo {title} {Long-range entanglement in
  the Dirac vacuum},\ }\href {https://doi.org/10.1103/PhysRevA.75.052307}
  {\bibfield  {journal} {\bibinfo  {journal} {Phys. Rev. A}\ }\textbf {\bibinfo
  {volume} {75}},\ \bibinfo {pages} {052307} (\bibinfo {year}
  {2007})}\BibitemShut {NoStop}%
\bibitem [{\citenamefont {Steeg}\ and\ \citenamefont
  {Menicucci}(2009)}]{PhysRevD.79.044027}%
  \BibitemOpen
  \bibfield  {author} {\bibinfo {author} {\bibfnamefont {G.~V.}\ \bibnamefont
  {Steeg}}\ and\ \bibinfo {author} {\bibfnamefont {N.~C.}\ \bibnamefont
  {Menicucci}},\ }\bibfield  {title} {\bibinfo {title} {Entangling power of an
  expanding universe},\ }\href {https://doi.org/10.1103/PhysRevD.79.044027}
  {\bibfield  {journal} {\bibinfo  {journal} {Phys. Rev. D}\ }\textbf {\bibinfo
  {volume} {79}},\ \bibinfo {pages} {044027} (\bibinfo {year}
  {2009})}\BibitemShut {NoStop}%
\bibitem [{\citenamefont {Salton}\ \emph {et~al.}(2015)\citenamefont {Salton},
  \citenamefont {Mann},\ and\ \citenamefont {Menicucci}}]{Salton_2015}%
  \BibitemOpen
  \bibfield  {author} {\bibinfo {author} {\bibfnamefont {G.}~\bibnamefont
  {Salton}}, \bibinfo {author} {\bibfnamefont {R.~B.}\ \bibnamefont {Mann}},\
  and\ \bibinfo {author} {\bibfnamefont {N.~C.}\ \bibnamefont {Menicucci}},\
  }\bibfield  {title} {\bibinfo {title} {Acceleration-assisted entanglement
  harvesting and rangefinding},\ }\href
  {https://doi.org/10.1088/1367-2630/17/3/035001} {\bibfield  {journal}
  {\bibinfo  {journal} {New Journal of Physics}\ }\textbf {\bibinfo {volume}
  {17}},\ \bibinfo {pages} {035001} (\bibinfo {year} {2015})}\BibitemShut
  {NoStop}%
\bibitem [{\citenamefont {Unruh}(1976)}]{PhysRevD.14.870}%
  \BibitemOpen
  \bibfield  {author} {\bibinfo {author} {\bibfnamefont {W.~G.}\ \bibnamefont
  {Unruh}},\ }\bibfield  {title} {\bibinfo {title} {Notes on black-hole
  evaporation},\ }\href {https://doi.org/10.1103/PhysRevD.14.870} {\bibfield
  {journal} {\bibinfo  {journal} {Phys. Rev. D}\ }\textbf {\bibinfo {volume}
  {14}},\ \bibinfo {pages} {870} (\bibinfo {year} {1976})}\BibitemShut
  {NoStop}%
\bibitem [{\citenamefont {Hotta}(2009)}]{doi:10.1143/JPSJ.78.034001}%
  \BibitemOpen
  \bibfield  {author} {\bibinfo {author} {\bibfnamefont {M.}~\bibnamefont
  {Hotta}},\ }\bibfield  {title} {\bibinfo {title} {Quantum energy
  teleportation in spin chain systems},\ }\href
  {https://doi.org/10.1143/JPSJ.78.034001} {\bibfield  {journal} {\bibinfo
  {journal} {Journal of the Physical Society of Japan}\ }\textbf {\bibinfo
  {volume} {78}},\ \bibinfo {pages} {034001} (\bibinfo {year} {2009})}\BibitemShut {NoStop}%
\bibitem [{\citenamefont {Hotta}(2011)}]{hotta2011}%
  \BibitemOpen
  \bibfield  {author} {\bibinfo {author} {\bibfnamefont {M.}~\bibnamefont
  {Hotta}},\ }\href {https://arxiv.org/abs/1101.3954} {\bibinfo {title}
  {Quantum energy teleportation: An introductory review}},\ \Eprint {https://arxiv.org/abs/1101.3954} {arXiv:1101.3954
  [quant-ph]} \BibitemShut {NoStop}%
\bibitem [{\citenamefont {Hotta}\ \emph {et~al.}(2014)\citenamefont {Hotta},
  \citenamefont {Matsumoto},\ and\ \citenamefont {Yusa}}]{PhysRevA.89.012311}%
  \BibitemOpen
  \bibfield  {author} {\bibinfo {author} {\bibfnamefont {M.}~\bibnamefont
  {Hotta}}, \bibinfo {author} {\bibfnamefont {J.}~\bibnamefont {Matsumoto}},\
  and\ \bibinfo {author} {\bibfnamefont {G.}~\bibnamefont {Yusa}},\ }\bibfield
  {title} {\bibinfo {title} {Quantum energy teleportation without a limit of
  distance},\ }\href {https://doi.org/10.1103/PhysRevA.89.012311} {\bibfield
  {journal} {\bibinfo  {journal} {Phys. Rev. A}\ }\textbf {\bibinfo {volume}
  {89}},\ \bibinfo {pages} {012311} (\bibinfo {year} {2014})}\BibitemShut
  {NoStop}%
\bibitem [{\citenamefont {Nambu}\ and\ \citenamefont
  {Hotta}(2010)}]{PhysRevA.82.042329}%
  \BibitemOpen
  \bibfield  {author} {\bibinfo {author} {\bibfnamefont {Y.}~\bibnamefont
  {Nambu}}\ and\ \bibinfo {author} {\bibfnamefont {M.}~\bibnamefont {Hotta}},\
  }\bibfield  {title} {\bibinfo {title} {Quantum energy teleportation with a
  linear harmonic chain},\ }\href {https://doi.org/10.1103/PhysRevA.82.042329}
  {\bibfield  {journal} {\bibinfo  {journal} {Phys. Rev. A}\ }\textbf {\bibinfo
  {volume} {82}},\ \bibinfo {pages} {042329} (\bibinfo {year}
  {2010})}\BibitemShut {NoStop}%
\bibitem [{\citenamefont {Rodr\'{\i}guez-Briones}\ \emph
  {et~al.}(2023)\citenamefont {Rodr\'{\i}guez-Briones}, \citenamefont
  {Katiyar}, \citenamefont {Mart\'{\i}n-Mart\'{\i}nez},\ and\ \citenamefont
  {Laflamme}}]{PhysRevLett.130.110801}%
  \BibitemOpen
  \bibfield  {author} {\bibinfo {author} {\bibfnamefont {N.~A.}\ \bibnamefont
  {Rodr\'{\i}guez-Briones}}, \bibinfo {author} {\bibfnamefont {H.}~\bibnamefont
  {Katiyar}}, \bibinfo {author} {\bibfnamefont {E.}~\bibnamefont
  {Mart\'{\i}n-Mart\'{\i}nez}},\ and\ \bibinfo {author} {\bibfnamefont
  {R.}~\bibnamefont {Laflamme}},\ }\bibfield  {title} {\bibinfo {title}
  {Experimental Activation of Strong Local Passive States with Quantum
  Information},\ }\href {https://doi.org/10.1103/PhysRevLett.130.110801}
  {\bibfield  {journal} {\bibinfo  {journal} {Phys. Rev. Lett.}\ }\textbf
  {\bibinfo {volume} {130}},\ \bibinfo {pages} {110801} (\bibinfo {year}
  {2023})}\BibitemShut {NoStop}%
\bibitem [{\citenamefont
  {Ikeda}(2023{\natexlab{a}})}]{PhysRevApplied.20.024051}%
  \BibitemOpen
  \bibfield  {author} {\bibinfo {author} {\bibfnamefont {K.}~\bibnamefont
  {Ikeda}},\ }\bibfield  {title} {\bibinfo {title} {Demonstration of quantum
  energy teleportation on superconducting quantum hardware},\ }\href
  {https://doi.org/10.1103/PhysRevApplied.20.024051} {\bibfield  {journal}
  {\bibinfo  {journal} {Phys. Rev. Appl.}\ }\textbf {\bibinfo {volume} {20}},\
  \bibinfo {pages} {024051} (\bibinfo {year} {2023}{\natexlab{a}})}\BibitemShut
  {NoStop}%
\bibitem [{\citenamefont {Ikeda}(2023{\natexlab{b}})}]{PhysRevD.107.L071502}%
  \BibitemOpen
  \bibfield  {author} {\bibinfo {author} {\bibfnamefont {K.}~\bibnamefont
  {Ikeda}},\ }\bibfield  {title} {\bibinfo {title} {Criticality of quantum
  energy teleportation at phase transition points in quantum field theory},\
  }\href {https://doi.org/10.1103/PhysRevD.107.L071502} {\bibfield  {journal}
  {\bibinfo  {journal} {Phys. Rev. D}\ }\textbf {\bibinfo {volume} {107}},\
  \bibinfo {pages} {L071502} (\bibinfo {year}
  {2023}{\natexlab{b}})}\BibitemShut {NoStop}%
\bibitem [{\citenamefont {Fan}\ \emph {et~al.}(2024)\citenamefont {Fan},
  \citenamefont {Wu}, \citenamefont {Wang}, \citenamefont {Liu},\ and\
  \citenamefont {Liu}}]{PhysRevA.110.052424}%
  \BibitemOpen
  \bibfield  {author} {\bibinfo {author} {\bibfnamefont {H.}~\bibnamefont
  {Fan}}, \bibinfo {author} {\bibfnamefont {F.-L.}\ \bibnamefont {Wu}},
  \bibinfo {author} {\bibfnamefont {L.}~\bibnamefont {Wang}}, \bibinfo {author}
  {\bibfnamefont {S.-Q.}\ \bibnamefont {Liu}},\ and\ \bibinfo {author}
  {\bibfnamefont {S.-Y.}\ \bibnamefont {Liu}},\ }\bibfield  {title} {\bibinfo
  {title} {Strong quantum energy teleportation},\ }\href
  {https://doi.org/10.1103/PhysRevA.110.052424} {\bibfield  {journal} {\bibinfo
   {journal} {Phys. Rev. A}\ }\textbf {\bibinfo {volume} {110}},\ \bibinfo
  {pages} {052424} (\bibinfo {year} {2024})}\BibitemShut {NoStop}%
\bibitem [{\citenamefont {Wang}\ and\ \citenamefont
  {Yao}(2024)}]{Wang2024quantumenergy}%
  \BibitemOpen
  \bibfield  {author} {\bibinfo {author} {\bibfnamefont {J.}~\bibnamefont
  {Wang}}\ and\ \bibinfo {author} {\bibfnamefont {S.}~\bibnamefont {Yao}},\
  }\bibfield  {title} {\bibinfo {title} {Quantum {E}nergy {T}eleportation
  versus {I}nformation {T}eleportation},\ }\href
  {https://doi.org/10.22331/q-2024-12-12-1564} {\bibfield  {journal} {\bibinfo
  {journal} {{Quantum}}\ }\textbf {\bibinfo {volume} {8}},\ \bibinfo {pages}
  {1564} (\bibinfo {year} {2024})}\BibitemShut {NoStop}%
\bibitem [{\citenamefont
  {Preskill}(1992)}]{preskill1992blackholesdestroyinformation}%
  \BibitemOpen
  \bibfield  {author} {\bibinfo {author} {\bibfnamefont {J.}~\bibnamefont
  {Preskill}},\ }\href {https://arxiv.org/abs/hep-th/9209058} {\bibinfo {title}
  {Do black holes destroy information?}},\ \Eprint
  {https://arxiv.org/abs/hep-th/9209058} {arXiv:hep-th/9209058 [hep-th]}
  \BibitemShut {NoStop}%
\bibitem [{\citenamefont {Giddings}(1995)}]{giddings1995black}%
  \BibitemOpen
  \bibfield  {author} {\bibinfo {author} {\bibfnamefont {S.~B.}\ \bibnamefont
  {Giddings}},\ }\bibfield  {title} {\bibinfo {title} {The black hole
  information paradox},\ }\href@noop {} {\bibfield  {journal} {\bibinfo
  {journal} {arXiv preprint hep-th/9508151}\ }}\BibitemShut {NoStop}%
\bibitem [{\citenamefont {Mathur}(2009)}]{Mathur_2009}%
  \BibitemOpen
  \bibfield  {author} {\bibinfo {author} {\bibfnamefont {S.~D.}\ \bibnamefont
  {Mathur}},\ }\bibfield  {title} {\bibinfo {title} {The information paradox: a
  pedagogical introduction},\ }\href
  {https://doi.org/10.1088/0264-9381/26/22/224001} {\bibfield  {journal}
  {\bibinfo  {journal} {Classical and Quantum Gravity}\ }\textbf {\bibinfo
  {volume} {26}},\ \bibinfo {pages} {224001} (\bibinfo {year}
  {2009})}\BibitemShut {NoStop}%
\bibitem [{\citenamefont {Raju}(2022)}]{RAJU20221}%
  \BibitemOpen
  \bibfield  {author} {\bibinfo {author} {\bibfnamefont {S.}~\bibnamefont
  {Raju}},\ }\bibfield  {title} {\bibinfo {title} {Lessons from the information
  paradox},\ }\href
  {https://doi.org/https://doi.org/10.1016/j.physrep.2021.10.001} {\bibfield
  {journal} {\bibinfo  {journal} {Physics Reports}\ }\textbf {\bibinfo {volume}
  {943}},\ \bibinfo {pages} {1} (\bibinfo {year} {2022})}\BibitemShut {NoStop}%
\bibitem [{\citenamefont {Susskind}\ \emph {et~al.}(1993)\citenamefont
  {Susskind}, \citenamefont {Thorlacius},\ and\ \citenamefont
  {Uglum}}]{PhysRevD.48.3743}%
  \BibitemOpen
  \bibfield  {author} {\bibinfo {author} {\bibfnamefont {L.}~\bibnamefont
  {Susskind}}, \bibinfo {author} {\bibfnamefont {L.}~\bibnamefont
  {Thorlacius}},\ and\ \bibinfo {author} {\bibfnamefont {J.}~\bibnamefont
  {Uglum}},\ }\bibfield  {title} {\bibinfo {title} {The stretched horizon and
  black hole complementarity},\ }\href
  {https://doi.org/10.1103/PhysRevD.48.3743} {\bibfield  {journal} {\bibinfo
  {journal} {Phys. Rev. D}\ }\textbf {\bibinfo {volume} {48}},\ \bibinfo
  {pages} {3743} (\bibinfo {year} {1993})}\BibitemShut {NoStop}%
\bibitem [{\citenamefont {Almheiri}\ \emph {et~al.}(2013)\citenamefont
  {Almheiri}, \citenamefont {Marolf}, \citenamefont {Polchinski},\ and\
  \citenamefont {Sully}}]{firewalls}%
  \BibitemOpen
  \bibfield  {author} {\bibinfo {author} {\bibfnamefont {A.}~\bibnamefont
  {Almheiri}}, \bibinfo {author} {\bibfnamefont {D.}~\bibnamefont {Marolf}},
  \bibinfo {author} {\bibfnamefont {J.}~\bibnamefont {Polchinski}},\ and\
  \bibinfo {author} {\bibfnamefont {J.}~\bibnamefont {Sully}},\ }\bibfield
  {title} {\bibinfo {title} {Black holes: complementarity or firewalls?},\
  }\href {https://doi.org/10.1007/JHEP02(2013)062} {\bibfield  {journal}
  {\bibinfo  {journal} {Journal of High Energy Physics}\ }\textbf {\bibinfo
  {volume} {2013}},\ \bibinfo {pages} {62} (\bibinfo {year}
  {2013})}\BibitemShut {NoStop}%
\bibitem [{\citenamefont {Braunstein}\ \emph {et~al.}(2013)\citenamefont
  {Braunstein}, \citenamefont {Pirandola},\ and\ \citenamefont
  {\ifmmode~\dot{Z}\else \.{Z}\fi{}yczkowski}}]{PhysRevLett.110.101301}%
  \BibitemOpen
  \bibfield  {author} {\bibinfo {author} {\bibfnamefont {S.~L.}\ \bibnamefont
  {Braunstein}}, \bibinfo {author} {\bibfnamefont {S.}~\bibnamefont
  {Pirandola}},\ and\ \bibinfo {author} {\bibfnamefont {K.}~\bibnamefont
  {\ifmmode~\dot{Z}\else \.{Z}\fi{}yczkowski}},\ }\bibfield  {title} {\bibinfo
  {title} {Better late than never: Information Retrieval from Black Holes},\
  }\href {https://doi.org/10.1103/PhysRevLett.110.101301} {\bibfield  {journal}
  {\bibinfo  {journal} {Phys. Rev. Lett.}\ }\textbf {\bibinfo {volume} {110}},\
  \bibinfo {pages} {101301} (\bibinfo {year} {2013})}\BibitemShut {NoStop}%
\bibitem [{\citenamefont {Pozas-Kerstjens}\ and\ \citenamefont
  {Mart\'{\i}n-Mart\'{\i}nez}(2015)}]{PhysRevD.92.064042}%
  \BibitemOpen
  \bibfield  {author} {\bibinfo {author} {\bibfnamefont {A.}~\bibnamefont
  {Pozas-Kerstjens}}\ and\ \bibinfo {author} {\bibfnamefont {E.}~\bibnamefont
  {Mart\'{\i}n-Mart\'{\i}nez}},\ }\bibfield  {title} {\bibinfo {title}
  {Harvesting correlations from the quantum vacuum},\ }\href
  {https://doi.org/10.1103/PhysRevD.92.064042} {\bibfield  {journal} {\bibinfo
  {journal} {Phys. Rev. D}\ }\textbf {\bibinfo {volume} {92}},\ \bibinfo
  {pages} {064042} (\bibinfo {year} {2015})}\BibitemShut {NoStop}%
\bibitem [{\citenamefont {Wu}\ \emph {et~al.}(2025)\citenamefont {Wu},
  \citenamefont {Wang}, \citenamefont {Huang},\ and\ \citenamefont
  {Wang}}]{Wu-etanglement}%
  \BibitemOpen
  \bibfield  {author} {\bibinfo {author} {\bibfnamefont {S.-M.}\ \bibnamefont
  {Wu}}, \bibinfo {author} {\bibfnamefont {R.-D.}\ \bibnamefont {Wang}},
  \bibinfo {author} {\bibfnamefont {X.-L.}\ \bibnamefont {Huang}},\ and\
  \bibinfo {author} {\bibfnamefont {Z.}~\bibnamefont {Wang}},\ }\bibfield
  {title} {\bibinfo {title} {Harvesting asymmetric steering via non-identical
  detectors},\ }\href {https://doi.org/10.1140/epjc/s10052-025-14414-4}
  {\bibfield  {journal} {\bibinfo  {journal} {The European Physical Journal C}\
  }\textbf {\bibinfo {volume} {85}},\ \bibinfo {pages} {708} (\bibinfo {year}
  {2025})}\BibitemShut {NoStop}%
\bibitem [{\citenamefont {Teixid\'o-Bonfill}\ and\ \citenamefont
  {Mart\'{\i}n-Mart\'{\i}nez}(2024)}]{PhysRevD.110.105016}%
  \BibitemOpen
  \bibfield  {author} {\bibinfo {author} {\bibfnamefont {A.}~\bibnamefont
  {Teixid\'o-Bonfill}}\ and\ \bibinfo {author} {\bibfnamefont {E.}~\bibnamefont
  {Mart\'{\i}n-Mart\'{\i}nez}},\ }\bibfield  {title} {\bibinfo {title}
  {Derivative coupling enables genuine entanglement harvesting in causal
  communication},\ }\href {https://doi.org/10.1103/PhysRevD.110.105016}
  {\bibfield  {journal} {\bibinfo  {journal} {Phys. Rev. D}\ }\textbf {\bibinfo
  {volume} {110}},\ \bibinfo {pages} {105016} (\bibinfo {year}
  {2024})}\BibitemShut {NoStop}%
\bibitem [{\citenamefont {Wu}\ \emph {et~al.}(2024)\citenamefont {Wu},
  \citenamefont {Wang}, \citenamefont {Huang},\ and\ \citenamefont
  {Wang}}]{Wu2}%
  \BibitemOpen
  \bibfield  {author} {\bibinfo {author} {\bibfnamefont {S.-M.}\ \bibnamefont
  {Wu}}, \bibinfo {author} {\bibfnamefont {R.-D.}\ \bibnamefont {Wang}},
  \bibinfo {author} {\bibfnamefont {X.-L.}\ \bibnamefont {Huang}},\ and\
  \bibinfo {author} {\bibfnamefont {Z.}~\bibnamefont {Wang}},\ }\bibfield
  {title} {\bibinfo {title} {Does gravitational wave assist vacuum steering and
  Bell nonlocality?},\ }\href {https://doi.org/10.1007/JHEP07(2024)155}
  {\bibfield  {journal} {\bibinfo  {journal} {Journal of High Energy Physics}\
  }\textbf {\bibinfo {volume} {2024}},\ \bibinfo {pages} {155} (\bibinfo {year}
  {2024})}\BibitemShut {NoStop}%
\bibitem [{\citenamefont {Bueley}\ \emph {et~al.}(2022)\citenamefont {Bueley},
  \citenamefont {Huang}, \citenamefont {Gallock-Yoshimura},\ and\ \citenamefont
  {Mann}}]{PhysRevD.106.025010}%
  \BibitemOpen
  \bibfield  {author} {\bibinfo {author} {\bibfnamefont {K.}~\bibnamefont
  {Bueley}}, \bibinfo {author} {\bibfnamefont {L.}~\bibnamefont {Huang}},
  \bibinfo {author} {\bibfnamefont {K.}~\bibnamefont {Gallock-Yoshimura}},\
  and\ \bibinfo {author} {\bibfnamefont {R.~B.}\ \bibnamefont {Mann}},\
  }\bibfield  {title} {\bibinfo {title} {Harvesting mutual information from BTZ
  black hole spacetime},\ }\href {https://doi.org/10.1103/PhysRevD.106.025010}
  {\bibfield  {journal} {\bibinfo  {journal} {Phys. Rev. D}\ }\textbf {\bibinfo
  {volume} {106}},\ \bibinfo {pages} {025010} (\bibinfo {year}
  {2022})}\BibitemShut {NoStop}%
\bibitem [{\citenamefont {Barman}\ \emph {et~al.}(2022)\citenamefont {Barman},
  \citenamefont {Barman},\ and\ \citenamefont {Majhi}}]{PhysRevD.106.045005}%
  \BibitemOpen
  \bibfield  {author} {\bibinfo {author} {\bibfnamefont {D.}~\bibnamefont
  {Barman}}, \bibinfo {author} {\bibfnamefont {S.}~\bibnamefont {Barman}},\
  and\ \bibinfo {author} {\bibfnamefont {B.~R.}\ \bibnamefont {Majhi}},\
  }\bibfield  {title} {\bibinfo {title} {Entanglement harvesting between two
  inertial Unruh-Dewitt detectors from nonvacuum quantum fluctuations},\ }\href
  {https://doi.org/10.1103/PhysRevD.106.045005} {\bibfield  {journal} {\bibinfo
   {journal} {Phys. Rev. D}\ }\textbf {\bibinfo {volume} {106}},\ \bibinfo
  {pages} {045005} (\bibinfo {year} {2022})}\BibitemShut {NoStop}%
\bibitem [{\citenamefont {Suryaatmadja}\ \emph {et~al.}(2022)\citenamefont
  {Suryaatmadja}, \citenamefont {Mann},\ and\ \citenamefont
  {Cong}}]{PhysRevD.106.076002}%
  \BibitemOpen
  \bibfield  {author} {\bibinfo {author} {\bibfnamefont {C.}~\bibnamefont
  {Suryaatmadja}}, \bibinfo {author} {\bibfnamefont {R.~B.}\ \bibnamefont
  {Mann}},\ and\ \bibinfo {author} {\bibfnamefont {W.}~\bibnamefont {Cong}},\
  }\bibfield  {title} {\bibinfo {title} {Entanglement harvesting of inertially
  moving Unruh-Dewitt detectors in minkowski spacetime},\ }\href
  {https://doi.org/10.1103/PhysRevD.106.076002} {\bibfield  {journal} {\bibinfo
   {journal} {Phys. Rev. D}\ }\textbf {\bibinfo {volume} {106}},\ \bibinfo
  {pages} {076002} (\bibinfo {year} {2022})}\BibitemShut {NoStop}%
\bibitem [{\citenamefont {Liu}\ \emph {et~al.}(2022)\citenamefont {Liu},
  \citenamefont {Zhang}, \citenamefont {Mann},\ and\ \citenamefont
  {Yu}}]{PhysRevD.105.085012}%
  \BibitemOpen
  \bibfield  {author} {\bibinfo {author} {\bibfnamefont {Z.}~\bibnamefont
  {Liu}}, \bibinfo {author} {\bibfnamefont {J.}~\bibnamefont {Zhang}}, \bibinfo
  {author} {\bibfnamefont {R.~B.}\ \bibnamefont {Mann}},\ and\ \bibinfo
  {author} {\bibfnamefont {H.}~\bibnamefont {Yu}},\ }\bibfield  {title}
  {\bibinfo {title} {Does acceleration assist entanglement harvesting?},\
  }\href {https://doi.org/10.1103/PhysRevD.105.085012} {\bibfield  {journal}
  {\bibinfo  {journal} {Phys. Rev. D}\ }\textbf {\bibinfo {volume} {105}},\
  \bibinfo {pages} {085012} (\bibinfo {year} {2022})}\BibitemShut {NoStop}%
\bibitem [{\citenamefont {Gallock-Yoshimura}\ \emph {et~al.}(2021)\citenamefont
  {Gallock-Yoshimura}, \citenamefont {Tjoa},\ and\ \citenamefont
  {Mann}}]{PhysRevD.104.025001}%
  \BibitemOpen
  \bibfield  {author} {\bibinfo {author} {\bibfnamefont {K.}~\bibnamefont
  {Gallock-Yoshimura}}, \bibinfo {author} {\bibfnamefont {E.}~\bibnamefont
  {Tjoa}},\ and\ \bibinfo {author} {\bibfnamefont {R.~B.}\ \bibnamefont
  {Mann}},\ }\bibfield  {title} {\bibinfo {title} {Harvesting entanglement with
  detectors freely falling into a black hole},\ }\href
  {https://doi.org/10.1103/PhysRevD.104.025001} {\bibfield  {journal} {\bibinfo
   {journal} {Phys. Rev. D}\ }\textbf {\bibinfo {volume} {104}},\ \bibinfo
  {pages} {025001} (\bibinfo {year} {2021})}\BibitemShut {NoStop}%
\bibitem [{\citenamefont {Foo}\ \emph {et~al.}(2021)\citenamefont {Foo},
  \citenamefont {Mann},\ and\ \citenamefont {Zych}}]{PhysRevD.103.065013}%
  \BibitemOpen
  \bibfield  {author} {\bibinfo {author} {\bibfnamefont {J.}~\bibnamefont
  {Foo}}, \bibinfo {author} {\bibfnamefont {R.~B.}\ \bibnamefont {Mann}},\ and\
  \bibinfo {author} {\bibfnamefont {M.}~\bibnamefont {Zych}},\ }\bibfield
  {title} {\bibinfo {title} {Entanglement amplification between superposed
  detectors in flat and curved spacetimes},\ }\href
  {https://doi.org/10.1103/PhysRevD.103.065013} {\bibfield  {journal} {\bibinfo
   {journal} {Phys. Rev. D}\ }\textbf {\bibinfo {volume} {103}},\ \bibinfo
  {pages} {065013} (\bibinfo {year} {2021})}\BibitemShut {NoStop}%
\bibitem [{\citenamefont {Ng}\ \emph {et~al.}(2018)\citenamefont {Ng},
  \citenamefont {Mann},\ and\ \citenamefont
  {Mart\'{\i}n-Mart\'{\i}nez}}]{PhysRevD.97.125011}%
  \BibitemOpen
  \bibfield  {author} {\bibinfo {author} {\bibfnamefont {K.~K.}\ \bibnamefont
  {Ng}}, \bibinfo {author} {\bibfnamefont {R.~B.}\ \bibnamefont {Mann}},\ and\
  \bibinfo {author} {\bibfnamefont {E.}~\bibnamefont
  {Mart\'{\i}n-Mart\'{\i}nez}},\ }\bibfield  {title} {\bibinfo {title} {New
  techniques for entanglement harvesting in flat and curved spacetimes},\
  }\href {https://doi.org/10.1103/PhysRevD.97.125011} {\bibfield  {journal}
  {\bibinfo  {journal} {Phys. Rev. D}\ }\textbf {\bibinfo {volume} {97}},\
  \bibinfo {pages} {125011} (\bibinfo {year} {2018})}\BibitemShut {NoStop}%
\bibitem [{\citenamefont {Li}\ and\ \citenamefont {Zhao}(2025)}]{zhao}%
  \BibitemOpen
  \bibfield  {author} {\bibinfo {author} {\bibfnamefont {R.}~\bibnamefont
  {Li}}\ and\ \bibinfo {author} {\bibfnamefont {Z.}~\bibnamefont {Zhao}},\
  }\bibfield  {title} {\bibinfo {title} {Entanglement harvesting of circularly
  accelerated detectors with a reflecting boundary},\ }\href
  {https://doi.org/10.1007/JHEP03(2025)185} {\bibfield  {journal} {\bibinfo
  {journal} {Journal of High Energy Physics}\ }\textbf {\bibinfo {volume}
  {2025}},\ \bibinfo {pages} {185} (\bibinfo {year} {2025})}\BibitemShut
  {NoStop}%
\bibitem [{\citenamefont {Mart\'{\i}n-Mart\'{\i}nez}\ \emph
  {et~al.}(2016)\citenamefont {Mart\'{\i}n-Mart\'{\i}nez}, \citenamefont
  {Smith},\ and\ \citenamefont {Terno}}]{PhysRevD.93.044001}%
  \BibitemOpen
  \bibfield  {author} {\bibinfo {author} {\bibfnamefont {E.}~\bibnamefont
  {Mart\'{\i}n-Mart\'{\i}nez}}, \bibinfo {author} {\bibfnamefont {A.~R.~H.}\
  \bibnamefont {Smith}},\ and\ \bibinfo {author} {\bibfnamefont {D.~R.}\
  \bibnamefont {Terno}},\ }\bibfield  {title} {\bibinfo {title} {Spacetime
  structure and vacuum entanglement},\ }\href
  {https://doi.org/10.1103/PhysRevD.93.044001} {\bibfield  {journal} {\bibinfo
  {journal} {Phys. Rev. D}\ }\textbf {\bibinfo {volume} {93}},\ \bibinfo
  {pages} {044001} (\bibinfo {year} {2016})}\BibitemShut {NoStop}%
\bibitem [{\citenamefont {Martín-Martínez}\ and\ \citenamefont
  {Menicucci}(2012)}]{Martin2012}%
  \BibitemOpen
  \bibfield  {author} {\bibinfo {author} {\bibfnamefont {E.}~\bibnamefont
  {Martín-Martínez}}\ and\ \bibinfo {author} {\bibfnamefont {N.~C.}\
  \bibnamefont {Menicucci}},\ }\bibfield  {title} {\bibinfo {title}
  {Cosmological quantum entanglement},\ }\href
  {https://doi.org/10.1088/0264-9381/29/22/224003} {\bibfield  {journal}
  {\bibinfo  {journal} {Classical and Quantum Gravity}\ }\textbf {\bibinfo
  {volume} {29}},\ \bibinfo {pages} {224003} (\bibinfo {year}
  {2012})}\BibitemShut {NoStop}%
\bibitem [{\citenamefont {Martín-Martínez}\ and\ \citenamefont
  {Menicucci}(2014)}]{Martin2014}%
  \BibitemOpen
  \bibfield  {author} {\bibinfo {author} {\bibfnamefont {E.}~\bibnamefont
  {Martín-Martínez}}\ and\ \bibinfo {author} {\bibfnamefont {N.~C.}\
  \bibnamefont {Menicucci}},\ }\bibfield  {title} {\bibinfo {title}
  {Entanglement in curved spacetimes and cosmology},\ }\href
  {https://doi.org/10.1088/0264-9381/31/21/214001} {\bibfield  {journal}
  {\bibinfo  {journal} {Classical and Quantum Gravity}\ }\textbf {\bibinfo
  {volume} {31}},\ \bibinfo {pages} {214001} (\bibinfo {year}
  {2014})}\BibitemShut {NoStop}%
\bibitem [{\citenamefont {Liu}\ \emph {et~al.}(2018)\citenamefont {Liu},
  \citenamefont {Tian}, \citenamefont {Wang},\ and\ \citenamefont
  {Jing}}]{PhysRevD.97.105030}%
  \BibitemOpen
  \bibfield  {author} {\bibinfo {author} {\bibfnamefont {X.}~\bibnamefont
  {Liu}}, \bibinfo {author} {\bibfnamefont {Z.}~\bibnamefont {Tian}}, \bibinfo
  {author} {\bibfnamefont {J.}~\bibnamefont {Wang}},\ and\ \bibinfo {author}
  {\bibfnamefont {J.}~\bibnamefont {Jing}},\ }\bibfield  {title} {\bibinfo
  {title} {Radiative process of two entanglement atoms in de Sitter
  spacetime},\ }\href {https://doi.org/10.1103/PhysRevD.97.105030} {\bibfield
  {journal} {\bibinfo  {journal} {Phys. Rev. D}\ }\textbf {\bibinfo {volume}
  {97}},\ \bibinfo {pages} {105030} (\bibinfo {year} {2018})}\BibitemShut
  {NoStop}%
\bibitem [{\citenamefont {Tian}\ \emph {et~al.}(2016)\citenamefont {Tian},
  \citenamefont {Wang}, \citenamefont {Jing},\ and\ \citenamefont
  {Dragan}}]{Tian-Casimir}%
  \BibitemOpen
  \bibfield  {author} {\bibinfo {author} {\bibfnamefont {Z.}~\bibnamefont
  {Tian}}, \bibinfo {author} {\bibfnamefont {J.}~\bibnamefont {Wang}}, \bibinfo
  {author} {\bibfnamefont {J.}~\bibnamefont {Jing}},\ and\ \bibinfo {author}
  {\bibfnamefont {A.}~\bibnamefont {Dragan}},\ }\bibfield  {title} {\bibinfo
  {title} {Detecting the curvature of de Sitter universe with two entangled
  atoms},\ }\href {https://doi.org/10.1038/srep35222} {\bibfield  {journal}
  {\bibinfo  {journal} {Scientific Reports}\ }\textbf {\bibinfo {volume} {6}},\
  \bibinfo {pages} {35222} (\bibinfo {year} {2016})}\BibitemShut {NoStop}%
\bibitem [{\citenamefont {Tian}\ and\ \citenamefont
  {Jing}(2014)}]{Tian-Casimir-Polder}%
  \BibitemOpen
  \bibfield  {author} {\bibinfo {author} {\bibfnamefont {Z.}~\bibnamefont
  {Tian}}\ and\ \bibinfo {author} {\bibfnamefont {J.}~\bibnamefont {Jing}},\
  }\bibfield  {title} {\bibinfo {title} {Distinguishing de Sitter universe from
  thermal Minkowski spacetime by Casimir-Polder-like force},\ }\href
  {https://doi.org/10.1007/JHEP07(2014)089} {\bibfield  {journal} {\bibinfo
  {journal} {Journal of High Energy Physics}\ }\textbf {\bibinfo {volume}
  {2014}},\ \bibinfo {pages} {89} (\bibinfo {year} {2014})}\BibitemShut
  {NoStop}%
\bibitem [{\citenamefont {Chakraborty}\ \emph {et~al.}(2025)\citenamefont
  {Chakraborty}, \citenamefont {Hackl},\ and\ \citenamefont
  {Zych}}]{PhysRevD.111.104052}%
  \BibitemOpen
  \bibfield  {author} {\bibinfo {author} {\bibfnamefont {A.}~\bibnamefont
  {Chakraborty}}, \bibinfo {author} {\bibfnamefont {L.}~\bibnamefont {Hackl}},\
  and\ \bibinfo {author} {\bibfnamefont {M.}~\bibnamefont {Zych}},\ }\bibfield
  {title} {\bibinfo {title} {Entanglement harvesting in quantum superposed
  spacetime},\ }\href {https://doi.org/10.1103/PhysRevD.111.104052} {\bibfield
  {journal} {\bibinfo  {journal} {Phys. Rev. D}\ }\textbf {\bibinfo {volume}
  {111}},\ \bibinfo {pages} {104052} (\bibinfo {year} {2025})}\BibitemShut
  {NoStop}%
\bibitem [{\citenamefont {Ji}\ \emph {et~al.}(2024)\citenamefont {Ji},
  \citenamefont {Zhang},\ and\ \citenamefont {Yu}}]{Yu-Zhang}%
  \BibitemOpen
  \bibfield  {author} {\bibinfo {author} {\bibfnamefont {Y.}~\bibnamefont
  {Ji}}, \bibinfo {author} {\bibfnamefont {J.}~\bibnamefont {Zhang}},\ and\
  \bibinfo {author} {\bibfnamefont {H.}~\bibnamefont {Yu}},\ }\bibfield
  {title} {\bibinfo {title} {Entanglement harvesting in cosmic string
  spacetime},\ }\href {https://doi.org/10.1007/JHEP06(2024)161} {\bibfield
  {journal} {\bibinfo  {journal} {Journal of High Energy Physics}\ }\textbf
  {\bibinfo {volume} {2024}},\ \bibinfo {pages} {161} (\bibinfo {year}
  {2024})}\BibitemShut {NoStop}%
\bibitem [{\citenamefont {Perche}\ \emph {et~al.}(2023)\citenamefont {Perche},
  \citenamefont {Ragula},\ and\ \citenamefont
  {Mart\'{\i}n-Mart\'{\i}nez}}]{PhysRevD.108.085025}%
  \BibitemOpen
  \bibfield  {author} {\bibinfo {author} {\bibfnamefont {T.~R.}\ \bibnamefont
  {Perche}}, \bibinfo {author} {\bibfnamefont {B.}~\bibnamefont {Ragula}},\
  and\ \bibinfo {author} {\bibfnamefont {E.}~\bibnamefont
  {Mart\'{\i}n-Mart\'{\i}nez}},\ }\bibfield  {title} {\bibinfo {title}
  {Harvesting entanglement from the gravitational vacuum},\ }\href
  {https://doi.org/10.1103/PhysRevD.108.085025} {\bibfield  {journal} {\bibinfo
   {journal} {Phys. Rev. D}\ }\textbf {\bibinfo {volume} {108}},\ \bibinfo
  {pages} {085025} (\bibinfo {year} {2023})}\BibitemShut {NoStop}%
\bibitem [{\citenamefont {Amelino-Camelia}(2013)}]{AmelinoCamelia:2008qg}%
  \BibitemOpen
  \bibfield  {author} {\bibinfo {author} {\bibfnamefont {G.}~\bibnamefont
  {Amelino-Camelia}},\ }\bibfield  {title} {\bibinfo {title}
  {{Quantum-Spacetime Phenomenology}},\ }\href
  {https://doi.org/10.12942/lrr-2013-5} {\bibfield  {journal} {\bibinfo
  {journal} {Living Rev. Rel.}\ }\textbf {\bibinfo {volume} {16}},\ \bibinfo
  {pages} {5} (\bibinfo {year} {2013})} \BibitemShut
  {NoStop}%
\bibitem [{\citenamefont {Chatterjee}\ \emph {et~al.}(2021)\citenamefont
  {Chatterjee}, \citenamefont {Gangopadhyay},\ and\ \citenamefont
  {Majumdar}}]{PhysRevD.104.124001}%
  \BibitemOpen
  \bibfield  {author} {\bibinfo {author} {\bibfnamefont {R.}~\bibnamefont
  {Chatterjee}}, \bibinfo {author} {\bibfnamefont {S.}~\bibnamefont
  {Gangopadhyay}},\ and\ \bibinfo {author} {\bibfnamefont {A.~S.}\ \bibnamefont
  {Majumdar}},\ }\bibfield  {title} {\bibinfo {title} {Violation of equivalence
  in an accelerating atom-mirror system in the generalized uncertainty
  principle framework},\ }\href {https://doi.org/10.1103/PhysRevD.104.124001}
  {\bibfield  {journal} {\bibinfo  {journal} {Phys. Rev. D}\ }\textbf {\bibinfo
  {volume} {104}},\ \bibinfo {pages} {124001} (\bibinfo {year}
  {2021})}\BibitemShut {NoStop}%
\bibitem [{\citenamefont {Kajuri}(2016)}]{Kajuri_2016}%
  \BibitemOpen
  \bibfield  {author} {\bibinfo {author} {\bibfnamefont {N.}~\bibnamefont
  {Kajuri}},\ }\bibfield  {title} {\bibinfo {title} {Polymer quantization
  predicts radiation in inertial frames},\ }\href
  {https://doi.org/10.1088/0264-9381/33/5/055007} {\bibfield  {journal}
  {\bibinfo  {journal} {Classical and Quantum Gravity}\ }\textbf {\bibinfo
  {volume} {33}},\ \bibinfo {pages} {055007} (\bibinfo {year}
  {2016})}\BibitemShut {NoStop}%
\bibitem [{\citenamefont {Husain}\ and\ \citenamefont
  {Louko}(2016)}]{PhysRevLett.116.061301}%
  \BibitemOpen
  \bibfield  {author} {\bibinfo {author} {\bibfnamefont {V.}~\bibnamefont
  {Husain}}\ and\ \bibinfo {author} {\bibfnamefont {J.}~\bibnamefont {Louko}},\
  }\bibfield  {title} {\bibinfo {title} {Low Energy Lorentz Violation from
  Modified Dispersion at High Energies},\ }\href
  {https://doi.org/10.1103/PhysRevLett.116.061301} {\bibfield  {journal}
  {\bibinfo  {journal} {Phys. Rev. Lett.}\ }\textbf {\bibinfo {volume} {116}},\
  \bibinfo {pages} {061301} (\bibinfo {year} {2016})}\BibitemShut {NoStop}%
\bibitem [{\citenamefont {Kajuri}\ and\ \citenamefont
  {Sardar}(2018)}]{KAJURI2018412}%
  \BibitemOpen
  \bibfield  {author} {\bibinfo {author} {\bibfnamefont {N.}~\bibnamefont
  {Kajuri}}\ and\ \bibinfo {author} {\bibfnamefont {G.}~\bibnamefont
  {Sardar}},\ }\bibfield  {title} {\bibinfo {title} {Low energy Lorentz
  violation in polymer quantization revisited},\ }\href
  {https://doi.org/https://doi.org/10.1016/j.physletb.2017.11.071} {\bibfield
  {journal} {\bibinfo  {journal} {Physics Letters B}\ }\textbf {\bibinfo
  {volume} {776}},\ \bibinfo {pages} {412} (\bibinfo {year}
  {2018})}\BibitemShut {NoStop}%
\bibitem [{\citenamefont {Louko}\ and\ \citenamefont
  {Upton}(2018)}]{PhysRevD.97.025008}%
  \BibitemOpen
  \bibfield  {author} {\bibinfo {author} {\bibfnamefont {J.}~\bibnamefont
  {Louko}}\ and\ \bibinfo {author} {\bibfnamefont {S.~D.}\ \bibnamefont
  {Upton}},\ }\bibfield  {title} {\bibinfo {title} {Low-energy Lorentz
  violation from high-energy modified dispersion in inertial and circular
  motion},\ }\href {https://doi.org/10.1103/PhysRevD.97.025008} {\bibfield
  {journal} {\bibinfo  {journal} {Phys. Rev. D}\ }\textbf {\bibinfo {volume}
  {97}},\ \bibinfo {pages} {025008} (\bibinfo {year} {2018})}\BibitemShut
  {NoStop}%
\bibitem [{\citenamefont {Tian}\ and\ \citenamefont
  {Du}(2021)}]{PhysRevD.103.085014}%
  \BibitemOpen
  \bibfield  {author} {\bibinfo {author} {\bibfnamefont {Z.}~\bibnamefont
  {Tian}}\ and\ \bibinfo {author} {\bibfnamefont {J.}~\bibnamefont {Du}},\
  }\bibfield  {title} {\bibinfo {title} {Probing low-energy Lorentz violation
  from high-energy modified dispersion in dipolar Bose-Einstein condensates},\
  }\href {https://doi.org/10.1103/PhysRevD.103.085014} {\bibfield  {journal}
  {\bibinfo  {journal} {Phys. Rev. D}\ }\textbf {\bibinfo {volume} {103}},\
  \bibinfo {pages} {085014} (\bibinfo {year} {2021})}\BibitemShut {NoStop}%
\bibitem [{\citenamefont {Wu}\ and\ \citenamefont
  {Tian}(2024)}]{wu2024geometricphaseassisteddetection}%
  \BibitemOpen
  \bibfield  {author} {\bibinfo {author} {\bibfnamefont {Y.}~\bibnamefont
  {Wu}}\ and\ \bibinfo {author} {\bibfnamefont {Z.}~\bibnamefont {Tian}},\
  }\href {https://arxiv.org/abs/2409.09257} {\bibinfo {title} {Geometric phase
  assisted detection of Lorentz-invariance violation from modified dispersion
  at high energies}},\ \Eprint
  {https://arxiv.org/abs/2409.09257} {arXiv:2409.09257 [hep-th]} \BibitemShut
  {NoStop}%
\bibitem [{\citenamefont {Agullo}\ \emph {et~al.}(2010)\citenamefont {Agullo},
  \citenamefont {Navarro-Salas}, \citenamefont {Olmo},\ and\ \citenamefont
  {Parker}}]{Agullo_2010}%
  \BibitemOpen
  \bibfield  {author} {\bibinfo {author} {\bibfnamefont {I.}~\bibnamefont
  {Agullo}}, \bibinfo {author} {\bibfnamefont {J.}~\bibnamefont
  {Navarro-Salas}}, \bibinfo {author} {\bibfnamefont {G.~J.}\ \bibnamefont
  {Olmo}},\ and\ \bibinfo {author} {\bibfnamefont {L.}~\bibnamefont {Parker}},\
  }\bibfield  {title} {\bibinfo {title} {Acceleration radiation, transition
  probabilities and trans-Planckian physics},\ }\href
  {https://doi.org/10.1088/1367-2630/12/9/095017} {\bibfield  {journal}
  {\bibinfo  {journal} {New Journal of Physics}\ }\textbf {\bibinfo {volume}
  {12}},\ \bibinfo {pages} {095017} (\bibinfo {year} {2010})}\BibitemShut
  {NoStop}%
\bibitem [{\citenamefont {Hossain}\ and\ \citenamefont
  {Sardar}(2015)}]{PhysRevD.92.024018}%
  \BibitemOpen
  \bibfield  {author} {\bibinfo {author} {\bibfnamefont {G.~M.}\ \bibnamefont
  {Hossain}}\ and\ \bibinfo {author} {\bibfnamefont {G.}~\bibnamefont
  {Sardar}},\ }\bibfield  {title} {\bibinfo {title} {Violation of the
  Kubo-Martin-Schwinger condition along a Rindler trajectory in polymer
  quantization},\ }\href {https://doi.org/10.1103/PhysRevD.92.024018}
  {\bibfield  {journal} {\bibinfo  {journal} {Phys. Rev. D}\ }\textbf {\bibinfo
  {volume} {92}},\ \bibinfo {pages} {024018} (\bibinfo {year}
  {2015})}\BibitemShut {NoStop}%
\bibitem [{\citenamefont {Carballo-Rubio}\ \emph {et~al.}(2019)\citenamefont
  {Carballo-Rubio}, \citenamefont {Garay}, \citenamefont
  {Mart\'{\i}n-Mart\'{\i}nez},\ and\ \citenamefont
  {de~Ram\'on}}]{PhysRevLett.123.041601}%
  \BibitemOpen
  \bibfield  {author} {\bibinfo {author} {\bibfnamefont {R.}~\bibnamefont
  {Carballo-Rubio}}, \bibinfo {author} {\bibfnamefont {L.~J.}\ \bibnamefont
  {Garay}}, \bibinfo {author} {\bibfnamefont {E.}~\bibnamefont
  {Mart\'{\i}n-Mart\'{\i}nez}},\ and\ \bibinfo {author} {\bibfnamefont
  {J.}~\bibnamefont {de~Ram\'on}},\ }\bibfield  {title} {\bibinfo {title}
  {Unruh Effect without Thermality},\ }\href
  {https://doi.org/10.1103/PhysRevLett.123.041601} {\bibfield  {journal}
  {\bibinfo  {journal} {Phys. Rev. Lett.}\ }\textbf {\bibinfo {volume} {123}},\
  \bibinfo {pages} {041601} (\bibinfo {year} {2019})}\BibitemShut {NoStop}%
\bibitem [{\citenamefont {Hossain}\ and\ \citenamefont
  {Sardar}(2016)}]{Hossain_2016}%
  \BibitemOpen
  \bibfield  {author} {\bibinfo {author} {\bibfnamefont {G.~M.}\ \bibnamefont
  {Hossain}}\ and\ \bibinfo {author} {\bibfnamefont {G.}~\bibnamefont
  {Sardar}},\ }\bibfield  {title} {\bibinfo {title} {Is there Unruh effect in
  polymer quantization?},\ }\href
  {https://doi.org/10.1088/0264-9381/33/24/245016} {\bibfield  {journal}
  {\bibinfo  {journal} {Classical and Quantum Gravity}\ }\textbf {\bibinfo
  {volume} {33}},\ \bibinfo {pages} {245016} (\bibinfo {year}
  {2016})}\BibitemShut {NoStop}%
\bibitem [{\citenamefont {Del~Porro}\ \emph {et~al.}(2025)\citenamefont
  {Del~Porro}, \citenamefont {Herrero-Valea}, \citenamefont {Liberati},\ and\
  \citenamefont {Schneider}}]{Unruh10}%
  \BibitemOpen
  \bibfield  {author} {\bibinfo {author} {\bibfnamefont {F.}~\bibnamefont
  {Del~Porro}}, \bibinfo {author} {\bibfnamefont {M.}~\bibnamefont
  {Herrero-Valea}}, \bibinfo {author} {\bibfnamefont {S.}~\bibnamefont
  {Liberati}},\ and\ \bibinfo {author} {\bibfnamefont {M.}~\bibnamefont
  {Schneider}},\ }\bibfield  {title} {\bibinfo {title} {Rescuing the Unruh
  effect in Lorentz violating gravity},\ }\href
  {https://doi.org/10.1140/epjc/s10052-025-14099-9} {\bibfield  {journal}
  {\bibinfo  {journal} {The European Physical Journal C}\ }\textbf {\bibinfo
  {volume} {85}},\ \bibinfo {pages} {386} (\bibinfo {year} {2025})}\BibitemShut
  {NoStop}%
\bibitem [{\citenamefont {Scardigli}\ \emph {et~al.}(2018)\citenamefont
  {Scardigli}, \citenamefont {Blasone}, \citenamefont {Luciano},\ and\
  \citenamefont {Casadio}}]{GUP}%
  \BibitemOpen
  \bibfield  {author} {\bibinfo {author} {\bibfnamefont {F.}~\bibnamefont
  {Scardigli}}, \bibinfo {author} {\bibfnamefont {M.}~\bibnamefont {Blasone}},
  \bibinfo {author} {\bibfnamefont {G.}~\bibnamefont {Luciano}},\ and\ \bibinfo
  {author} {\bibfnamefont {R.}~\bibnamefont {Casadio}},\ }\bibfield  {title}
  {\bibinfo {title} {Modified Unruh effect from generalized uncertainty
  principle},\ }\href {https://doi.org/10.1140/epjc/s10052-018-6209-y}
  {\bibfield  {journal} {\bibinfo  {journal} {The European Physical Journal C}\
  }\textbf {\bibinfo {volume} {78}},\ \bibinfo {pages} {728} (\bibinfo {year}
  {2018})}\BibitemShut {NoStop}%
\bibitem [{\citenamefont {Tian}\ \emph {et~al.}(2022)\citenamefont {Tian},
  \citenamefont {Wu}, \citenamefont {Zhang}, \citenamefont {Jing},\ and\
  \citenamefont {Du}}]{PhysRevD.106.L061701}%
  \BibitemOpen
  \bibfield  {author} {\bibinfo {author} {\bibfnamefont {Z.}~\bibnamefont
  {Tian}}, \bibinfo {author} {\bibfnamefont {L.}~\bibnamefont {Wu}}, \bibinfo
  {author} {\bibfnamefont {L.}~\bibnamefont {Zhang}}, \bibinfo {author}
  {\bibfnamefont {J.}~\bibnamefont {Jing}},\ and\ \bibinfo {author}
  {\bibfnamefont {J.}~\bibnamefont {Du}},\ }\bibfield  {title} {\bibinfo
  {title} {Probing Lorentz-invariance-violation-induced nonthermal Unruh effect
  in quasi-two-dimensional dipolar condensates},\ }\href
  {https://doi.org/10.1103/PhysRevD.106.L061701} {\bibfield  {journal}
  {\bibinfo  {journal} {Phys. Rev. D}\ }\textbf {\bibinfo {volume} {106}},\
  \bibinfo {pages} {L061701} (\bibinfo {year} {2022})}\BibitemShut {NoStop}%
\bibitem [{\citenamefont
  {Xu}(2025)}]{xu2025momentumresolvedprobinglorentzviolatingdispersion}%
  \BibitemOpen
  \bibfield  {author} {\bibinfo {author} {\bibfnamefont {H.}~\bibnamefont
  {Xu}},\ }\href {https://arxiv.org/abs/2503.17757} {\bibinfo {title}
  {Momentum-resolved probing of Lorentz-violating dispersion relations via
  Unruh-Dewitt detector}},\ \Eprint
  {https://arxiv.org/abs/2503.17757} {arXiv:2503.17757 [gr-qc]} \BibitemShut
  {NoStop}%
\bibitem [{\citenamefont {Colladay}\ and\ \citenamefont
  {Kosteleck\'y}(1997)}]{PhysRevD.55.6760}%
  \BibitemOpen
  \bibfield  {author} {\bibinfo {author} {\bibfnamefont {D.}~\bibnamefont
  {Colladay}}\ and\ \bibinfo {author} {\bibfnamefont {V.~A.}\ \bibnamefont
  {Kosteleck\'y}},\ }\bibfield  {title} {\bibinfo {title} {$\mathrm{CPT}$
  violation and the standard model},\ }\href
  {https://doi.org/10.1103/PhysRevD.55.6760} {\bibfield  {journal} {\bibinfo
  {journal} {Phys. Rev. D}\ }\textbf {\bibinfo {volume} {55}},\ \bibinfo
  {pages} {6760} (\bibinfo {year} {1997})}\BibitemShut {NoStop}%
\bibitem [{\citenamefont {Zhang}\ \emph {et~al.}(2023)\citenamefont {Zhang},
  \citenamefont {Wang},\ and\ \citenamefont {Jing}}]{Zhang:2023wwk}%
  \BibitemOpen
  \bibfield  {author} {\bibinfo {author} {\bibfnamefont {X.}~\bibnamefont
  {Zhang}}, \bibinfo {author} {\bibfnamefont {M.}~\bibnamefont {Wang}},\ and\
  \bibinfo {author} {\bibfnamefont {J.}~\bibnamefont {Jing}},\ }\bibfield
  {title} {\bibinfo {title} {{Quasinormal modes and late time tails of
  perturbation fields on a Schwarzschild-like black hole with a global monopole
  in the Einstein-bumblebee theory}},\ }\href
  {https://doi.org/10.1007/s11433-023-2153-6} {\bibfield  {journal} {\bibinfo
  {journal} {Sci. China Phys. Mech. Astron.}\ }\textbf {\bibinfo {volume}
  {66}},\ \bibinfo {pages} {100411} (\bibinfo {year} {2023})}, \BibitemShut
  {NoStop}%
\bibitem [{\citenamefont {Quan}\ \emph {et~al.}(2025)\citenamefont {Quan},
  \citenamefont {Li}, \citenamefont {Pan}, \citenamefont {Wang},\ and\
  \citenamefont {Jing}}]{Quan:2025tgz}%
  \BibitemOpen
  \bibfield  {author} {\bibinfo {author} {\bibfnamefont {F.}~\bibnamefont
  {Quan}}, \bibinfo {author} {\bibfnamefont {F.}~\bibnamefont {Li}}, \bibinfo
  {author} {\bibfnamefont {Q.}~\bibnamefont {Pan}}, \bibinfo {author}
  {\bibfnamefont {M.}~\bibnamefont {Wang}},\ and\ \bibinfo {author}
  {\bibfnamefont {J.}~\bibnamefont {Jing}},\ }\bibfield  {title} {\bibinfo
  {title} {{Stationary scalar clouds around a rotating BTZ-like black hole in
  the Einstein-bumblebee gravity}},\ }\ \Eprint {https://arxiv.org/abs/2501.15759} {arXiv:2501.15759
  [gr-qc]} \BibitemShut {NoStop}%
\bibitem [{\citenamefont {Colladay}\ and\ \citenamefont
  {Kosteleck\'y}(1998)}]{PhysRevD.58.116002}%
  \BibitemOpen
  \bibfield  {author} {\bibinfo {author} {\bibfnamefont {D.}~\bibnamefont
  {Colladay}}\ and\ \bibinfo {author} {\bibfnamefont {V.~A.}\ \bibnamefont
  {Kosteleck\'y}},\ }\bibfield  {title} {\bibinfo {title} {Lorentz-violating
  extension of the standard model},\ }\href
  {https://doi.org/10.1103/PhysRevD.58.116002} {\bibfield  {journal} {\bibinfo
  {journal} {Phys. Rev. D}\ }\textbf {\bibinfo {volume} {58}},\ \bibinfo
  {pages} {116002} (\bibinfo {year} {1998})}\BibitemShut {NoStop}%
\bibitem [{\citenamefont {Mattingly}(2005)}]{Mattingly:2005re}%
  \BibitemOpen
  \bibfield  {author} {\bibinfo {author} {\bibfnamefont {D.}~\bibnamefont
  {Mattingly}},\ }\bibfield  {title} {\bibinfo {title} {{Modern tests of
  Lorentz invariance}},\ }\href {https://doi.org/10.12942/lrr-2005-5}
  {\bibfield  {journal} {\bibinfo  {journal} {Living Rev. Rel.}\ }\textbf
  {\bibinfo {volume} {8}},\ \bibinfo {pages} {5} (\bibinfo {year} {2005})},\
  \Eprint {https://arxiv.org/abs/gr-qc/0502097} {arXiv:gr-qc/0502097}
  \BibitemShut {NoStop}%
\bibitem [{\citenamefont {Kosteleck\'y}\ and\ \citenamefont
  {Russell}(2011)}]{RevModPhys.83.11}%
  \BibitemOpen
  \bibfield  {author} {\bibinfo {author} {\bibfnamefont {V.~A.}\ \bibnamefont
  {Kosteleck\'y}}\ and\ \bibinfo {author} {\bibfnamefont {N.}~\bibnamefont
  {Russell}},\ }\bibfield  {title} {\bibinfo {title} {Data Tables for Lorentz
  and $CPT$ Violation},\ }\href {https://doi.org/10.1103/RevModPhys.83.11}
  {\bibfield  {journal} {\bibinfo  {journal} {Rev. Mod. Phys.}\ }\textbf
  {\bibinfo {volume} {83}},\ \bibinfo {pages} {11} (\bibinfo {year}
  {2011})}\BibitemShut {NoStop}%
\bibitem [{\citenamefont {Birrell}\ and\ \citenamefont
  {Davies}(1982)}]{birrell_davies_1982}%
  \BibitemOpen
  \bibfield  {author} {\bibinfo {author} {\bibfnamefont {N.~D.}\ \bibnamefont
  {Birrell}}\ and\ \bibinfo {author} {\bibfnamefont {P.~C.~W.}\ \bibnamefont
  {Davies}},\ }\href {https://doi.org/10.1017/CBO9780511622632} {\emph
  {\bibinfo {title} {Quantum Fields in Curved Space}}},\ Cambridge Monographs
  on Mathematical Physics\ (\bibinfo  {publisher} {Cambridge University
  Press},\ \bibinfo {year} {1982})\BibitemShut {NoStop}%
\bibitem [{\citenamefont {Takagi}(1986)}]{10.1143/PTP.88.1}%
  \BibitemOpen
  \bibfield  {author} {\bibinfo {author} {\bibfnamefont {S.}~\bibnamefont
  {Takagi}},\ }\bibfield  {title} {\bibinfo {title} {{Vacuum Noise and Stress
  Induced by Uniform Acceleration: Hawking-Unruh Effect in Rindler Manifold of
  Arbitrary Dimension}},\ }\href {https://doi.org/10.1143/PTP.88.1} {\bibfield
  {journal} {\bibinfo  {journal} {Progress of Theoretical Physics Supplement}\
  }\textbf {\bibinfo {volume} {88}},\ \bibinfo {pages} {1} (\bibinfo {year}
  {1986})}\BibitemShut {NoStop}%
\bibitem [{\citenamefont {Crispino}\ \emph {et~al.}(2008)\citenamefont
  {Crispino}, \citenamefont {Higuchi},\ and\ \citenamefont
  {Matsas}}]{RevModPhys.80.787}%
  \BibitemOpen
  \bibfield  {author} {\bibinfo {author} {\bibfnamefont {L.~C.~B.}\
  \bibnamefont {Crispino}}, \bibinfo {author} {\bibfnamefont {A.}~\bibnamefont
  {Higuchi}},\ and\ \bibinfo {author} {\bibfnamefont {G.~E.~A.}\ \bibnamefont
  {Matsas}},\ }\bibfield  {title} {\bibinfo {title} {The Unruh Effect and its
  Applications},\ }\href {https://doi.org/10.1103/RevModPhys.80.787} {\bibfield
   {journal} {\bibinfo  {journal} {Rev. Mod. Phys.}\ }\textbf {\bibinfo
  {volume} {80}},\ \bibinfo {pages} {787} (\bibinfo {year} {2008})}\BibitemShut
  {NoStop}%
\bibitem [{\citenamefont {Hu}\ \emph {et~al.}(2012)\citenamefont {Hu},
  \citenamefont {Lin},\ and\ \citenamefont {Louko}}]{Hu_2012}%
  \BibitemOpen
  \bibfield  {author} {\bibinfo {author} {\bibfnamefont {B.~L.}\ \bibnamefont
  {Hu}}, \bibinfo {author} {\bibfnamefont {S.-Y.}\ \bibnamefont {Lin}},\ and\
  \bibinfo {author} {\bibfnamefont {J.}~\bibnamefont {Louko}},\ }\bibfield
  {title} {\bibinfo {title} {Relativistic quantum information in
  detectors{\textendash}field interactions},\ }\href
  {https://doi.org/10.1088/0264-9381/29/22/224005} {\bibfield  {journal}
  {\bibinfo  {journal} {Classical and Quantum Gravity}\ }\textbf {\bibinfo
  {volume} {29}},\ \bibinfo {pages} {224005} (\bibinfo {year}
  {2012})}\BibitemShut {NoStop}%
\bibitem [{\citenamefont {Liu}\ \emph {et~al.}(2025)\citenamefont {Liu},
  \citenamefont {Liu}, \citenamefont {Liu},\ and\ \citenamefont
  {Wang}}]{liu2025harvestingcorrelationsbtzblack}%
  \BibitemOpen
  \bibfield  {author} {\bibinfo {author} {\bibfnamefont {X.}~\bibnamefont
  {Liu}}, \bibinfo {author} {\bibfnamefont {W.}~\bibnamefont {Liu}}, \bibinfo
  {author} {\bibfnamefont {Z.}~\bibnamefont {Liu}},\ and\ \bibinfo {author}
  {\bibfnamefont {J.}~\bibnamefont {Wang}},\ }\href
  {https://arxiv.org/abs/2503.06404} {\bibinfo {title} {Harvesting correlations
  from BTZ black hole coupled to a Lorentz-violating vector field}},\ \Eprint {https://arxiv.org/abs/2503.06404} {arXiv:2503.06404
  [gr-qc]} \BibitemShut {NoStop}%
\bibitem [{\citenamefont {Baranov}(2008)}]{BARANOV200871}%
  \BibitemOpen
  \bibfield  {author} {\bibinfo {author} {\bibfnamefont {M.}~\bibnamefont
  {Baranov}},\ }\bibfield  {title} {\bibinfo {title} {Theoretical progress in
  many-body physics with ultracold dipolar gases},\ }\href
  {https://doi.org/https://doi.org/10.1016/j.physrep.2008.04.007} {\bibfield
  {journal} {\bibinfo  {journal} {Physics Reports}\ }\textbf {\bibinfo {volume}
  {464}},\ \bibinfo {pages} {71 } (\bibinfo {year} {2008})}\BibitemShut
  {NoStop}%
\bibitem [{\citenamefont {Recati}\ \emph {et~al.}(2005)\citenamefont {Recati},
  \citenamefont {Fedichev}, \citenamefont {Zwerger}, \citenamefont {von
  Delft},\ and\ \citenamefont {Zoller}}]{PhysRevLett.94.040404}%
  \BibitemOpen
  \bibfield  {author} {\bibinfo {author} {\bibfnamefont {A.}~\bibnamefont
  {Recati}}, \bibinfo {author} {\bibfnamefont {P.~O.}\ \bibnamefont
  {Fedichev}}, \bibinfo {author} {\bibfnamefont {W.}~\bibnamefont {Zwerger}},
  \bibinfo {author} {\bibfnamefont {J.}~\bibnamefont {von Delft}},\ and\
  \bibinfo {author} {\bibfnamefont {P.}~\bibnamefont {Zoller}},\ }\bibfield
  {title} {\bibinfo {title} {Atomic Quantum Dots Coupled to a Reservoir of a
  Superfluid Bose-Einstein Condensate},\ }\href
  {https://doi.org/10.1103/PhysRevLett.94.040404} {\bibfield  {journal}
  {\bibinfo  {journal} {Phys. Rev. Lett.}\ }\textbf {\bibinfo {volume} {94}},\
  \bibinfo {pages} {040404} (\bibinfo {year} {2005})}\BibitemShut {NoStop}%
\bibitem [{\citenamefont {Fedichev}\ and\ \citenamefont
  {Fischer}(2003)}]{PhysRevLett.91.240407}%
  \BibitemOpen
  \bibfield  {author} {\bibinfo {author} {\bibfnamefont {P.~O.}\ \bibnamefont
  {Fedichev}}\ and\ \bibinfo {author} {\bibfnamefont {U.~R.}\ \bibnamefont
  {Fischer}},\ }\bibfield  {title} {\bibinfo {title} {Gibbons-Hawking Effect in
  the Sonic de Sitter Space-time of an Expanding Bose-Einstein-Condensed Gas},\
  }\href {https://doi.org/10.1103/PhysRevLett.91.240407} {\bibfield  {journal}
  {\bibinfo  {journal} {Phys. Rev. Lett.}\ }\textbf {\bibinfo {volume} {91}},\
  \bibinfo {pages} {240407} (\bibinfo {year} {2003})}\BibitemShut {NoStop}%
\bibitem [{\citenamefont {Ronen}\ \emph {et~al.}(2007)\citenamefont {Ronen},
  \citenamefont {Bortolotti},\ and\ \citenamefont
  {Bohn}}]{PhysRevLett.98.030406}%
  \BibitemOpen
  \bibfield  {author} {\bibinfo {author} {\bibfnamefont {S.}~\bibnamefont
  {Ronen}}, \bibinfo {author} {\bibfnamefont {D.~C.~E.}\ \bibnamefont
  {Bortolotti}},\ and\ \bibinfo {author} {\bibfnamefont {J.~L.}\ \bibnamefont
  {Bohn}},\ }\bibfield  {title} {\bibinfo {title} {Radial and Angular Rotons in
  Trapped Dipolar Gases},\ }\href
  {https://doi.org/10.1103/PhysRevLett.98.030406} {\bibfield  {journal}
  {\bibinfo  {journal} {Phys. Rev. Lett.}\ }\textbf {\bibinfo {volume} {98}},\
  \bibinfo {pages} {030406} (\bibinfo {year} {2007})}\BibitemShut {NoStop}%
\bibitem [{\citenamefont {Tian}\ \emph {et~al.}(2018)\citenamefont {Tian},
  \citenamefont {Ch\"a},\ and\ \citenamefont {Fischer}}]{PhysRevA.97.063611}%
  \BibitemOpen
  \bibfield  {author} {\bibinfo {author} {\bibfnamefont {Z.}~\bibnamefont
  {Tian}}, \bibinfo {author} {\bibfnamefont {S.-Y.}\ \bibnamefont {Ch\"a}},\
  and\ \bibinfo {author} {\bibfnamefont {U.~R.}\ \bibnamefont {Fischer}},\
  }\bibfield  {title} {\bibinfo {title} {Roton entanglement in quenched dipolar
  Bose-Einstein condensates},\ }\href
  {https://doi.org/10.1103/PhysRevA.97.063611} {\bibfield  {journal} {\bibinfo
  {journal} {Phys. Rev. A}\ }\textbf {\bibinfo {volume} {97}},\ \bibinfo
  {pages} {063611} (\bibinfo {year} {2018})}\BibitemShut {NoStop}%
\bibitem [{\citenamefont {Ch\"a}\ and\ \citenamefont
  {Fischer}(2017)}]{PhysRevLett.118.130404}%
  \BibitemOpen
  \bibfield  {author} {\bibinfo {author} {\bibfnamefont {S.-Y.}\ \bibnamefont
  {Ch\"a}}\ and\ \bibinfo {author} {\bibfnamefont {U.~R.}\ \bibnamefont
  {Fischer}},\ }\bibfield  {title} {\bibinfo {title} {Probing the Scale
  Invariance of the Inflationary Power Spectrum in Expanding
  Quasi-Two-Dimensional Dipolar Condensates},\ }\href
  {https://doi.org/10.1103/PhysRevLett.118.130404} {\bibfield  {journal}
  {\bibinfo  {journal} {Phys. Rev. Lett.}\ }\textbf {\bibinfo {volume} {118}},\
  \bibinfo {pages} {130404} (\bibinfo {year} {2017})}\BibitemShut {NoStop}%
\bibitem [{\citenamefont {Fischer}(2006)}]{PhysRevA.73.031602}%
  \BibitemOpen
  \bibfield  {author} {\bibinfo {author} {\bibfnamefont {U.~R.}\ \bibnamefont
  {Fischer}},\ }\bibfield  {title} {\bibinfo {title} {Stability of
  quasi-two-dimensional Bose-Einstein condensates with dominant dipole-dipole
  interactions},\ }\href {https://doi.org/10.1103/PhysRevA.73.031602}
  {\bibfield  {journal} {\bibinfo  {journal} {Phys. Rev. A}\ }\textbf {\bibinfo
  {volume} {73}},\ \bibinfo {pages} {031602} (\bibinfo {year}
  {2006})}\BibitemShut {NoStop}%
\bibitem [{\citenamefont {Santos}\ \emph {et~al.}(2003)\citenamefont {Santos},
  \citenamefont {Shlyapnikov},\ and\ \citenamefont
  {Lewenstein}}]{PhysRevLett.90.250403}%
  \BibitemOpen
  \bibfield  {author} {\bibinfo {author} {\bibfnamefont {L.}~\bibnamefont
  {Santos}}, \bibinfo {author} {\bibfnamefont {G.~V.}\ \bibnamefont
  {Shlyapnikov}},\ and\ \bibinfo {author} {\bibfnamefont {M.}~\bibnamefont
  {Lewenstein}},\ }\bibfield  {title} {\bibinfo {title} {Roton-maxon Spectrum
  and Stability of Trapped Dipolar Bose-Einstein Condensates},\ }\href
  {https://doi.org/10.1103/PhysRevLett.90.250403} {\bibfield  {journal}
  {\bibinfo  {journal} {Phys. Rev. Lett.}\ }\textbf {\bibinfo {volume} {90}},\
  \bibinfo {pages} {250403} (\bibinfo {year} {2003})}\BibitemShut {NoStop}%
\bibitem [{\citenamefont {Edwards}\ and\ \citenamefont
  {Kostelecký}(2018)}]{EDWARDS2018319}%
  \BibitemOpen
  \bibfield  {author} {\bibinfo {author} {\bibfnamefont {B.~R.}\ \bibnamefont
  {Edwards}}\ and\ \bibinfo {author} {\bibfnamefont {V.~A.}\ \bibnamefont
  {Kostelecký}},\ }\bibfield  {title} {\bibinfo {title} {Riemann–Finsler
  geometry and Lorentz-violating scalar fields},\ }\href
  {https://doi.org/https://doi.org/10.1016/j.physletb.2018.10.011} {\bibfield
  {journal} {\bibinfo  {journal} {Physics Letters B}\ }\textbf {\bibinfo
  {volume} {786}},\ \bibinfo {pages} {319 } (\bibinfo {year}
  {2018})}\BibitemShut {NoStop}%
\bibitem [{\citenamefont {{Castin}}(2001)}]{2001camw.book1C}%
  \BibitemOpen
  \bibfield  {author} {\bibinfo {author} {\bibfnamefont {Y.}~\bibnamefont
  {{Castin}}},\ }\bibinfo {title} {{{Bose-Einstein Condensates in Atomic Gases:
  Simple Theoretical Results}}},\ in\ \href
  {https://doi.org/10.1007/3-540-45338-5_1} {\emph {\bibinfo {booktitle}
  {Coherent atomic matter waves}}},\ \bibinfo {series and number} {Les Houches
  Session LXXII},\ \bibinfo {editor} {edited by\ \bibinfo {editor}
  {\bibfnamefont {R.}~\bibnamefont {{Kaiser}}}, \bibinfo {editor}
  {\bibfnamefont {C.}~\bibnamefont {{Westbrook}}},\ and\ \bibinfo {editor}
  {\bibfnamefont {F.}~\bibnamefont {{David}}}}\ (\bibinfo  {publisher}
  {Springer, Berlin},\ \bibinfo {year} {2001})\BibitemShut {NoStop}%
\bibitem [{\citenamefont {Courteille}\ \emph {et~al.}(1998)\citenamefont
  {Courteille}, \citenamefont {Freeland}, \citenamefont {Heinzen},
  \citenamefont {van Abeelen},\ and\ \citenamefont
  {Verhaar}}]{PhysRevLett.81.69}%
  \BibitemOpen
  \bibfield  {author} {\bibinfo {author} {\bibfnamefont {P.}~\bibnamefont
  {Courteille}}, \bibinfo {author} {\bibfnamefont {R.~S.}\ \bibnamefont
  {Freeland}}, \bibinfo {author} {\bibfnamefont {D.~J.}\ \bibnamefont
  {Heinzen}}, \bibinfo {author} {\bibfnamefont {F.~A.}\ \bibnamefont {van
  Abeelen}},\ and\ \bibinfo {author} {\bibfnamefont {B.~J.}\ \bibnamefont
  {Verhaar}},\ }\bibfield  {title} {\bibinfo {title} {Observation of a Feshbach
  Resonance in Cold Atom Scattering},\ }\href
  {https://doi.org/10.1103/PhysRevLett.81.69} {\bibfield  {journal} {\bibinfo
  {journal} {Phys. Rev. Lett.}\ }\textbf {\bibinfo {volume} {81}},\ \bibinfo
  {pages} {69} (\bibinfo {year} {1998})}\BibitemShut {NoStop}%
\bibitem [{\citenamefont {Inouye}\ \emph {et~al.}(1998)\citenamefont {Inouye},
  \citenamefont {Andrews}, \citenamefont {Stenger}, \citenamefont {Miesner},
  \citenamefont {Stamper-Kurn},\ and\ \citenamefont
  {Ketterle}}]{Inouye1998Observation}%
  \BibitemOpen
  \bibfield  {author} {\bibinfo {author} {\bibfnamefont {S.}~\bibnamefont
  {Inouye}}, \bibinfo {author} {\bibfnamefont {M.~R.}\ \bibnamefont {Andrews}},
  \bibinfo {author} {\bibfnamefont {J.}~\bibnamefont {Stenger}}, \bibinfo
  {author} {\bibfnamefont {H.~J.}\ \bibnamefont {Miesner}}, \bibinfo {author}
  {\bibfnamefont {D.~M.}\ \bibnamefont {Stamper-Kurn}},\ and\ \bibinfo {author}
  {\bibfnamefont {W.}~\bibnamefont {Ketterle}},\ }\bibfield  {title} {\bibinfo
  {title} {Observation of Feshbach resonances in a Bose-Einstein condensate},\
  }\href {https://doi.org/10.1038/32354} {\bibfield  {journal} {\bibinfo
  {journal} {Nature}\ }\textbf {\bibinfo {volume} {392}},\ \bibinfo {pages}
  {151} (\bibinfo {year} {1998})}\BibitemShut {NoStop}%
\bibitem [{\citenamefont {Giovanazzi}\ \emph {et~al.}(2002)\citenamefont
  {Giovanazzi}, \citenamefont {G\"orlitz},\ and\ \citenamefont
  {Pfau}}]{PhysRevLett.89.130401}%
  \BibitemOpen
  \bibfield  {author} {\bibinfo {author} {\bibfnamefont {S.}~\bibnamefont
  {Giovanazzi}}, \bibinfo {author} {\bibfnamefont {A.}~\bibnamefont
  {G\"orlitz}},\ and\ \bibinfo {author} {\bibfnamefont {T.}~\bibnamefont
  {Pfau}},\ }\bibfield  {title} {\bibinfo {title} {Tuning the Dipolar
  Interaction in Quantum Gases},\ }\href
  {https://doi.org/10.1103/PhysRevLett.89.130401} {\bibfield  {journal}
  {\bibinfo  {journal} {Phys. Rev. Lett.}\ }\textbf {\bibinfo {volume} {89}},\
  \bibinfo {pages} {130401} (\bibinfo {year} {2002})}\BibitemShut {NoStop}%
\bibitem [{\citenamefont {Holanda~Ribeiro}\ and\ \citenamefont
  {Fischer}(2023)}]{PhysRevD.107.L121502}%
  \BibitemOpen
  \bibfield  {author} {\bibinfo {author} {\bibfnamefont {C.~C.}\ \bibnamefont
  {Holanda~Ribeiro}}\ and\ \bibinfo {author} {\bibfnamefont {U.~R.}\
  \bibnamefont {Fischer}},\ }\bibfield  {title} {\bibinfo {title} {Impact of
  trans-Planckian excitations on black-hole radiation in dipolar condensates},\
  }\href {https://doi.org/10.1103/PhysRevD.107.L121502} {\bibfield  {journal}
  {\bibinfo  {journal} {Phys. Rev. D}\ }\textbf {\bibinfo {volume} {107}},\
  \bibinfo {pages} {L121502} (\bibinfo {year} {2023})}\BibitemShut {NoStop}%
\bibitem [{\citenamefont {Rana}\ \emph {et~al.}(2023)\citenamefont {Rana},
  \citenamefont {Pendse}, \citenamefont {W{\"u}ster},\ and\ \citenamefont
  {Panda}}]{Rana_2023}%
  \BibitemOpen
  \bibfield  {author} {\bibinfo {author} {\bibfnamefont {A.}~\bibnamefont
  {Rana}}, \bibinfo {author} {\bibfnamefont {A.}~\bibnamefont {Pendse}},
  \bibinfo {author} {\bibfnamefont {S.}~\bibnamefont {W{\"u}ster}},\ and\
  \bibinfo {author} {\bibfnamefont {S.}~\bibnamefont {Panda}},\ }\bibfield
  {title} {\bibinfo {title} {Anisotropic inflation in dipolar Bose-Einstein
  condensates},\ }\href {https://doi.org/10.1088/1367-2630/ad091f} {\bibfield
  {journal} {\bibinfo  {journal} {New Journal of Physics}\ }\textbf {\bibinfo
  {volume} {25}},\ \bibinfo {pages} {113040} (\bibinfo {year}
  {2023})}\BibitemShut {NoStop}%
\bibitem [{\citenamefont {Ribeiro}\ and\ \citenamefont
  {Fischer}(2023)}]{10.21468/SciPostPhysCore.6.1.003}%
  \BibitemOpen
  \bibfield  {author} {\bibinfo {author} {\bibfnamefont {C.~C.~H.}\
  \bibnamefont {Ribeiro}}\ and\ \bibinfo {author} {\bibfnamefont {U.~R.}\
  \bibnamefont {Fischer}},\ }\bibfield  {title} {\bibinfo {title} {{Nonlocal
  field theory of quasiparticle scattering in dipolar Bose-Einstein
  condensates}},\ }\href {https://doi.org/10.21468/SciPostPhysCore.6.1.003}
  {\bibfield  {journal} {\bibinfo  {journal} {SciPost Phys. Core}\ }\textbf
  {\bibinfo {volume} {6}},\ \bibinfo {pages} {003} (\bibinfo {year}
  {2023})}\BibitemShut {NoStop}%
\bibitem [{\citenamefont {Chandran}\ and\ \citenamefont
  {Fischer}(2025)}]{chandran2025expansioncontractiondualitybreakingplanckscale}%
  \BibitemOpen
  \bibfield  {author} {\bibinfo {author} {\bibfnamefont {S.~M.}\ \bibnamefont
  {Chandran}}\ and\ \bibinfo {author} {\bibfnamefont {U.~R.}\ \bibnamefont
  {Fischer}},\ }\href {https://arxiv.org/abs/2506.02719} {\bibinfo {title}
  {Expansion-contraction duality breaking in a planck-scale sensitive
  cosmological quantum simulator}},\ \Eprint
  {https://arxiv.org/abs/2506.02719} {arXiv:2506.02719 [gr-qc]} \BibitemShut
  {NoStop}%
\bibitem [{\citenamefont {Dewitt}(1979)}]{Dewitt1979General}%
  \BibitemOpen
  \bibfield  {author} {\bibinfo {author} {\bibfnamefont {B.}~\bibnamefont
  {Dewitt}},\ }\href@noop {} {\emph {\bibinfo {title} {General relativity:an
  Einstein centenary survey}}}\ (\bibinfo  {publisher} {Cambridge University
  Press},\ \bibinfo {year} {1979})\BibitemShut {NoStop}%
\bibitem [{\citenamefont {Mart\'{\i}n-Mart\'{\i}nez}\ \emph
  {et~al.}(2013)\citenamefont {Mart\'{\i}n-Mart\'{\i}nez}, \citenamefont
  {Montero},\ and\ \citenamefont {del Rey}}]{PhysRevD.87.064038}%
  \BibitemOpen
  \bibfield  {author} {\bibinfo {author} {\bibfnamefont {E.}~\bibnamefont
  {Mart\'{\i}n-Mart\'{\i}nez}}, \bibinfo {author} {\bibfnamefont
  {M.}~\bibnamefont {Montero}},\ and\ \bibinfo {author} {\bibfnamefont
  {M.}~\bibnamefont {del Rey}},\ }\bibfield  {title} {\bibinfo {title}
  {Wavepacket detection with the Unruh-Dewitt model},\ }\href
  {https://doi.org/10.1103/PhysRevD.87.064038} {\bibfield  {journal} {\bibinfo
  {journal} {Phys. Rev. D}\ }\textbf {\bibinfo {volume} {87}},\ \bibinfo
  {pages} {064038} (\bibinfo {year} {2013})}\BibitemShut {NoStop}%
\bibitem [{\citenamefont {Alhambra}\ \emph {et~al.}(2014)\citenamefont
  {Alhambra}, \citenamefont {Kempf},\ and\ \citenamefont
  {Mart\'{\i}n-Mart\'{\i}nez}}]{PhysRevA.89.033835}%
  \BibitemOpen
  \bibfield  {author} {\bibinfo {author} {\bibfnamefont {A.~M.}\ \bibnamefont
  {Alhambra}}, \bibinfo {author} {\bibfnamefont {A.}~\bibnamefont {Kempf}},\
  and\ \bibinfo {author} {\bibfnamefont {E.}~\bibnamefont
  {Mart\'{\i}n-Mart\'{\i}nez}},\ }\bibfield  {title} {\bibinfo {title} {Casimir
  forces on atoms in optical cavities},\ }\href
  {https://doi.org/10.1103/PhysRevA.89.033835} {\bibfield  {journal} {\bibinfo
  {journal} {Phys. Rev. A}\ }\textbf {\bibinfo {volume} {89}},\ \bibinfo
  {pages} {033835} (\bibinfo {year} {2014})}\BibitemShut {NoStop}%
\bibitem [{\citenamefont {Marino}\ \emph {et~al.}(2017)\citenamefont {Marino},
  \citenamefont {Recati},\ and\ \citenamefont
  {Carusotto}}]{PhysRevLett.118.045301}%
  \BibitemOpen
  \bibfield  {author} {\bibinfo {author} {\bibfnamefont {J.}~\bibnamefont
  {Marino}}, \bibinfo {author} {\bibfnamefont {A.}~\bibnamefont {Recati}},\
  and\ \bibinfo {author} {\bibfnamefont {I.}~\bibnamefont {Carusotto}},\
  }\bibfield  {title} {\bibinfo {title} {Casimir Forces and Quantum Friction
  from Ginzburg Radiation in Atomic Bose-Einstein Condensates},\ }\href
  {https://doi.org/10.1103/PhysRevLett.118.045301} {\bibfield  {journal}
  {\bibinfo  {journal} {Phys. Rev. Lett.}\ }\textbf {\bibinfo {volume} {118}},\
  \bibinfo {pages} {045301} (\bibinfo {year} {2017})}\BibitemShut {NoStop}%
  \bibitem [{\citenamefont {Marino}\ \emph {et~al.}(2020)\citenamefont {Marino},
  \citenamefont {Menezes},\ and\ \citenamefont
  {Carusotto}}]{PhysRevResearch.2.042009}%
  \BibitemOpen
  \bibfield  {author} {\bibinfo {author} {\bibfnamefont {Jamir}\ \bibnamefont
  {Marino}}, \bibinfo {author} {\bibfnamefont {Gabriel}\ \bibnamefont
  {Menezes}}, \ and\ \bibinfo {author} {\bibfnamefont {Iacopo}\ \bibnamefont
  {Carusotto}},\ }\bibfield  {title} {\enquote {\bibinfo {title} {Zero-point
  excitation of a circularly moving detector in an atomic condensate and phonon
  laser dynamical instabilities},}\ }\href {\doibase
  10.1103/PhysRevResearch.2.042009} {\bibfield  {journal} {\bibinfo  {journal}
  {Phys. Rev. Res.}\ }\textbf {\bibinfo {volume} {2}},\ \bibinfo {pages}
  {042009} (\bibinfo {year} {2020})}\BibitemShut {NoStop}%
\bibitem [{Sup()}]{Supplemental-Material}%
  \BibitemOpen
  \href@noop {} {\emph {\bibinfo {title} {See Supplemental Material for
  detailed calculations of the density matrix for the final state of the
  detectors}}}\BibitemShut {NoStop}%
\bibitem [{\citenamefont {Wootters}(1998)}]{PhysRevLett.80.2245}%
  \BibitemOpen
  \bibfield  {author} {\bibinfo {author} {\bibfnamefont {W.~K.}\ \bibnamefont
  {Wootters}},\ }\bibfield  {title} {\bibinfo {title} {Entanglement of
  Formation of an Arbitrary State of Two Qubits},\ }\href
  {https://doi.org/10.1103/PhysRevLett.80.2245} {\bibfield  {journal} {\bibinfo
   {journal} {Phys. Rev. Lett.}\ }\textbf {\bibinfo {volume} {80}},\ \bibinfo
  {pages} {2245} (\bibinfo {year} {1998})}\BibitemShut {NoStop}%
\bibitem [{\citenamefont {Retzker}\ \emph {et~al.}(2008)\citenamefont
  {Retzker}, \citenamefont {Cirac}, \citenamefont {Plenio},\ and\ \citenamefont
  {Reznik}}]{PhysRevLett.101.110402}%
  \BibitemOpen
  \bibfield  {author} {\bibinfo {author} {\bibfnamefont {A.}~\bibnamefont
  {Retzker}}, \bibinfo {author} {\bibfnamefont {J.~I.}\ \bibnamefont {Cirac}},
  \bibinfo {author} {\bibfnamefont {M.~B.}\ \bibnamefont {Plenio}},\ and\
  \bibinfo {author} {\bibfnamefont {B.}~\bibnamefont {Reznik}},\ }\bibfield
  {title} {\bibinfo {title} {Methods for Detecting Acceleration Radiation in a
  Bose-Einstein Condensate},\ }\href
  {https://doi.org/10.1103/PhysRevLett.101.110402} {\bibfield  {journal}
  {\bibinfo  {journal} {Phys. Rev. Lett.}\ }\textbf {\bibinfo {volume} {101}},\
  \bibinfo {pages} {110402} (\bibinfo {year} {2008})}\BibitemShut {NoStop}%
\bibitem [{\citenamefont {Gooding}\ \emph {et~al.}(2020)\citenamefont
  {Gooding}, \citenamefont {Biermann}, \citenamefont {Erne}, \citenamefont
  {Louko}, \citenamefont {Unruh}, \citenamefont {Schmiedmayer},\ and\
  \citenamefont {Weinfurtner}}]{PhysRevLett.125.213603}%
  \BibitemOpen
  \bibfield  {author} {\bibinfo {author} {\bibfnamefont {C.}~\bibnamefont
  {Gooding}}, \bibinfo {author} {\bibfnamefont {S.}~\bibnamefont {Biermann}},
  \bibinfo {author} {\bibfnamefont {S.}~\bibnamefont {Erne}}, \bibinfo {author}
  {\bibfnamefont {J.}~\bibnamefont {Louko}}, \bibinfo {author} {\bibfnamefont
  {W.~G.}\ \bibnamefont {Unruh}}, \bibinfo {author} {\bibfnamefont
  {J.}~\bibnamefont {Schmiedmayer}},\ and\ \bibinfo {author} {\bibfnamefont
  {S.}~\bibnamefont {Weinfurtner}},\ }\bibfield  {title} {\bibinfo {title}
  {Interferometric Unruh Detectors for Bose-Einstein Condensates},\ }\href
  {https://doi.org/10.1103/PhysRevLett.125.213603} {\bibfield  {journal}
  {\bibinfo  {journal} {Phys. Rev. Lett.}\ }\textbf {\bibinfo {volume} {125}},\
  \bibinfo {pages} {213603} (\bibinfo {year} {2020})}\BibitemShut {NoStop}%
\bibitem [{\citenamefont {Tian}\ and\ \citenamefont {Jing}(2023)}]{tianzehua}%
  \BibitemOpen
  \bibfield  {author} {\bibinfo {author} {\bibfnamefont {Z.}~\bibnamefont
  {Tian}}\ and\ \bibinfo {author} {\bibfnamefont {J.}~\bibnamefont {Jing}},\
  }\bibfield  {title} {\bibinfo {title} {Using nanoKelvin quantum thermometry
  to detect timelike Unruh effect in a Bose--Einstein condensate},\ }\href
  {https://doi.org/10.1140/epjc/s10052-023-12191-6} {\bibfield  {journal}
  {\bibinfo  {journal} {The European Physical Journal C}\ }\textbf {\bibinfo
  {volume} {83}},\ \bibinfo {pages} {1022} (\bibinfo {year}
  {2023})}\BibitemShut {NoStop}%
\bibitem [{\citenamefont {Lahaye}\ \emph {et~al.}(2007)\citenamefont {Lahaye},
  \citenamefont {Koch}, \citenamefont {Fr\"ohlich}, \citenamefont {Fattori},
  \citenamefont {Metz}, \citenamefont {Griesmaier}, \citenamefont
  {Giovanazzi},\ and\ \citenamefont {Pfau}}]{Chromium}%
  \BibitemOpen
  \bibfield  {author} {\bibinfo {author} {\bibfnamefont {T.}~\bibnamefont
  {Lahaye}}, \bibinfo {author} {\bibfnamefont {T.}~\bibnamefont {Koch}},
  \bibinfo {author} {\bibfnamefont {B.}~\bibnamefont {Fr\"ohlich}}, \bibinfo
  {author} {\bibfnamefont {M.}~\bibnamefont {Fattori}}, \bibinfo {author}
  {\bibfnamefont {J.}~\bibnamefont {Metz}}, \bibinfo {author} {\bibfnamefont
  {A.}~\bibnamefont {Griesmaier}}, \bibinfo {author} {\bibfnamefont
  {S.}~\bibnamefont {Giovanazzi}},\ and\ \bibinfo {author} {\bibfnamefont
  {T.}~\bibnamefont {Pfau}},\ }\bibfield  {title} {\bibinfo {title} {Strong
  dipolar effects in a quantum ferrofluid},\ }\href
  {http://dx.doi.org/10.1038/nature06036} {\bibfield  {journal} {\bibinfo
  {journal} {Nature}\ }\textbf {\bibinfo {volume} {448}},\ \bibinfo {pages}
  {672} (\bibinfo {year} {2007})}\BibitemShut {NoStop}%
\bibitem [{\citenamefont {Aikawa}\ \emph {et~al.}(2012)\citenamefont {Aikawa},
  \citenamefont {Frisch}, \citenamefont {Mark}, \citenamefont {Baier},
  \citenamefont {Rietzler}, \citenamefont {Grimm},\ and\ \citenamefont
  {Ferlaino}}]{PhysRevLett.108.210401}%
  \BibitemOpen
  \bibfield  {author} {\bibinfo {author} {\bibfnamefont {K.}~\bibnamefont
  {Aikawa}}, \bibinfo {author} {\bibfnamefont {A.}~\bibnamefont {Frisch}},
  \bibinfo {author} {\bibfnamefont {M.}~\bibnamefont {Mark}}, \bibinfo {author}
  {\bibfnamefont {S.}~\bibnamefont {Baier}}, \bibinfo {author} {\bibfnamefont
  {A.}~\bibnamefont {Rietzler}}, \bibinfo {author} {\bibfnamefont
  {R.}~\bibnamefont {Grimm}},\ and\ \bibinfo {author} {\bibfnamefont
  {F.}~\bibnamefont {Ferlaino}},\ }\bibfield  {title} {\bibinfo {title}
  {{Bose-Einstein Condensation of Erbium}},\ }\href
  {https://doi.org/10.1103/PhysRevLett.108.210401} {\bibfield  {journal}
  {\bibinfo  {journal} {Phys. Rev. Lett.}\ }\textbf {\bibinfo {volume} {108}},\
  \bibinfo {pages} {210401} (\bibinfo {year} {2012})}\BibitemShut {NoStop}%
\bibitem [{\citenamefont {Lu}\ \emph {et~al.}(2011)\citenamefont {Lu},
  \citenamefont {Burdick}, \citenamefont {Youn},\ and\ \citenamefont
  {Lev}}]{PhysRevLett.107.190401}%
  \BibitemOpen
  \bibfield  {author} {\bibinfo {author} {\bibfnamefont {M.}~\bibnamefont
  {Lu}}, \bibinfo {author} {\bibfnamefont {N.~Q.}\ \bibnamefont {Burdick}},
  \bibinfo {author} {\bibfnamefont {S.~H.}\ \bibnamefont {Youn}},\ and\
  \bibinfo {author} {\bibfnamefont {B.~L.}\ \bibnamefont {Lev}},\ }\bibfield
  {title} {\bibinfo {title} {{Strongly Dipolar Bose-Einstein Condensate of
  Dysprosium}},\ }\href {https://doi.org/10.1103/PhysRevLett.107.190401}
  {\bibfield  {journal} {\bibinfo  {journal} {Phys. Rev. Lett.}\ }\textbf
  {\bibinfo {volume} {107}},\ \bibinfo {pages} {190401} (\bibinfo {year}
  {2011})}\BibitemShut {NoStop}%
\bibitem [{\citenamefont {Klaus}\ \emph {et~al.}(2022)\citenamefont {Klaus},
  \citenamefont {Bland}, \citenamefont {Poli}, \citenamefont {Politi},
  \citenamefont {Lamporesi}, \citenamefont {Casotti}, \citenamefont {Bisset},
  \citenamefont {Mark},\ and\ \citenamefont {Ferlaino}}]{NP1}%
  \BibitemOpen
  \bibfield  {author} {\bibinfo {author} {\bibfnamefont {L.}~\bibnamefont
  {Klaus}}, \bibinfo {author} {\bibfnamefont {T.}~\bibnamefont {Bland}},
  \bibinfo {author} {\bibfnamefont {E.}~\bibnamefont {Poli}}, \bibinfo {author}
  {\bibfnamefont {C.}~\bibnamefont {Politi}}, \bibinfo {author} {\bibfnamefont
  {G.}~\bibnamefont {Lamporesi}}, \bibinfo {author} {\bibfnamefont
  {E.}~\bibnamefont {Casotti}}, \bibinfo {author} {\bibfnamefont {R.~N.}\
  \bibnamefont {Bisset}}, \bibinfo {author} {\bibfnamefont {M.~J.}\
  \bibnamefont {Mark}},\ and\ \bibinfo {author} {\bibfnamefont
  {F.}~\bibnamefont {Ferlaino}},\ }\bibfield  {title} {\bibinfo {title}
  {Observation of vortices and vortex stripes in a dipolar condensate},\ }\href
  {https://doi.org/10.1038/s41567-022-01793-8} {\bibfield  {journal} {\bibinfo
  {journal} {Nature Physics}\ }\textbf {\bibinfo {volume} {18}},\ \bibinfo
  {pages} {1453} (\bibinfo {year} {2022})}\BibitemShut {NoStop}%
\bibitem [{\citenamefont {Casotti}\ \emph {et~al.}(2024)\citenamefont
  {Casotti}, \citenamefont {Poli}, \citenamefont {Klaus}, \citenamefont
  {Litvinov}, \citenamefont {Ulm}, \citenamefont {Politi}, \citenamefont
  {Mark}, \citenamefont {Bland},\ and\ \citenamefont {Ferlaino}}]{N1}%
  \BibitemOpen
  \bibfield  {author} {\bibinfo {author} {\bibfnamefont {E.}~\bibnamefont
  {Casotti}}, \bibinfo {author} {\bibfnamefont {E.}~\bibnamefont {Poli}},
  \bibinfo {author} {\bibfnamefont {L.}~\bibnamefont {Klaus}}, \bibinfo
  {author} {\bibfnamefont {A.}~\bibnamefont {Litvinov}}, \bibinfo {author}
  {\bibfnamefont {C.}~\bibnamefont {Ulm}}, \bibinfo {author} {\bibfnamefont
  {C.}~\bibnamefont {Politi}}, \bibinfo {author} {\bibfnamefont {M.~J.}\
  \bibnamefont {Mark}}, \bibinfo {author} {\bibfnamefont {T.}~\bibnamefont
  {Bland}},\ and\ \bibinfo {author} {\bibfnamefont {F.}~\bibnamefont
  {Ferlaino}},\ }\bibfield  {title} {\bibinfo {title} {Observation of vortices
  in a dipolar supersolid},\ }\href
  {https://doi.org/10.1038/s41586-024-08149-7} {\bibfield  {journal} {\bibinfo
  {journal} {Nature}\ }\textbf {\bibinfo {volume} {635}},\ \bibinfo {pages}
  {327} (\bibinfo {year} {2024})}\BibitemShut {NoStop}%
\bibitem [{\citenamefont {Chomaz}\ \emph {et~al.}(2022)\citenamefont {Chomaz},
  \citenamefont {Ferrier-Barbut}, \citenamefont {Ferlaino}, \citenamefont
  {Laburthe-Tolra}, \citenamefont {Lev},\ and\ \citenamefont
  {Pfau}}]{chomaz2022dipolar}%
  \BibitemOpen
  \bibfield  {author} {\bibinfo {author} {\bibfnamefont {L.}~\bibnamefont
  {Chomaz}}, \bibinfo {author} {\bibfnamefont {I.}~\bibnamefont
  {Ferrier-Barbut}}, \bibinfo {author} {\bibfnamefont {F.}~\bibnamefont
  {Ferlaino}}, \bibinfo {author} {\bibfnamefont {B.}~\bibnamefont
  {Laburthe-Tolra}}, \bibinfo {author} {\bibfnamefont {B.~L.}\ \bibnamefont
  {Lev}},\ and\ \bibinfo {author} {\bibfnamefont {T.}~\bibnamefont {Pfau}},\
  }\bibfield  {title} {\bibinfo {title} {Dipolar physics: a review of
  experiments with magnetic quantum gases},\ }\href@noop {} {\bibfield
  {journal} {\bibinfo  {journal} {Reports on Progress in Physics}\ }\textbf
  {\bibinfo {volume} {86}},\ \bibinfo {pages} {026401} (\bibinfo {year}
  {2022})}\BibitemShut {NoStop}%
\bibitem [{\citenamefont {Qu\'em\'ener}\ and\ \citenamefont
  {Julienne}(2012)}]{doi:10.1021/cr300092g}%
  \BibitemOpen
  \bibfield  {author} {\bibinfo {author} {\bibfnamefont {G.}~\bibnamefont
  {Qu\'em\'ener}}\ and\ \bibinfo {author} {\bibfnamefont {P.~S.}\ \bibnamefont
  {Julienne}},\ }\bibfield  {title} {\bibinfo {title} {{Ultracold Molecules
  under Control!}},\ }\href {https://doi.org/10.1021/cr300092g} {\bibfield
  {journal} {\bibinfo  {journal} {Chemical Reviews}\ }\textbf {\bibinfo
  {volume} {112}},\ \bibinfo {pages} {4949} (\bibinfo {year}
  {2012})}\BibitemShut {NoStop}%
\bibitem [{\citenamefont {Park}\ \emph {et~al.}(2015)\citenamefont {Park},
  \citenamefont {Will},\ and\ \citenamefont
  {Zwierlein}}]{PhysRevLett.114.205302}%
  \BibitemOpen
  \bibfield  {author} {\bibinfo {author} {\bibfnamefont {J.~W.}\ \bibnamefont
  {Park}}, \bibinfo {author} {\bibfnamefont {S.~A.}\ \bibnamefont {Will}},\
  and\ \bibinfo {author} {\bibfnamefont {M.~W.}\ \bibnamefont {Zwierlein}},\
  }\bibfield  {title} {\bibinfo {title} {Ultracold Dipolar Gas of Fermionic
  $^{23}\mathrm{Na}^{40}\mathrm{K}$ Molecules in Their Absolute Ground State},\
  }\href {https://doi.org/10.1103/PhysRevLett.114.205302} {\bibfield  {journal}
  {\bibinfo  {journal} {Phys. Rev. Lett.}\ }\textbf {\bibinfo {volume} {114}},\
  \bibinfo {pages} {205302} (\bibinfo {year} {2015})}\BibitemShut {NoStop}%
\bibitem [{\citenamefont {Guo}\ \emph {et~al.}(2016)\citenamefont {Guo},
  \citenamefont {Zhu}, \citenamefont {Lu}, \citenamefont {Ye}, \citenamefont
  {Wang}, \citenamefont {Vexiau}, \citenamefont {Bouloufa-Maafa}, \citenamefont
  {Qu\'em\'ener}, \citenamefont {Dulieu},\ and\ \citenamefont
  {Wang}}]{PhysRevLett.116.205303}%
  \BibitemOpen
  \bibfield  {author} {\bibinfo {author} {\bibfnamefont {M.}~\bibnamefont
  {Guo}}, \bibinfo {author} {\bibfnamefont {B.}~\bibnamefont {Zhu}}, \bibinfo
  {author} {\bibfnamefont {B.}~\bibnamefont {Lu}}, \bibinfo {author}
  {\bibfnamefont {X.}~\bibnamefont {Ye}}, \bibinfo {author} {\bibfnamefont
  {F.}~\bibnamefont {Wang}}, \bibinfo {author} {\bibfnamefont {R.}~\bibnamefont
  {Vexiau}}, \bibinfo {author} {\bibfnamefont {N.}~\bibnamefont
  {Bouloufa-Maafa}}, \bibinfo {author} {\bibfnamefont {G.}~\bibnamefont
  {Qu\'em\'ener}}, \bibinfo {author} {\bibfnamefont {O.}~\bibnamefont
  {Dulieu}},\ and\ \bibinfo {author} {\bibfnamefont {D.}~\bibnamefont {Wang}},\
  }\bibfield  {title} {\bibinfo {title} {Creation of an Ultracold Gas of
  Ground-State Dipolar $^{23}\mathrm{Na}^{87}\mathrm{Rb}$ Molecules},\ }\href
  {https://doi.org/10.1103/PhysRevLett.116.205303} {\bibfield  {journal}
  {\bibinfo  {journal} {Phys. Rev. Lett.}\ }\textbf {\bibinfo {volume} {116}},\
  \bibinfo {pages} {205303} (\bibinfo {year} {2016})}\BibitemShut {NoStop}%
\bibitem [{\citenamefont {Rvachov}\ \emph {et~al.}(2017)\citenamefont
  {Rvachov}, \citenamefont {Son}, \citenamefont {Sommer}, \citenamefont
  {Ebadi}, \citenamefont {Park}, \citenamefont {Zwierlein}, \citenamefont
  {Ketterle},\ and\ \citenamefont {Jamison}}]{PhysRevLett.119.143001}%
  \BibitemOpen
  \bibfield  {author} {\bibinfo {author} {\bibfnamefont {T.~M.}\ \bibnamefont
  {Rvachov}}, \bibinfo {author} {\bibfnamefont {H.}~\bibnamefont {Son}},
  \bibinfo {author} {\bibfnamefont {A.~T.}\ \bibnamefont {Sommer}}, \bibinfo
  {author} {\bibfnamefont {S.}~\bibnamefont {Ebadi}}, \bibinfo {author}
  {\bibfnamefont {J.~J.}\ \bibnamefont {Park}}, \bibinfo {author}
  {\bibfnamefont {M.~W.}\ \bibnamefont {Zwierlein}}, \bibinfo {author}
  {\bibfnamefont {W.}~\bibnamefont {Ketterle}},\ and\ \bibinfo {author}
  {\bibfnamefont {A.~O.}\ \bibnamefont {Jamison}},\ }\bibfield  {title}
  {\bibinfo {title} {Long-Lived Ultracold Molecules with Electric and Magnetic
  Dipole Moments},\ }\href {https://doi.org/10.1103/PhysRevLett.119.143001}
  {\bibfield  {journal} {\bibinfo  {journal} {Phys. Rev. Lett.}\ }\textbf
  {\bibinfo {volume} {119}},\ \bibinfo {pages} {143001} (\bibinfo {year}
  {2017})}\BibitemShut {NoStop}%
\bibitem [{\citenamefont {De~Marco}\ \emph {et~al.}(2019)\citenamefont
  {De~Marco}, \citenamefont {Valtolina}, \citenamefont {Matsuda}, \citenamefont
  {Tobias}, \citenamefont {Covey},\ and\ \citenamefont {Ye}}]{De-Marco853}%
  \BibitemOpen
  \bibfield  {author} {\bibinfo {author} {\bibfnamefont {L.}~\bibnamefont
  {De~Marco}}, \bibinfo {author} {\bibfnamefont {G.}~\bibnamefont {Valtolina}},
  \bibinfo {author} {\bibfnamefont {K.}~\bibnamefont {Matsuda}}, \bibinfo
  {author} {\bibfnamefont {W.~G.}\ \bibnamefont {Tobias}}, \bibinfo {author}
  {\bibfnamefont {J.~P.}\ \bibnamefont {Covey}},\ and\ \bibinfo {author}
  {\bibfnamefont {J.}~\bibnamefont {Ye}},\ }\bibfield  {title} {\bibinfo
  {title} {A degenerate fermi gas of polar molecules},\ }\href
  {https://doi.org/10.1126/science.aau7230} {\bibfield  {journal} {\bibinfo
  {journal} {Science}\ }\textbf {\bibinfo {volume} {363}},\ \bibinfo {pages}
  {853} (\bibinfo {year} {2019})}\BibitemShut
  {NoStop}%
\bibitem [{\citenamefont {Son}\ \emph {et~al.}(2020)\citenamefont {Son},
  \citenamefont {Park}, \citenamefont {Ketterle},\ and\ \citenamefont
  {Jamison}}]{Son}%
  \BibitemOpen
  \bibfield  {author} {\bibinfo {author} {\bibfnamefont {H.}~\bibnamefont
  {Son}}, \bibinfo {author} {\bibfnamefont {J.~J.}\ \bibnamefont {Park}},
  \bibinfo {author} {\bibfnamefont {W.}~\bibnamefont {Ketterle}},\ and\
  \bibinfo {author} {\bibfnamefont {A.~O.}\ \bibnamefont {Jamison}},\
  }\bibfield  {title} {\bibinfo {title} {Collisional cooling of ultracold
  molecules},\ }\href {https://doi.org/10.1038/s41586-020-2141-z} {\bibfield
  {journal} {\bibinfo  {journal} {Nature}\ }\textbf {\bibinfo {volume} {580}},\
  \bibinfo {pages} {197} (\bibinfo {year} {2020})}\BibitemShut {NoStop}%
\bibitem [{\citenamefont {Bigagli}\ \emph {et~al.}(2024)\citenamefont
  {Bigagli}, \citenamefont {Yuan}, \citenamefont {Zhang}, \citenamefont
  {Bulatovic}, \citenamefont {Karman}, \citenamefont {Stevenson},\ and\
  \citenamefont {Will}}]{N2}%
  \BibitemOpen
  \bibfield  {author} {\bibinfo {author} {\bibfnamefont {N.}~\bibnamefont
  {Bigagli}}, \bibinfo {author} {\bibfnamefont {W.}~\bibnamefont {Yuan}},
  \bibinfo {author} {\bibfnamefont {S.}~\bibnamefont {Zhang}}, \bibinfo
  {author} {\bibfnamefont {B.}~\bibnamefont {Bulatovic}}, \bibinfo {author}
  {\bibfnamefont {T.}~\bibnamefont {Karman}}, \bibinfo {author} {\bibfnamefont
  {I.}~\bibnamefont {Stevenson}},\ and\ \bibinfo {author} {\bibfnamefont
  {S.}~\bibnamefont {Will}},\ }\bibfield  {title} {\bibinfo {title}
  {Observation of Bose--Einstein condensation of dipolar molecules},\ }\href
  {https://doi.org/10.1038/s41586-024-07492-z} {\bibfield  {journal} {\bibinfo
  {journal} {Nature}\ }\textbf {\bibinfo {volume} {631}},\ \bibinfo {pages}
  {289} (\bibinfo {year} {2024})}\BibitemShut {NoStop}%
\bibitem [{\citenamefont {Chomaz}\ \emph {et~al.}(2018)\citenamefont {Chomaz},
  \citenamefont {van Bijnen}, \citenamefont {Petter}, \citenamefont {Faraoni},
  \citenamefont {Baier}, \citenamefont {Becher}, \citenamefont {Mark},
  \citenamefont {W{\"a}chtler}, \citenamefont {Santos},\ and\ \citenamefont
  {Ferlaino}}]{Chomaz}%
  \BibitemOpen
  \bibfield  {author} {\bibinfo {author} {\bibfnamefont {L.}~\bibnamefont
  {Chomaz}}, \bibinfo {author} {\bibfnamefont {R.~M.~W.}\ \bibnamefont {van
  Bijnen}}, \bibinfo {author} {\bibfnamefont {D.}~\bibnamefont {Petter}},
  \bibinfo {author} {\bibfnamefont {G.}~\bibnamefont {Faraoni}}, \bibinfo
  {author} {\bibfnamefont {S.}~\bibnamefont {Baier}}, \bibinfo {author}
  {\bibfnamefont {J.~H.}\ \bibnamefont {Becher}}, \bibinfo {author}
  {\bibfnamefont {M.~J.}\ \bibnamefont {Mark}}, \bibinfo {author}
  {\bibfnamefont {F.}~\bibnamefont {W{\"a}chtler}}, \bibinfo {author}
  {\bibfnamefont {L.}~\bibnamefont {Santos}},\ and\ \bibinfo {author}
  {\bibfnamefont {F.}~\bibnamefont {Ferlaino}},\ }\bibfield  {title} {\bibinfo
  {title} {Observation of roton mode population in a dipolar quantum gas},\
  }\href {https://doi.org/10.1038/s41567-018-0054-7} {\bibfield  {journal}
  {\bibinfo  {journal} {Nature Physics}\ }\textbf {\bibinfo {volume} {14}},\
  \bibinfo {pages} {442} (\bibinfo {year} {2018})}\BibitemShut {NoStop}%
\bibitem [{\citenamefont {Kadau}\ \emph {et~al.}(2016)\citenamefont {Kadau},
  \citenamefont {Schmitt}, \citenamefont {Wenzel}, \citenamefont {Wink},
  \citenamefont {Maier}, \citenamefont {Ferrier-Barbut},\ and\ \citenamefont
  {Pfau}}]{DDIexperiment}%
  \BibitemOpen
  \bibfield  {author} {\bibinfo {author} {\bibfnamefont {H.}~\bibnamefont
  {Kadau}}, \bibinfo {author} {\bibfnamefont {M.}~\bibnamefont {Schmitt}},
  \bibinfo {author} {\bibfnamefont {M.}~\bibnamefont {Wenzel}}, \bibinfo
  {author} {\bibfnamefont {C.}~\bibnamefont {Wink}}, \bibinfo {author}
  {\bibfnamefont {T.}~\bibnamefont {Maier}}, \bibinfo {author} {\bibfnamefont
  {I.}~\bibnamefont {Ferrier-Barbut}},\ and\ \bibinfo {author} {\bibfnamefont
  {T.}~\bibnamefont {Pfau}},\ }\bibfield  {title} {\bibinfo {title} {{Observing
  the Rosensweig instability of a quantum ferrofluid}},\ }\href
  {http://dx.doi.org/10.1038/nature16485} {\bibfield  {journal} {\bibinfo
  {journal} {Nature}\ }\textbf {\bibinfo {volume} {530}},\ \bibinfo {pages}
  {194} (\bibinfo {year} {2016})}\BibitemShut {NoStop}%
\bibitem [{\citenamefont {Natale}\ \emph {et~al.}(2019)\citenamefont {Natale},
  \citenamefont {van Bijnen}, \citenamefont {Patscheider}, \citenamefont
  {Petter}, \citenamefont {Mark}, \citenamefont {Chomaz},\ and\ \citenamefont
  {Ferlaino}}]{PhysRevLett.123.050402}%
  \BibitemOpen
  \bibfield  {author} {\bibinfo {author} {\bibfnamefont {G.}~\bibnamefont
  {Natale}}, \bibinfo {author} {\bibfnamefont {R.~M.~W.}\ \bibnamefont {van
  Bijnen}}, \bibinfo {author} {\bibfnamefont {A.}~\bibnamefont {Patscheider}},
  \bibinfo {author} {\bibfnamefont {D.}~\bibnamefont {Petter}}, \bibinfo
  {author} {\bibfnamefont {M.~J.}\ \bibnamefont {Mark}}, \bibinfo {author}
  {\bibfnamefont {L.}~\bibnamefont {Chomaz}},\ and\ \bibinfo {author}
  {\bibfnamefont {F.}~\bibnamefont {Ferlaino}},\ }\bibfield  {title} {\bibinfo
  {title} {Excitation Spectrum of a Trapped Dipolar Supersolid and Its
  Experimental Evidence},\ }\href
  {https://doi.org/10.1103/PhysRevLett.123.050402} {\bibfield  {journal}
  {\bibinfo  {journal} {Phys. Rev. Lett.}\ }\textbf {\bibinfo {volume} {123}},\
  \bibinfo {pages} {050402} (\bibinfo {year} {2019})}\BibitemShut {NoStop}%
\bibitem [{\citenamefont {Petter}\ \emph {et~al.}(2019)\citenamefont {Petter},
  \citenamefont {Natale}, \citenamefont {van Bijnen}, \citenamefont
  {Patscheider}, \citenamefont {Mark}, \citenamefont {Chomaz},\ and\
  \citenamefont {Ferlaino}}]{PhysRevLett.122.183401}%
  \BibitemOpen
  \bibfield  {author} {\bibinfo {author} {\bibfnamefont {D.}~\bibnamefont
  {Petter}}, \bibinfo {author} {\bibfnamefont {G.}~\bibnamefont {Natale}},
  \bibinfo {author} {\bibfnamefont {R.~M.~W.}\ \bibnamefont {van Bijnen}},
  \bibinfo {author} {\bibfnamefont {A.}~\bibnamefont {Patscheider}}, \bibinfo
  {author} {\bibfnamefont {M.~J.}\ \bibnamefont {Mark}}, \bibinfo {author}
  {\bibfnamefont {L.}~\bibnamefont {Chomaz}},\ and\ \bibinfo {author}
  {\bibfnamefont {F.}~\bibnamefont {Ferlaino}},\ }\bibfield  {title} {\bibinfo
  {title} {Probing the Roton Excitation Spectrum of a Stable Dipolar Bose
  Gas},\ }\href {https://doi.org/10.1103/PhysRevLett.122.183401} {\bibfield
  {journal} {\bibinfo  {journal} {Phys. Rev. Lett.}\ }\textbf {\bibinfo
  {volume} {122}},\ \bibinfo {pages} {183401} (\bibinfo {year}
  {2019})}\BibitemShut {NoStop}%
\bibitem [{\citenamefont {Recati}\ and\ \citenamefont {Stringari}(2023)}]{NP2}%
  \BibitemOpen
  \bibfield  {author} {\bibinfo {author} {\bibfnamefont {A.}~\bibnamefont
  {Recati}}\ and\ \bibinfo {author} {\bibfnamefont {S.}~\bibnamefont
  {Stringari}},\ }\bibfield  {title} {\bibinfo {title} {Supersolidity in
  ultracold dipolar gases},\ }\href
  {https://doi.org/10.1038/s42254-023-00648-2} {\bibfield  {journal} {\bibinfo
  {journal} {Nature Reviews Physics}\ }\textbf {\bibinfo {volume} {5}},\
  \bibinfo {pages} {735} (\bibinfo {year} {2023})}\BibitemShut {NoStop}%
\bibitem [{\citenamefont {Hung}\ \emph {et~al.}(2013)\citenamefont {Hung},
  \citenamefont {Gurarie},\ and\ \citenamefont {Chin}}]{PMID:23907531}%
  \BibitemOpen
  \bibfield  {author} {\bibinfo {author} {\bibfnamefont {C.-L.}\ \bibnamefont
  {Hung}}, \bibinfo {author} {\bibfnamefont {V.}~\bibnamefont {Gurarie}},\ and\
  \bibinfo {author} {\bibfnamefont {C.}~\bibnamefont {Chin}},\ }\bibfield
  {title} {\bibinfo {title} {From cosmology to cold atoms: observation of
  Sakharov oscillations in a quenched atomic superfluid},\ }\href
  {https://doi.org/10.1126/science.1237557} {\bibfield  {journal} {\bibinfo
  {journal} {Science (New York, N.Y.)}\ }\textbf {\bibinfo {volume} {341}},\
  \bibinfo {pages} {1213—1215} (\bibinfo {year} {2013})}\BibitemShut
  {NoStop}%
\bibitem [{\citenamefont {Hodgman}\ \emph {et~al.}(2011)\citenamefont
  {Hodgman}, \citenamefont {Dall}, \citenamefont {Manning}, \citenamefont
  {Baldwin},\ and\ \citenamefont {Truscott}}]{PMID:21350171}%
  \BibitemOpen
  \bibfield  {author} {\bibinfo {author} {\bibfnamefont {S.}~\bibnamefont
  {Hodgman}}, \bibinfo {author} {\bibfnamefont {R.}~\bibnamefont {Dall}},
  \bibinfo {author} {\bibfnamefont {A.}~\bibnamefont {Manning}}, \bibinfo
  {author} {\bibfnamefont {K.~G.~H.}\ \bibnamefont {Baldwin}},\ and\ \bibinfo
  {author} {\bibfnamefont {A.}~\bibnamefont {Truscott}},\ }\bibfield  {title}
  {\bibinfo {title} {Direct measurement of long-range third-order coherence in
  Bose-Einstein condensates},\ }\href {https://doi.org/10.1126/science.1198481}
  {\bibfield  {journal} {\bibinfo  {journal} {Science (New York, N.Y.)}\
  }\textbf {\bibinfo {volume} {331}},\ \bibinfo {pages} {1046—1049} (\bibinfo
  {year} {2011})}\BibitemShut {NoStop}%
\bibitem [{\citenamefont {Steinhauer}(2016)}]{thermal-Hawking-radiation1}%
  \BibitemOpen
  \bibfield  {author} {\bibinfo {author} {\bibfnamefont {J.}~\bibnamefont
  {Steinhauer}},\ }\bibfield  {title} {\bibinfo {title} {Observation of quantum
  Hawking radiation and its entanglement in an analogue black hole},\ }\href
  {https://doi.org/10.1038/nphys3863} {\bibfield  {journal} {\bibinfo
  {journal} {Nature Physics}\ }\textbf {\bibinfo {volume} {12}},\ \bibinfo
  {pages} {959} (\bibinfo {year} {2016})}\BibitemShut {NoStop}%
\bibitem [{\citenamefont {Mu{\~n}oz~de Nova}\ \emph {et~al.}(2019)\citenamefont
  {Mu{\~n}oz~de Nova}, \citenamefont {Golubkov}, \citenamefont {Kolobov},\ and\
  \citenamefont {Steinhauer}}]{thermal-Hawking-radiation2}%
  \BibitemOpen
  \bibfield  {author} {\bibinfo {author} {\bibfnamefont {J.~R.}\ \bibnamefont
  {Mu{\~n}oz~de Nova}}, \bibinfo {author} {\bibfnamefont {K.}~\bibnamefont
  {Golubkov}}, \bibinfo {author} {\bibfnamefont {V.~I.}\ \bibnamefont
  {Kolobov}},\ and\ \bibinfo {author} {\bibfnamefont {J.}~\bibnamefont
  {Steinhauer}},\ }\bibfield  {title} {\bibinfo {title} {Observation of thermal
  Hawking radiation and its temperature in an analogue black hole},\ }\href
  {https://doi.org/10.1038/s41586-019-1241-0} {\bibfield  {journal} {\bibinfo
  {journal} {Nature}\ }\textbf {\bibinfo {volume} {569}},\ \bibinfo {pages}
  {688} (\bibinfo {year} {2019})}\BibitemShut {NoStop}%
\bibitem [{\citenamefont {Chen}\ \emph {et~al.}(2021)\citenamefont {Chen},
  \citenamefont {Khlebnikov},\ and\ \citenamefont
  {Hung}}]{PhysRevLett.127.060404}%
  \BibitemOpen
  \bibfield  {author} {\bibinfo {author} {\bibfnamefont {C.-A.}\ \bibnamefont
  {Chen}}, \bibinfo {author} {\bibfnamefont {S.}~\bibnamefont {Khlebnikov}},\
  and\ \bibinfo {author} {\bibfnamefont {C.-L.}\ \bibnamefont {Hung}},\
  }\bibfield  {title} {\bibinfo {title} {Observation of Quasiparticle Pair
  Production and Quantum Entanglement in Atomic Quantum Gases Quenched to an
  Attractive Interaction},\ }\href
  {https://doi.org/10.1103/PhysRevLett.127.060404} {\bibfield  {journal}
  {\bibinfo  {journal} {Phys. Rev. Lett.}\ }\textbf {\bibinfo {volume} {127}},\
  \bibinfo {pages} {060404} (\bibinfo {year} {2021})}\BibitemShut {NoStop}%
\bibitem [{\citenamefont {Sparn}\ \emph {et~al.}(2024)\citenamefont {Sparn},
  \citenamefont {Kath}, \citenamefont {Liebster}, \citenamefont {Duchene},
  \citenamefont {Schmidt}, \citenamefont {Tolosa-Sime\'on}, \citenamefont
  {Parra-L\'opez}, \citenamefont {Floerchinger}, \citenamefont {Strobel},\ and\
  \citenamefont {Oberthaler}}]{PhysRevLett.133.260201}%
  \BibitemOpen
  \bibfield  {author} {\bibinfo {author} {\bibfnamefont {M.}~\bibnamefont
  {Sparn}}, \bibinfo {author} {\bibfnamefont {E.}~\bibnamefont {Kath}},
  \bibinfo {author} {\bibfnamefont {N.}~\bibnamefont {Liebster}}, \bibinfo
  {author} {\bibfnamefont {J.}~\bibnamefont {Duchene}}, \bibinfo {author}
  {\bibfnamefont {C.~F.}\ \bibnamefont {Schmidt}}, \bibinfo {author}
  {\bibfnamefont {M.}~\bibnamefont {Tolosa-Sime\'on}}, \bibinfo {author}
  {\bibfnamefont {A.}~\bibnamefont {Parra-L\'opez}}, \bibinfo {author}
  {\bibfnamefont {S.}~\bibnamefont {Floerchinger}}, \bibinfo {author}
  {\bibfnamefont {H.}~\bibnamefont {Strobel}},\ and\ \bibinfo {author}
  {\bibfnamefont {M.~K.}\ \bibnamefont {Oberthaler}},\ }\bibfield  {title}
  {\bibinfo {title} {Experimental Particle Production in Time-Dependent
  Apacetimes: A one-Dimensional Scattering Problem},\ }\href
  {https://doi.org/10.1103/PhysRevLett.133.260201} {\bibfield  {journal}
  {\bibinfo  {journal} {Phys. Rev. Lett.}\ }\textbf {\bibinfo {volume} {133}},\
  \bibinfo {pages} {260201} (\bibinfo {year} {2024})}\BibitemShut {NoStop}%
\bibitem [{\citenamefont {Viermann}\ \emph {et~al.}(2022)\citenamefont
  {Viermann}, \citenamefont {Sparn}, \citenamefont {Liebster}, \citenamefont
  {Hans}, \citenamefont {Kath}, \citenamefont {Parra-L{\'o}pez}, \citenamefont
  {Tolosa-Sime{\'o}n}, \citenamefont {S{\'a}nchez-Kuntz}, \citenamefont {Haas},
  \citenamefont {Strobel}, \citenamefont {Floerchinger},\ and\ \citenamefont
  {Oberthaler}}]{N3}%
  \BibitemOpen
  \bibfield  {author} {\bibinfo {author} {\bibfnamefont {C.}~\bibnamefont
  {Viermann}}, \bibinfo {author} {\bibfnamefont {M.}~\bibnamefont {Sparn}},
  \bibinfo {author} {\bibfnamefont {N.}~\bibnamefont {Liebster}}, \bibinfo
  {author} {\bibfnamefont {M.}~\bibnamefont {Hans}}, \bibinfo {author}
  {\bibfnamefont {E.}~\bibnamefont {Kath}}, \bibinfo {author} {\bibfnamefont
  {{\'A}.}~\bibnamefont {Parra-L{\'o}pez}}, \bibinfo {author} {\bibfnamefont
  {M.}~\bibnamefont {Tolosa-Sime{\'o}n}}, \bibinfo {author} {\bibfnamefont
  {N.}~\bibnamefont {S{\'a}nchez-Kuntz}}, \bibinfo {author} {\bibfnamefont
  {T.}~\bibnamefont {Haas}}, \bibinfo {author} {\bibfnamefont {H.}~\bibnamefont
  {Strobel}}, \bibinfo {author} {\bibfnamefont {S.}~\bibnamefont
  {Floerchinger}},\ and\ \bibinfo {author} {\bibfnamefont {M.~K.}\ \bibnamefont
  {Oberthaler}},\ }\bibfield  {title} {\bibinfo {title} {Quantum field
  simulator for dynamics in curved spacetime},\ }\href
  {https://doi.org/10.1038/s41586-022-05313-9} {\bibfield  {journal} {\bibinfo
  {journal} {Nature}\ }\textbf {\bibinfo {volume} {611}},\ \bibinfo {pages}
  {260} (\bibinfo {year} {2022})}\BibitemShut {NoStop}%
\bibitem [{\citenamefont {Tamura}\ \emph {et~al.}(2025)\citenamefont {Tamura},
  \citenamefont {Banerjee}, \citenamefont {Li}, \citenamefont {Kevrekidis},
  \citenamefont {Mistakidis},\ and\ \citenamefont
  {Hung}}]{tamura2025observationmanybodycoherencequasionedimensional}%
  \BibitemOpen
  \bibfield  {author} {\bibinfo {author} {\bibfnamefont {H.}~\bibnamefont
  {Tamura}}, \bibinfo {author} {\bibfnamefont {S.}~\bibnamefont {Banerjee}},
  \bibinfo {author} {\bibfnamefont {R.}~\bibnamefont {Li}}, \bibinfo {author}
  {\bibfnamefont {P.}~\bibnamefont {Kevrekidis}}, \bibinfo {author}
  {\bibfnamefont {S.~I.}\ \bibnamefont {Mistakidis}},\ and\ \bibinfo {author}
  {\bibfnamefont {C.-L.}\ \bibnamefont {Hung}},\ }\href
  {https://arxiv.org/abs/2506.13597} {\bibinfo {title} {Observation of
  many-body coherence in quasi-one-dimensional attractive Bose gases}},\ \Eprint {https://arxiv.org/abs/2506.13597}
  {arXiv:2506.13597 [cond-mat.quant-gas]} \BibitemShut {NoStop}%
\bibitem [{\citenamefont {Hu}\ \emph {et~al.}(2019)\citenamefont {Hu},
  \citenamefont {Feng}, \citenamefont {Zhang},\ and\ \citenamefont
  {Chin}}]{Unruh-Simulation1}%
  \BibitemOpen
  \bibfield  {author} {\bibinfo {author} {\bibfnamefont {J.}~\bibnamefont
  {Hu}}, \bibinfo {author} {\bibfnamefont {L.}~\bibnamefont {Feng}}, \bibinfo
  {author} {\bibfnamefont {Z.}~\bibnamefont {Zhang}},\ and\ \bibinfo {author}
  {\bibfnamefont {C.}~\bibnamefont {Chin}},\ }\bibfield  {title} {\bibinfo
  {title} {Quantum simulation of Unruh radiation},\ }\href
  {https://doi.org/10.1038/s41567-019-0537-1} {\bibfield  {journal} {\bibinfo
  {journal} {Nature Physics}\ }\textbf {\bibinfo {volume} {15}},\ \bibinfo
  {pages} {785} (\bibinfo {year} {2019})}\BibitemShut {NoStop}%
\bibitem [{\citenamefont {Tomza}\ \emph {et~al.}(2019)\citenamefont {Tomza},
  \citenamefont {Jachymski}, \citenamefont {Gerritsma}, \citenamefont
  {Negretti}, \citenamefont {Calarco}, \citenamefont {Idziaszek},\ and\
  \citenamefont {Julienne}}]{RevModPhys.91.035001}%
  \BibitemOpen
  \bibfield  {author} {\bibinfo {author} {\bibfnamefont {M.}~\bibnamefont
  {Tomza}}, \bibinfo {author} {\bibfnamefont {K.}~\bibnamefont {Jachymski}},
  \bibinfo {author} {\bibfnamefont {R.}~\bibnamefont {Gerritsma}}, \bibinfo
  {author} {\bibfnamefont {A.}~\bibnamefont {Negretti}}, \bibinfo {author}
  {\bibfnamefont {T.}~\bibnamefont {Calarco}}, \bibinfo {author} {\bibfnamefont
  {Z.}~\bibnamefont {Idziaszek}},\ and\ \bibinfo {author} {\bibfnamefont
  {P.~S.}\ \bibnamefont {Julienne}},\ }\bibfield  {title} {\bibinfo {title}
  {Cold Hybrid Ion-Atom Systems},\ }\href
  {https://doi.org/10.1103/RevModPhys.91.035001} {\bibfield  {journal}
  {\bibinfo  {journal} {Rev. Mod. Phys.}\ }\textbf {\bibinfo {volume} {91}},\
  \bibinfo {pages} {035001} (\bibinfo {year} {2019})}\BibitemShut {NoStop}%
\bibitem [{\citenamefont {Grusdt}\ \emph {et~al.}(2025)\citenamefont {Grusdt},
  \citenamefont {Mostaan}, \citenamefont {Demler},\ and\ \citenamefont
  {Ardila}}]{Grusdt_2025}%
  \BibitemOpen
  \bibfield  {author} {\bibinfo {author} {\bibfnamefont {F.}~\bibnamefont
  {Grusdt}}, \bibinfo {author} {\bibfnamefont {N.}~\bibnamefont {Mostaan}},
  \bibinfo {author} {\bibfnamefont {E.}~\bibnamefont {Demler}},\ and\ \bibinfo
  {author} {\bibfnamefont {L.~A.~P.}\ \bibnamefont {Ardila}},\ }\bibfield
  {title} {\bibinfo {title} {Impurities and polarons in bosonic quantum gases:
  a review on recent progress},\ }\href
  {https://doi.org/10.1088/1361-6633/add94b} {\bibfield  {journal} {\bibinfo
  {journal} {Reports on Progress in Physics}\ }\textbf {\bibinfo {volume}
  {88}},\ \bibinfo {pages} {066401} (\bibinfo {year} {2025})}\BibitemShut
  {NoStop}%
\bibitem [{\citenamefont {{Zipkes}}\ \emph {et~al.}(2010)\citenamefont
  {{Zipkes}}, \citenamefont {{Palzer}}, \citenamefont {{Sias}},\ and\
  \citenamefont {{K{\"o}hl}}}]{2010Natur.464..388Z}%
  \BibitemOpen
  \bibfield  {author} {\bibinfo {author} {\bibfnamefont {C.}~\bibnamefont
  {{Zipkes}}}, \bibinfo {author} {\bibfnamefont {S.}~\bibnamefont {{Palzer}}},
  \bibinfo {author} {\bibfnamefont {C.}~\bibnamefont {{Sias}}},\ and\ \bibinfo
  {author} {\bibfnamefont {M.}~\bibnamefont {{K{\"o}hl}}},\ }\bibfield  {title}
  {\bibinfo {title} {{A trapped single ion inside a Bose-Einstein
  condensate}},\ }\href {https://doi.org/10.1038/nature08865} {\bibfield
  {journal} {\bibinfo  {journal} {\nat}\ }\textbf {\bibinfo {volume} {464}},\
  \bibinfo {pages} {388} (\bibinfo {year} {2010})}
  \BibitemShut {NoStop}%
\bibitem [{\citenamefont {Weckesser}\ \emph {et~al.}(2021)\citenamefont
  {Weckesser}, \citenamefont {Thielemann}, \citenamefont {Wiater},
  \citenamefont {Wojciechowska}, \citenamefont {Karpa}, \citenamefont
  {Jachymski}, \citenamefont {Tomza}, \citenamefont {Walker},\ and\
  \citenamefont {Schaetz}}]{N4}%
  \BibitemOpen
  \bibfield  {author} {\bibinfo {author} {\bibfnamefont {P.}~\bibnamefont
  {Weckesser}}, \bibinfo {author} {\bibfnamefont {F.}~\bibnamefont
  {Thielemann}}, \bibinfo {author} {\bibfnamefont {D.}~\bibnamefont {Wiater}},
  \bibinfo {author} {\bibfnamefont {A.}~\bibnamefont {Wojciechowska}}, \bibinfo
  {author} {\bibfnamefont {L.}~\bibnamefont {Karpa}}, \bibinfo {author}
  {\bibfnamefont {K.}~\bibnamefont {Jachymski}}, \bibinfo {author}
  {\bibfnamefont {M.}~\bibnamefont {Tomza}}, \bibinfo {author} {\bibfnamefont
  {T.}~\bibnamefont {Walker}},\ and\ \bibinfo {author} {\bibfnamefont
  {T.}~\bibnamefont {Schaetz}},\ }\bibfield  {title} {\bibinfo {title}
  {Observation of Feshbach resonances between a single ion and ultracold
  atoms},\ }\href {https://doi.org/10.1038/s41586-021-04112-y} {\bibfield
  {journal} {\bibinfo  {journal} {Nature}\ }\textbf {\bibinfo {volume} {600}},\
  \bibinfo {pages} {429} (\bibinfo {year} {2021})}\BibitemShut {NoStop}%
\bibitem [{\citenamefont {{Balewski}}\ \emph {et~al.}(2013)\citenamefont
  {{Balewski}}, \citenamefont {{Krupp}}, \citenamefont {{Gaj}}, \citenamefont
  {{Peter}}, \citenamefont {{B{\"u}chler}}, \citenamefont {{L{\"o}w}},
  \citenamefont {{Hofferberth}},\ and\ \citenamefont
  {{Pfau}}}]{2013Natur.502..664B}%
  \BibitemOpen
  \bibfield  {author} {\bibinfo {author} {\bibfnamefont {J.~B.}\ \bibnamefont
  {{Balewski}}}, \bibinfo {author} {\bibfnamefont {A.~T.}\ \bibnamefont
  {{Krupp}}}, \bibinfo {author} {\bibfnamefont {A.}~\bibnamefont {{Gaj}}},
  \bibinfo {author} {\bibfnamefont {D.}~\bibnamefont {{Peter}}}, \bibinfo
  {author} {\bibfnamefont {H.~P.}\ \bibnamefont {{B{\"u}chler}}}, \bibinfo
  {author} {\bibfnamefont {R.}~\bibnamefont {{L{\"o}w}}}, \bibinfo {author}
  {\bibfnamefont {S.}~\bibnamefont {{Hofferberth}}},\ and\ \bibinfo {author}
  {\bibfnamefont {T.}~\bibnamefont {{Pfau}}},\ }\bibfield  {title} {\bibinfo
  {title} {{Coupling a single electron to a Bose-Einstein condensate}},\ }\href
  {https://doi.org/10.1038/nature12592} {\bibfield  {journal} {\bibinfo
  {journal} {\nat}\ }\textbf {\bibinfo {volume} {502}},\ \bibinfo {pages} {664}
  (\bibinfo {year} {2013})}\BibitemShut {NoStop}%
\bibitem [{\citenamefont {Schmid}\ \emph {et~al.}(2010)\citenamefont {Schmid},
  \citenamefont {H\"arter},\ and\ \citenamefont
  {Denschlag}}]{PhysRevLett.105.133202}%
  \BibitemOpen
  \bibfield  {author} {\bibinfo {author} {\bibfnamefont {S.}~\bibnamefont
  {Schmid}}, \bibinfo {author} {\bibfnamefont {A.}~\bibnamefont {H\"arter}},\
  and\ \bibinfo {author} {\bibfnamefont {J.~H.}\ \bibnamefont {Denschlag}},\
  }\bibfield  {title} {\bibinfo {title} {Dynamics of a Cold Trapped Ion in a
  Bose-Einstein Condensate},\ }\href
  {https://doi.org/10.1103/PhysRevLett.105.133202} {\bibfield  {journal}
  {\bibinfo  {journal} {Phys. Rev. Lett.}\ }\textbf {\bibinfo {volume} {105}},\
  \bibinfo {pages} {133202} (\bibinfo {year} {2010})}\BibitemShut {NoStop}%
\bibitem [{\citenamefont {Scelle}\ \emph {et~al.}(2013)\citenamefont {Scelle},
  \citenamefont {Rentrop}, \citenamefont {Trautmann}, \citenamefont
  {Schuster},\ and\ \citenamefont {Oberthaler}}]{PhysRevLett.111.070401}%
  \BibitemOpen
  \bibfield  {author} {\bibinfo {author} {\bibfnamefont {R.}~\bibnamefont
  {Scelle}}, \bibinfo {author} {\bibfnamefont {T.}~\bibnamefont {Rentrop}},
  \bibinfo {author} {\bibfnamefont {A.}~\bibnamefont {Trautmann}}, \bibinfo
  {author} {\bibfnamefont {T.}~\bibnamefont {Schuster}},\ and\ \bibinfo
  {author} {\bibfnamefont {M.~K.}\ \bibnamefont {Oberthaler}},\ }\bibfield
  {title} {\bibinfo {title} {Motional Coherence of Fermions Immersed in a Bose
  Gas},\ }\href {https://doi.org/10.1103/PhysRevLett.111.070401} {\bibfield
  {journal} {\bibinfo  {journal} {Phys. Rev. Lett.}\ }\textbf {\bibinfo
  {volume} {111}},\ \bibinfo {pages} {070401} (\bibinfo {year}
  {2013})}\BibitemShut {NoStop}%
\bibitem [{\citenamefont {Spethmann}\ \emph {et~al.}(2012)\citenamefont
  {Spethmann}, \citenamefont {Kindermann}, \citenamefont {John}, \citenamefont
  {Weber}, \citenamefont {Meschede},\ and\ \citenamefont
  {Widera}}]{PhysRevLett.109.235301}%
  \BibitemOpen
  \bibfield  {author} {\bibinfo {author} {\bibfnamefont {N.}~\bibnamefont
  {Spethmann}}, \bibinfo {author} {\bibfnamefont {F.}~\bibnamefont
  {Kindermann}}, \bibinfo {author} {\bibfnamefont {S.}~\bibnamefont {John}},
  \bibinfo {author} {\bibfnamefont {C.}~\bibnamefont {Weber}}, \bibinfo
  {author} {\bibfnamefont {D.}~\bibnamefont {Meschede}},\ and\ \bibinfo
  {author} {\bibfnamefont {A.}~\bibnamefont {Widera}},\ }\bibfield  {title}
  {\bibinfo {title} {Dynamics of Single Neutral Impurity Atoms Immersed in an
  Ultracold Gas},\ }\href {https://doi.org/10.1103/PhysRevLett.109.235301}
  {\bibfield  {journal} {\bibinfo  {journal} {Phys. Rev. Lett.}\ }\textbf
  {\bibinfo {volume} {109}},\ \bibinfo {pages} {235301} (\bibinfo {year}
  {2012})}\BibitemShut {NoStop}%
\bibitem [{\citenamefont {Schmidt}\ \emph {et~al.}(2018)\citenamefont
  {Schmidt}, \citenamefont {Mayer}, \citenamefont {Bouton}, \citenamefont
  {Adam}, \citenamefont {Lausch}, \citenamefont {Spethmann},\ and\
  \citenamefont {Widera}}]{PhysRevLett.121.130403}%
  \BibitemOpen
  \bibfield  {author} {\bibinfo {author} {\bibfnamefont {F.}~\bibnamefont
  {Schmidt}}, \bibinfo {author} {\bibfnamefont {D.}~\bibnamefont {Mayer}},
  \bibinfo {author} {\bibfnamefont {Q.}~\bibnamefont {Bouton}}, \bibinfo
  {author} {\bibfnamefont {D.}~\bibnamefont {Adam}}, \bibinfo {author}
  {\bibfnamefont {T.}~\bibnamefont {Lausch}}, \bibinfo {author} {\bibfnamefont
  {N.}~\bibnamefont {Spethmann}},\ and\ \bibinfo {author} {\bibfnamefont
  {A.}~\bibnamefont {Widera}},\ }\bibfield  {title} {\bibinfo {title} {Quantum
  Spin Dynamics of Individual Neutral Impurities Coupled to a Bose-Einstein
  Condensate},\ }\href {https://doi.org/10.1103/PhysRevLett.121.130403}
  {\bibfield  {journal} {\bibinfo  {journal} {Phys. Rev. Lett.}\ }\textbf
  {\bibinfo {volume} {121}},\ \bibinfo {pages} {130403} (\bibinfo {year}
  {2018})}\BibitemShut {NoStop}%
\bibitem [{\citenamefont {Kleinbach}\ \emph {et~al.}(2018)\citenamefont
  {Kleinbach}, \citenamefont {Engel}, \citenamefont {Dieterle}, \citenamefont
  {L\"ow}, \citenamefont {Pfau},\ and\ \citenamefont
  {Meinert}}]{PhysRevLett.120.193401}%
  \BibitemOpen
  \bibfield  {author} {\bibinfo {author} {\bibfnamefont {K.~S.}\ \bibnamefont
  {Kleinbach}}, \bibinfo {author} {\bibfnamefont {F.}~\bibnamefont {Engel}},
  \bibinfo {author} {\bibfnamefont {T.}~\bibnamefont {Dieterle}}, \bibinfo
  {author} {\bibfnamefont {R.}~\bibnamefont {L\"ow}}, \bibinfo {author}
  {\bibfnamefont {T.}~\bibnamefont {Pfau}},\ and\ \bibinfo {author}
  {\bibfnamefont {F.}~\bibnamefont {Meinert}},\ }\bibfield  {title} {\bibinfo
  {title} {Ionic Impurity in a Bose-Einstein Condensate at SubmicroKelvin
  Temperatures},\ }\href {https://doi.org/10.1103/PhysRevLett.120.193401}
  {\bibfield  {journal} {\bibinfo  {journal} {Phys. Rev. Lett.}\ }\textbf
  {\bibinfo {volume} {120}},\ \bibinfo {pages} {193401} (\bibinfo {year}
  {2018})}\BibitemShut {NoStop}%
\bibitem [{\citenamefont {Dieterle}\ \emph {et~al.}(2021)\citenamefont
  {Dieterle}, \citenamefont {Berngruber}, \citenamefont {H\"olzl},
  \citenamefont {L\"ow}, \citenamefont {Jachymski}, \citenamefont {Pfau},\ and\
  \citenamefont {Meinert}}]{PhysRevLett.126.033401}%
  \BibitemOpen
  \bibfield  {author} {\bibinfo {author} {\bibfnamefont {T.}~\bibnamefont
  {Dieterle}}, \bibinfo {author} {\bibfnamefont {M.}~\bibnamefont
  {Berngruber}}, \bibinfo {author} {\bibfnamefont {C.}~\bibnamefont {H\"olzl}},
  \bibinfo {author} {\bibfnamefont {R.}~\bibnamefont {L\"ow}}, \bibinfo
  {author} {\bibfnamefont {K.}~\bibnamefont {Jachymski}}, \bibinfo {author}
  {\bibfnamefont {T.}~\bibnamefont {Pfau}},\ and\ \bibinfo {author}
  {\bibfnamefont {F.}~\bibnamefont {Meinert}},\ }\bibfield  {title} {\bibinfo
  {title} {Transport of a Single Cold Ion Immersed in a Bose-Einstein
  Condensate},\ }\href {https://doi.org/10.1103/PhysRevLett.126.033401}
  {\bibfield  {journal} {\bibinfo  {journal} {Phys. Rev. Lett.}\ }\textbf
  {\bibinfo {volume} {126}},\ \bibinfo {pages} {033401} (\bibinfo {year}
  {2021})}\BibitemShut {NoStop}%
\bibitem [{\citenamefont {Georgescu}\ \emph {et~al.}(2014)\citenamefont
  {Georgescu}, \citenamefont {Ashhab},\ and\ \citenamefont
  {Nori}}]{RevModPhys.86.153}%
  \BibitemOpen
  \bibfield  {author} {\bibinfo {author} {\bibfnamefont {I.~M.}\ \bibnamefont
  {Georgescu}}, \bibinfo {author} {\bibfnamefont {S.}~\bibnamefont {Ashhab}},\
  and\ \bibinfo {author} {\bibfnamefont {F.}~\bibnamefont {Nori}},\ }\bibfield
  {title} {\bibinfo {title} {Quantum Simulation},\ }\href
  {https://doi.org/10.1103/RevModPhys.86.153} {\bibfield  {journal} {\bibinfo
  {journal} {Rev. Mod. Phys.}\ }\textbf {\bibinfo {volume} {86}},\ \bibinfo
  {pages} {153} (\bibinfo {year} {2014})}\BibitemShut {NoStop}%
\end{thebibliography}%

\newpage
\pagebreak
\clearpage
\widetext

\begin{center}
\textbf{\large Supplementary Material}
\end{center}

\setcounter{equation}{0}
\setcounter{section}{0}
\setcounter{page}{1}
\makeatletter
\renewcommand{\theequation}{S\arabic{equation}}

\section{The detectors' density matrix} 
Let us begin with the interaction Hamiltonian of two Unruh-DeWitt detectors interacting with the analogue Lorentz-violating quantum field in the interaction picture,
\begin{eqnarray}\label{Hamiltonian-1}
\hat{H}_I(\tau)=\sum_{\text{D}=\text{A}, \text{B}}\big[\lambda\chi(\tau)(\sigma^+_\text{D}e^{i\Omega\tau}+\sigma^-_\text{D}e^{-i\Omega\tau})\delta\hat{\rho}(t_\text{D}(\tau), \br_\text{D}(\tau))\big],
\end{eqnarray}
where $\lambda$ is the coupling parameter and $\chi(\tau)$ is similar to a real-valued smooth switching function that specifies how the interaction is 
turned on and off. With the Hamiltonian \eqref{Hamiltonian-1}, the interaction-picture time evolution operator for any initial state can be written as 
\begin{eqnarray}
\hat{U}=\mathcal{T}\bigg[\exp\bigg(-i\int^\infty_{-\infty}d\tau\hat{H}_I(\tau)\bigg)\bigg].
\end{eqnarray}
In the weak coupling regime, the evolution operator $\hat{U}$ can be approximated using the Dyson series. After this, one can find the evolution operator, to 
the second order in the coupling constant, reads
\begin{eqnarray}
\hat{U}=\hat{U}^{(0)}+\hat{U}^{(1)}+\hat{U}^{(2)}+\mathcal{O}(\lambda^3)
\end{eqnarray}
with 
\begin{eqnarray}
\hat{U}^{(0)}&=&\mathbf{I},
\\
\hat{U}^{(1)}&=&-i\int^\infty_{-\infty}d\tau\hat{H}_I(\tau),
\\
\hat{U}^{(2)}&=&-\int^\infty_{-\infty}d\tau\int^\tau_{-\infty}d\tau^\prime\hat{H}_I(\tau)\hat{H}_I(\tau^\prime),
\end{eqnarray}
where $\mathbf{I}$ is the identity matrix.

At the beginning, the detectors and the field are prepared at the state 
\begin{eqnarray}
\hat{\rho}_0=|g_\text{A}\rangle\langle g_\text{A}|\otimes|g_\text{B}\rangle\langle g_\text{B}|\otimes|\psi\rangle\langle\psi|,
\end{eqnarray}
where $|g\rangle_\text{D}$ with $\text{D}\in\{\text{A}, \text{B}\}$ denotes the ground state of the detector and $|\psi\rangle\langle\psi|$ denotes the initial state of the field. The time-evolved state yields $\hat{\rho}_T=\hat{U}\hat{\rho}_0\hat{U}^\dagger$. To the second order, it is found to be 
\begin{eqnarray}
\hat{\rho}_T&=&\hat{\rho}_0+\hat{\rho}^{(1,0)}+\hat{\rho}^{(0,1)}+\hat{\rho}^{(1,1)}+\hat{\rho}^{(2,0)}+\hat{\rho}^{(0,2)}+\mathcal{O}(\lambda^3)
\\
&=&\hat{\rho}_0+\hat{U}^{(1)}\hat{\rho}_0+\hat{\rho}_0\hat{U}^{\dagger(1)}+\hat{U}^{(1)}\hat{\rho}_0\hat{U}^{\dagger(1)}+\hat{U}^{(2)}\hat{\rho}_0+\hat{\rho}_0\hat{U}^{\dagger(2)}+\mathcal{O}(\lambda^3).
\end{eqnarray}
Since we are interested in the detectors' state, we will trace over the degree of freedom of the quantum field. Therefore, the evolved state of the detector can be written as 
\begin{eqnarray}\label{RDM}
\nonumber
\hat{\rho}_\text{AB}&=&\mathrm{Tr}_{\delta\hat{\rho}}[\hat{\rho}_T]=|g_\text{A}\rangle\langle g_\text{A}|\otimes|g_\text{B}\rangle\langle g_\text{B}|+\mathrm{Tr}_{\delta\hat{\rho}}[\hat{U}^{(1)}\hat{\rho}_0]+\mathrm{Tr}_{\delta\hat{\rho}}[\hat{\rho}_0\hat{U}^{\dagger(1)}]
\\
&&+\mathrm{Tr}_{\delta\hat{\rho}}[\hat{U}^{(1)}\hat{\rho}_0\hat{U}^{\dagger(1)}]+\mathrm{Tr}_{\delta\hat{\rho}}[\hat{U}^{(2)}\hat{\rho}_0]+\mathrm{Tr}_{\delta\hat{\rho}}[\hat{\rho}_0\hat{U}^{\dagger(2)}].
\end{eqnarray} 
Note that the terms $\mathrm{Tr}_{\delta\hat{\rho}}[\hat{U}^{(1)}\hat{\rho}_0]$ and $\mathrm{Tr}_{\delta\hat{\rho}}[\hat{\rho}_0\hat{U}^{\dagger(1)}]$ are dependent 
on the one-point correlator of the field, which for the vaccum and thermal state is zero. Therefore, we can find that the leading order corrections to the state
of the detectors should be at the second order, which depend on the two-point correlation function of the field,
\begin{eqnarray}
\mathcal{W}(t, \br, t^\prime, \br^\prime)=\langle\psi|\delta\hat{\rho}(t,\br)\delta\hat{\rho}(t^\prime,\br^\prime)|\psi\rangle,
\end{eqnarray}
also known as the Wightman function. In our case, the analogue quantum field is the density fluctuations of dipolar BEC, which  is of the form,
\begin{eqnarray}
\delta\hat{\rho}(t,\br)=\sqrt{\rho_0}\int\frac{d^2\bk}{(2\pi)^2}(u_\bk+v_\bk)\big[\hat{b}_\bk(t)e^{i\bk\cdot\br}+\hat{b}^\dagger_\bk(t)e^{-i\bk\cdot\br}\big].
\end{eqnarray}
This leads to the corresponding Wightman function as
\begin{eqnarray}
\mathcal{W}(t, \br, t^\prime, \br^\prime)=\frac{\rho_0}{(2\pi)^2}\int\,d^2\bk(u_{k}+v_{k})^2e^{-i\omega_\bk(t-t^\prime)+i\bk\cdot(\br-\br^\prime)}.
\end{eqnarray}

In the basis $\{|g_\text{A}\rangle|g_\text{B}\rangle, |g_\text{A}\rangle|e_\text{B}\rangle, |e_\text{A}\rangle|g_\text{B}\rangle, |e_\text{A}\rangle|e_\text{B}\rangle\}$,
the reduced density matrix of the detectors shown in \eqref{RDM} can be written as 
\begin{eqnarray}
\rho_\text{AB}=\begin{pmatrix}
1-P_\text{A}-P_\text{B}  & 0 & 0 & X    \\
0& P_\text{B} & C   &   0      \\
0& C^\ast & P_\text{A}   &   0    \\
X^\ast     &0    &0 &     0
\end{pmatrix}+\mathcal{O}(g^4),
\end{eqnarray}
where
the transition probability $P_\text{D}$ reads 
\begin{eqnarray}
\nonumber
P_\text{D}=\lambda^2\iint\,d\tau\,d\tau^\prime\chi(\tau)\chi(\tau^\prime)e^{-i\Omega(\tau-\tau^\prime)}\mathcal{W}(t_\text{D},\br_\text{D}, t^\prime_\text{D}, \br^\prime_\text{D}),~~~\text{D}\in\{\text{A}, \text{B}\},
\\
\end{eqnarray}
and the quantities $C$ and $X$, which characterize correlations, are respectively given by 
\begin{eqnarray}
C=\lambda^2\iint d\tau\chi(\tau^\prime)\chi(\tau^\prime)e^{-i\Omega(\tau-\tau^\prime)}\mathcal{W}(t_\text{A},\br_\text{A}, t^\prime_\text{B}, \br^\prime_\text{B})
\end{eqnarray}
and
\begin{eqnarray}
X=-\lambda^2\iint d\tau d\tau^\prime\chi(\tau)\chi(\tau^\prime)e^{-i\Omega(\tau+\tau^\prime)}\big[\theta(t^\prime-t)\mathcal{W}(t_\text{A}, \br_\text{A}, t_\text{B}^\prime, \br^\prime_\text{B})+\theta(t-t^\prime)\mathcal{W}(t_\text{B}^\prime, \br^\prime_\text{B}, t_\text{A}, \br_\text{A})\big].
\end{eqnarray}
Here $\theta$ represents the Heaviside theta function. Note that the detector's coordinate time is a function of its proper time in the above equations, i.e., $t=t(\tau)$.

\end{document}